\documentclass[11pt,a4paper]{article}
\usepackage{epsfig,axodraw}
\usepackage{array,cite,amsmath,amssymb}
\usepackage[flushleft]{threeparttable}
\usepackage{subcaption}
\usepackage{kantlipsum}
\usepackage{float}
\usepackage[usenames,dvipsnames]{xcolor}
\usepackage[breakable, theorems, skins]{tcolorbox}
\tcbset{enhanced}
\usepackage{booktabs}
\usepackage{caption}
\usepackage{threeparttable}
\usepackage{multirow}
\usepackage{listings}
\definecolor{DarkMagenta}{rgb}{0.55,0.0,0.55}
\newcommand{\newc}{\newcommand}
\newc{\gev}{\,GeV}
\newcolumntype{M}[1]{>{\centering\arraybackslash}m{#1}}
\newcolumntype{N}{@{}m{0pt}@{}}
\newc{\mev}{\,MeV}
\newc{\ra}{\rightarrow}
\newc{\rpv}{$\mathrm{\not\!R_p}$}
\newc{\rp}{$\mathrm{R_p}$}
\newc{\real}{\mathcal{R}e}
\newc{\alsm}{{\displaystyle \sum_{\alpha=1,2}}}
\newc{\besm}{{\displaystyle \sum_{\beta=1,2}}}
\newc{\al}{\alpha}
\newc{\sgn}{\mr{sgn}\,}
\newc{\be}{\beta}
\newc{\ga}{\gamma}
\newc{\de}{\delta}
\newc{\sla}{\!\!\!\!\!\not\:\:\!}
\newc{\slab}{\!\!\!\!\!\not\,\,\,}
\newc{\slac}{\!\!\!\!\!\!\!\not\,\,\,\,}
\newc{\met}{$\not\!\!E_T$}
\newc{\cw}{\cos\theta_W}
\newc{\sw}{\sin\theta_W}
\newc{\ssw}{\sin^2\theta_W}
\newc{\ccw}{\cos^2\theta_W}
\newc{\cbe}{\cos\beta}
\newc{\sbe}{\sin\beta}
\newc{\ort}{\frac1{\sqrt{2}}}
\newc{\sh}{\hat{s}}
\newc{\uh}{\hat{u}}
\newc{\tha}{\hat{t}}
\newc{\sa}{\sin\al}
\newc{\ca}{\cos\al}
\newc{\mz}{M_{\mr{Z}}}
\newc{\mw}{M_{\mr{W}}}
\newc{\bv}{$\mathrm{\not\!B}$}
\newc{\lv}{$\mathrm{\not\!L}$}
\newc{\beq}{\begin{equation}}
\newc{\eeq}{\end{equation}}
\newc{\ie}{{\it i.e.\/}\ }
\newc{\lam}{\lambda}
\newc{\cht}{\tilde{\chi}}
\newc{\glt}{\tilde{g}}
\newc{\upt}{\tilde{u}}
\newc{\qkt}{\tilde{q}}
\newc{\elt}{\tilde{\ell}}
\newc{\hgt}{\tilde{H}}
\newc{\nut}{\tilde{\nu}}
\newc{\dnt}{\tilde{d}}
\newc{\ftl}{\mr{\tilde{f}}}
\newc{\psb}{\bar{\psi}}
\newc{\rtt}{2^{1/2}}
\newc{\mut}{\tilde{\mu}}
\newc{\mr}{\mathrm}
\newc{\bath}{\bar{\theta}}
\newc{\tht}{\theta}
\newc{\JC}{{\bf J}}
\newc{\lra}{\longrightarrow}
\newc{\eg}{{\it e.g.\  }}
\newc{\barr}{\begin{eqnarray}}
\newc{\earr}{\end{eqnarray}}
\newc{\me}{\mathcal{M}}
\newc{\dbm}{\partial_\mu}
\newc{\dbmu}{\stackrel{\leftrightarrow\  }{\partial^\mu}}
\newc{\sgm}{\sigma_\mu}
\newc{\captionB}[2]{\caption[{#1}]{{\small {#2}}}}
\newc{\ahref}[2]{#2}
\pdfoutput=1 

\usepackage{array,cite,amsmath,color}
\usepackage{xspace}
\usepackage{bbm,subcaption}
\usepackage{slashed}

\usepackage{jheppub} 
\title{\boldmath Quarkonium Parton Shower in Herwig~7}

\author{M.R.~Masouminia,}
\author{P.~Richardson}

\affiliation{Institute for Particle Physics Phenomenology, Durham University, Durham, UK}

\emailAdd{mohammad.r.masouminia@durham.ac.uk}
\emailAdd{peter.richardson@durham.ac.uk}

\abstract{ \small
We present the implementation of a fully automated quarkonium parton shower in \textsf{Herwig~7}, based on non-relativistic QCD (NRQCD) factorisation with spin-colour projections. The framework systematically incorporates colour-singlet and colour-octet production mechanisms, gluon fragmentation, and diquark production processes. Perturbative short-distance coefficients are combined with non-perturbative NRQCD matrix elements to simulate heavy-quark bound state formation. Splitting functions for $S$-, $P$- and $D$-wave states are explicitly derived and integrated into the angular-ordered shower evolution. The implementation preserves spin correlations and polarisation effects while accurately accounting for feed-down contributions. Results demonstrate improved agreement with the existing LHC data. This quarkonium parton shower will become publicly available with the release of \textsf{Herwig-7.4.0}.
}

\begin{document}
\noindent{\hfill \small IPPP/25/06 \\[0.1in]}
\maketitle
\flushbottom

\section{Introduction}
\label{sec:intro}

Quarkonia, as flavourless bound states of heavy quarks and their corresponding antiquarks, are typically characterised by large binding energies and non-relativistic velocities. Such mesonic states are essential in probing the dynamics of QCD owing to their dual nature, encapsulating both perturbative and non-perturbative facets of the strong interaction~\cite{Lansberg:2006dh, Brambilla:2010cs}. Studying quarkonia provides critical insights into the nature of the QCD potential, particularly in the intermediate regime where neither purely perturbative approaches nor straightforward non-perturbative methods are fully reliable~\cite{Eichten:1979ms,Godfrey:1985xj}. The rich spectroscopy of quarkonium states, manifested by their narrow widths and a clear mass hierarchy, facilitates stringent tests of QCD-based predictions, including spin-dependent forces and the effects of gluon exchange~\cite{Kwong:1986ty,Buchmuller:1981aj}. Furthermore, quarkonium systems serve as a valuable testing ground for effective field theories such as Non-Relativistic QCD (NRQCD), which systematically incorporates relativistic corrections and higher-order QCD contributions, thereby providing a comprehensive framework for investigating heavy-quark dynamics~\cite{Eichten:1980mw,Bodwin:1994jh,Brambilla:2004jw}.

From an experimental perspective, quarkonium studies have long been a cornerstone of high-energy physics, offering key tests of QCD and insights into the strong interaction. The discovery of the $J/\psi$ meson at SLAC~\cite{SLAC-SP-017:1974ind} and Brookhaven~\cite{E598:1974sol, Aubert:1976oaa} in the 1970s was a seminal event, demonstrating the existence of the charm quark and ushering in the so-called ``November Revolution" in particle physics. Subsequent experiments at the Large Hadron Collider (LHC), Fermilab’s Tevatron, and electron-positron colliders such as LEP and Belle have comprehensively investigated various quarkonium states, covering both the charmonium and bottomonium families~\cite{CMS:2015lbl,ATLAS:2015zdw,D0:2008axd}. These endeavours include measurements of production cross-sections, decay rates, and polarisation observables, furnishing stringent tests of theoretical frameworks~\cite{LHCb:2013izl,ATLAS:2015zdw}. High-precision spectroscopy has exposed a myriad of states, encompassing singlet and triplet states, $S$-, $P$- and $D$-wave levels, and higher orbital excitations~\cite{Klopfenstein:1983nx,Aaij:2014bga}. The study of quarkonium production mechanisms in hadronic collisions, mediated through gluon-gluon fusion and quark-antiquark annihilation, has underscored the importance of both colour-singlet and colour-octet components as envisaged by NRQCD~\cite{Bodwin:1994jh}. Quarkonium decays into light hadrons, photons, and leptons further illuminate hadronisation processes and the gluon content of the proton~\cite{Abe:2002rb,Choi:2003ue}. Recent data from LHCb, ALICE, ATLAS, and CMS continue to refine our understanding, unveiling subtle QCD effects that challenge existing theoretical models~\cite{Lyubushkina:2020vro, CMS:2024wgu, LHCb:2024hrk, ALICE:2023hou}.

Despite significant progress, simulating quarkonium production within QCD-based event generators remains challenging due to the intricate interplay between perturbative and non-perturbative dynamics. NRQCD provides a systematic framework, where short-distance coefficients are computed perturbatively, while long-distance matrix elements (LDMEs) encode hadronisation effects~\cite{Bodwin:1994jh}\footnote{An alternative approach is the Colour Evaporation Model (CEM), which postulates that heavy quark pairs are produced perturbatively and subsequently neutralise their colour through soft-gluon emissions before hadronisation~\cite{Fritzsch:1977ay, Halzen:1977rs}. While CEM provides a simple, process-independent framework for estimating quarkonium yields, it historically lacked predictive power for spin and kinematic distributions. However, recent advancements have improved its accuracy by incorporating NLO corrections and scale-dependent fragmentation functions~\cite{Guiot:2023lkn,Lee:2022anw}, leading to more precise predictions of transverse momentum distributions and inclusive production rates. These refinements bring CEM closer to NRQCD-based approaches in certain kinematic regimes, enhancing its viability for phenomenology.}. To reconcile discrepancies between theoretical predictions and observed production rates, both colour-singlet and colour-octet contributions must be incorporated~\cite{Brambilla:2010cs}. While the colour-singlet mechanism captures direct production, it underestimates cross-sections, particularly at high transverse momentum, where gluon fragmentation and higher-order corrections become dominant~\cite{Braaten:1994vv}. The colour-octet mechanism accounts for these effects by allowing the heavy quark-antiquark pair to be produced in a coloured state, subsequently neutralising its colour through soft gluon emissions. This process enhances production rates, improving agreement with experimental data, as first observed at the Tevatron~\cite{CDF:1997ykw} and later confirmed at the LHC~\cite{CMS:2015lbl, ATLAS:2015zdw}. Embedding NRQCD within general-purpose Monte Carlo generators, particularly through parton showers that evolve directly from the hard process to hadronisation, enables a consistent treatment of quarkonium production in the broader context of hadronic events. This approach naturally accounts for feed-down from excited states, multiple parton interactions, and other complex QCD effects~\cite{CMS:2012tqw, Aaij:2014bga}. By providing fully differential predictions that can be directly compared with data, such simulations are needed for improving theoretical precision and refining searches for new physics~\cite{Andronic:2015wma}. 

In this paper, we present the implementation of a process-independent quarkonium parton shower in \textsf{Herwig~7}~\cite{Bahr:2008pv, Bellm:2015jjp, Bellm:2017bvx, Bellm:2019zci, Bewick:2023tfi}, designed within the NRQCD framework and incorporating both colour-singlet and colour-octet contributions. Perturbatively calculated short-distance coefficients are systematically combined with non-perturbative LDMEs to model heavy-quark bound states. The production amplitude accounts for spin and angular momentum states, with the Bethe-Salpeter wavefunction describing the bound state structure~\cite{Eichten:1980mw, Bodwin:1994jh}. Decays include both electromagnetic and strong processes, maintaining spin correlations and polarisation effects via helicity amplitudes and splitting functions within the parton shower algorithm. Feed-down effects from higher excited states are also considered. This parton shower includes a broad range of quarkonium production mechanisms: colour-singlet splittings such as $q \to q'\, \mathcal{O}_1$ and $g \to g\, \mathcal{O}_1$, colour-octet transitions via $g \to \mathcal{O}_8$, as well as diquark-producing splittings. This implementation supports the formation of mesonic and baryonic heavy-quark bound states in both colour-neutral and colour-charged configurations. It builds upon earlier developments of process-independent showers in \textsf{Herwig}, including the fully automated electroweak parton shower~\cite{Masouminia:2021kne} (extensively studied in~\cite{Darvishi:2021het,Darvishi:2020paz,Feng:2022inv,Darvishi:2022gqt,Campbell:2022qmc,Frixione:2022ofv}), its generalisation to a BSM parton shower~\cite{Lee:2023hef}, and the implementation of a dark sector shower based on the Hidden Valley model~\cite{Kulkarni:2024okx}. Recently, based on a similar approach, a process-independent quarkonium shower has been implemented in \textsf{Pythia8}~\cite{Sjostrand:2007gs, Sjostrand:2014zea}, facilitating comprehensive NRQCD-based simulations~\cite{Cooke:2023ldt}. We will provide comparisons between the \textsf{Herwig~7} quarkonium shower and the \textsf{Pythia8} implementation to assess the relative performance and theoretical consistency of both approaches.

The outline of this paper is as follows:  
Section~\ref{sec:formalism} describes the theoretical framework of NRQCD factorisation and its implementation in a parton shower context, including spin-colour projections and Bethe-Salpeter wavefunctions.  
Section~\ref{sec:splitting} presents the explicit derivation and kinematic structure of quarkonium splitting functions for various spin-orbital ${}^nL_J$ states, integrated into the angular-ordered evolution of the \textsf{Herwig~7} shower algorithm.  
Section~\ref{sec:matrixelementTuning} discusses the tuning of non-perturbative parameters to experimental data, where we use the perturbatively computed matrix elements (MEs) and measured decay widths to extract effective LDMEs for different quarkonium states.  
Section~\ref{sec:results} compares simulation outputs with LHC measurements across multiple final states and kinematic regions, highlighting significant improvements in both normalisation and differential distributions.  
Section~\ref{sec:conc} summarises the main developments, with an emphasis on the process-independent nature of the implementation and ongoing extensions within the \textsf{Herwig~7} framework.  
Appendix~\ref{sec:AppA} details the specific datasets and cross-section measurements used for tuning, covering both charmonium and bottomonium systems across LHC experiments.  
Appendix~\ref{sec:AppB} provides the full configuration interface within \textsf{Herwig~7}, outlining parameters, module dependencies, and customisation options for quarkonium parton shower evolution.

\section{NRQCD Factorisation}
\label{sec:formalism}

The production and decay of bound states, such as heavy quarkonia, can be systematically investigated through a combination of NRQCD and perturbative QCD. In the context of the \textsf{Herwig~7} general-purpose event generators, we aim to develop a framework that incorporates both the short-distance production mechanisms and the non-perturbative binding effects into a consistent framework. Below, we outline the formalism used to construct the amplitude for quarkonium production and decay, focusing on the role of the Bethe-Salpeter wavefunction and the necessary spin-colour projections.

Consider the production of a quarkonium bound state $\mathcal{O}_c({}^nL_J)$, characterised by principal quantum number $n$, spin $S$, orbital angular momentum $L$, total angular momentum $J$ and $c=1,8$ denoting colour-singlet and octet states, respectively. In the non-relativistic approximation, assuming the heavy-quark velocity $v \ll 1$ where the relativistic corrections are subleading, the factorised production amplitude can be expressed as~\cite{Bodwin:1994jh, Eichten:1995ch, Petrelli:1997ge}
\begin{equation}
  \mathcal{M}(p) \;=\; \sum_{L_Z,S_Z}\,\int \frac{\mathrm{d}^3\mathbf{k}}{(2\pi)^3} \,\psi_{LL_Z}(\mathbf{k}) \;\mathcal{A}(p,k)\,,
  \label{eq:prodAmplitude}
\end{equation}
where $p$ is the momentum of the bound state, and $L_Z$, $S_Z$, and $J_Z$ denote the projections of the orbital, spin, and total angular momenta, respectively. The integral is performed over the relative three-momentum $\mathbf{k}$ of the constituent quark and antiquark, while $\psi_{LL_Z}(\mathbf{k})$ represents the Bethe-Salpeter wavefunction in momentum space. This wavefunction is assumed to vary smoothly and slowly with $\mathbf{k}$, consistent with the heavy-quark non-relativistic regime. 

The short-distance amplitude, $\mathcal{A}(p,k)$ can be defined as:
\begin{equation}
  \mathcal{A}(p,k) \;=\; \bigl\langle L\,L_Z;\,S\,S_Z\,\big|\;J\,J_Z \bigr\rangle \; \Gamma_{S\,S_Z}(p,k)\,,
  \label{eq:shortDistance}
\end{equation}
which encodes the perturbative production of the quark-antiquark pair with a given Dirac structure $\Gamma$. Here, $\langle LL_Z;SS_Z|JJ_Z\rangle$ is the standard Clebsch-Gordan coefficient that combines the orbital and spin angular momentum states into a total angular momentum eigenstate. $\Gamma_{S,S_Z}(p,k)$ represents the spinor structure of the heavy quark and antiquark in the non-relativistic limit, reflecting the spin alignment of the quark pair and depending explicitly on the kinematics and quantum numbers of the produced quarkonium state. The explicit form of $\Gamma_{S,S_Z}(p,k)$ can be derived by projecting the spin states of the quark and antiquark. For $S$-wave quarkonium production, it reads
\begin{eqnarray}
  \Gamma_{S\,S_Z}^{\rm prod}(p,k) 
  &\;=\;& \sqrt{\frac{m_1 + m_2}{2\,m_1\,m_2}}
  \sum_{s, \bar{s}}
  \biggl\langle \tfrac{1}{2} s;\,\tfrac{1}{2} \bar{s} \;\bigg|\; S\,S_Z \biggr\rangle \;
  v(a_2\,p + k,\;\bar{s}) \;\overline{u}(a_1\,p - k,\;s)
  \nonumber\\[6pt]
  &\approx& \frac{\sqrt{m_1 + m_2}}{4\,m_1\,m_2}\;
  \bigl(a_2 \,\slashed{p} + \slashed{k} \;-\; m_2\bigr)\;
  \begin{pmatrix}
     \gamma_5 \\[4pt]
     -\,\slashed{\epsilon}\bigl(p,\,S_Z\bigr)
  \end{pmatrix}
  \bigl(a_1 \,\slashed{p} \;-\; \slashed{k} \;+\; m_1\bigr),
  \label{eq:gammaProd}
\end{eqnarray}
where $u$ and $v$ are the Dirac spinors for the heavy quark ($q_1$) and antiquark ($\bar{q}_2$) with masses $m_1$ and $m_2$, respectively. The parameters $a_1$ and $a_2$ represent the momentum fractions carried by the quark and antiquark, satisfying $a_1 + a_2 = 1$ in the centre-of-mass frame. $\epsilon(p,S_Z)$ is the polarisation vector for a spin-1 state; the lower (upper) entry in the column vector accommodates pseudoscalar (vector) states. In the non-relativistic limit, one typically neglects components of $k$ that are of higher order in $v$, simplifying the expression further. Meanwhile, for the decay of an $S$-wave quarkonium state, where the roles of the quark and antiquark spinors are exchanged compared to production, the analogous spinor structure becomes
\begin{eqnarray}
  \Gamma_{S\,S_Z}^{\rm decay}(p,k)
  &\;=\;& \sqrt{\frac{m_1 + m_2}{2\,m_1\,m_2}}
  \sum_{s, \bar{s}}
  \biggl\langle \tfrac{1}{2} s;\,\tfrac{1}{2} \bar{s} \;\bigg|\; S\,S_Z \biggr\rangle \;
  u(a_1\,p + k,\;\bar{s}) \;\overline{v}(a_2\,p - k,\;s)
  \nonumber\\[6pt]
  &\approx& \frac{\sqrt{m_1 + m_2}}{4\,m_1\,m_2}\;
  \bigl(a_2 \,\slashed{p} + \slashed{k} \;+\; m_2\bigr)\;
  \begin{pmatrix}
     \gamma_5 \\[4pt]
     -\,\slashed{\epsilon}\bigl(p,\,S_Z\bigr)
  \end{pmatrix}
  \bigl(a_1 \,\slashed{p} \;-\; \slashed{k} \;-\; m_1\bigr).
  \label{eq:gammaDecay}
\end{eqnarray}
This expression appears when the bound state dissociates into final-state quarks or partons, which then undergo further parton shower evolution and hadronisation in a Monte Carlo simulation.

Beyond the $S$-wave contribution, one must also include appropriate projections for higher orbital angular momentum ($L\!=\!1,2,\dots$) states,
\begin{subequations}
  \begin{eqnarray}
    \sum_{L_Z, S_Z} \bigl\langle 1\,L_Z;\,1\,S_Z\,\big|\;0\,0\bigr\rangle \;\epsilon^\alpha(L_Z)\,\epsilon^\beta(S_Z)
    &=& \sqrt{\frac{1}{3}}\;\mathcal{P}^{\alpha\beta},
    \\[4pt]
    \sum_{L_Z, S_Z} \bigl\langle 1\,L_Z;\,1\,S_Z\,\big|\;1\,J_Z\bigr\rangle \;\epsilon^\alpha(L_Z)\,\epsilon^\beta(S_Z)
    &=& \mathrm{i}\,\sqrt{\frac{1}{2}}\;\epsilon^{\alpha\beta\gamma\delta}\,\frac{p_\gamma}{(m_1 + m_2)}\;\epsilon_\delta(J_Z),
    \\[4pt]
    \sum_{L_Z, S_Z} \bigl\langle 1\,L_Z;\,1\,S_Z\,\big|\;2\,J_Z\bigr\rangle \;\epsilon^\alpha(L_Z)\,\epsilon^\beta(S_Z)
    &=& \epsilon^{\mu\nu}(J_Z).
  \end{eqnarray}
\end{subequations}
Similarly for the ${}^3D_J$ states, one can write
\begin{subequations}
  \begin{eqnarray}
    \sum_{L_Z,S_Z} \bigl\langle 2\,L_Z;1\,S_Z \big|1\,J_Z \bigr\rangle \epsilon^{\alpha\beta}(L_Z)\epsilon^\rho(S_Z) &=&
    -\sqrt{\frac{3}{20}}\left[\frac23\mathcal{P}^{\alpha\beta}\epsilon^\rho(J_Z)
    -\mathcal{P}^{\alpha\rho}\epsilon^\beta(J_Z)-\mathcal{P}^{\beta\rho}\epsilon^\alpha(J_Z)\right],\nonumber \\ \\
    \sum_{L_Z,S_Z} \bigl\langle 2\,L_Z;1\,S_Z \big|2\,J_Z \bigr\rangle \epsilon^{\alpha\beta}(L_Z)\epsilon^\rho(S_Z) &=& \frac{i}{\sqrt{6}(m_1+m_2)}
    \left[\epsilon^{\alpha\sigma}(J_Z)\epsilon^{\tau\beta\rho\mu}p_\tau g_{\sigma\mu} \right. \nonumber \\
      && \left. +\epsilon^{\beta\sigma}(J_Z)\epsilon^{\tau\alpha\rho\mu}p_\tau g_{\sigma\mu}   \right], \\
    \sum_{L_Z,S_Z} \bigl\langle 2\,L_Z;1\,S_Z \big|3\,J_Z \bigr\rangle \epsilon^{\alpha\beta}(L_Z)\epsilon^\rho(S_Z) &=& \epsilon^{\alpha\beta\rho}(J_Z).
  \end{eqnarray}
\end{subequations}
where $\mathcal{P}^{\alpha\beta}=g^{\alpha\beta}-p^\alpha p^\beta/(m_1+m_2)^2$ is the projection operator onto transverse Lorentz components. $\epsilon^\alpha(L_Z)$, $\epsilon^{\alpha\beta}(J_Z)$ and $\epsilon^{\alpha\beta\sigma}(J_Z)$ are symmetric and traceless tensors of ranks $1$, $2$ and $3$ respectively, while $\epsilon^{\alpha\beta\gamma\delta}$ is the four-dimensional Levi-Civita symbol. Analogous expressions exist for higher orbital angular momentum states, involving higher-rank polarisation tensors~\cite{Brambilla:2004jw, Bodwin:2007fz, Kuhn:1992qw}. These projection identities play a crucial role in NRQCD matching conditions, ensuring the separation of short-distance coefficients from the non-perturbative LDMEs.

The Bethe-Salpeter wavefunction $\psi_{L\,L_Z}(\mathbf{k})$ in Eq.~\eqref{eq:prodAmplitude} encapsulates the non-perturbative binding energy of the heavy quark-antiquark pair. In the non-relativistic limit, one typically writes this wavefunction in terms of partial waves ($S$-, $P$-, $D$-waves, etc.) with a radial part $R^{(n)}(0)$ evaluated at the origin. This is because the most relevant contributions to production and decay amplitudes come from small spatial separations of the quark-antiquark pair. One can write:
\begin{subequations}
\begin{eqnarray}
  \int \!\!\frac{\mathrm{d}^3\mathbf{k}}{(2\pi)^3}\;\psi_{0\,0}(\mathbf{k})
  &\;=\;& \frac{R(0)}{\sqrt{4\pi}},
  \label{eq:wavefunS}\\[6pt]
  \int \!\!\frac{\mathrm{d}^3\mathbf{k}}{(2\pi)^3}\;k^\alpha\,\psi_{1\,L_Z}(\mathbf{k})
  &\;=\;& -\,\mathrm{i}\,\sqrt{\frac{3}{4\pi}}\;R'(0)\;\epsilon^\alpha\!\bigl(p,L_Z\bigr),
  \label{eq:wavefunP}\\[6pt]
  \int \!\!\frac{\mathrm{d}^3\mathbf{k}}{(2\pi)^3}\;k^\alpha\,k^\beta\;\psi_{2\,L_Z}(\mathbf{k})
  &\;=\;& \sqrt{\frac{15}{8\pi}}\;R''(0)\;\epsilon^{\alpha\beta}\!\bigl(p,L_Z\bigr),
  \label{eq:wavefunD}
\end{eqnarray}
\end{subequations}
with $R(0)$, $R'(0)$, and $R''(0)$ (with the general notation of $R_{{}^nL}(0)$) denoting the radial wavefunction and its first and second derivatives, respectively, evaluated at the origin. These correspond to the $S$-, $P$-, and $D$-wave states, respectively, and govern the probability amplitude for the heavy quark pair to be found at small spatial separation in each case. These are closely related to the non-perturbative LDMEs in NRQCD and can be extracted from potential models~\cite{Buchmuller:1981aj, Eichten:1995ch}, lattice calculations~\cite{Lepage:1992tx, Bodwin:2007fz}, or global fits to experimental data~\cite{Cho:1995vh, Beneke:1996tk}. In physical processes, $R(0)$ is most relevant for $S$-wave states (e.g.\ $\eta_c$, $J/\psi$, $\Upsilon$), $R'(0)$ for $P$-wave states (e.g.\ $\chi_{cJ}$, $\chi_{bJ}$), and so forth.

Since heavy quarkonia of physical interest are colour-singlets, the projection onto the singlet channel is performed using $\delta_{ij}/\sqrt{N_C}$, where $i$ and $j$ are colour indices and $N_C=3$ is the number of colour charges in QCD. For colour-octet configurations, the projection operator becomes $\lambda^a_{ij}/\sqrt{2}$, where $\lambda^a$ are the Gell-Mann matrices in the fundamental representation. In NRQCD, such octet contributions are crucial to explain quarkonium production yields observed in experiments, particularly at high transverse momentum. Within \textsf{Herwig~7}, both singlet and octet channels can be implemented, where the short-distance coefficients are computed perturbatively, and the non-perturbative LDMEs are extracted from experimental fits, see Section~\ref{sec:matrixelementTuning}.

To leading order in the relative velocity $v$, the non-perturbative LDMEs in NRQCD, $\bigl\langle\mathcal{O}_1({}^nL_J)\bigr\rangle$, are related to the wavefunction at the origin (or its derivatives) via
\begin{subequations}
  \begin{eqnarray}
    \bigl\langle 0 \big|\; \mathcal{O}_1({}^3S_1) \;\big| 0 \bigr\rangle
    &\;=\;& \frac{3\,N_C}{2\pi}\;\bigl|R(0)\bigr|^2,
    \\[4pt]
    \bigl\langle 0 \big|\; \mathcal{O}_1({}^3P_J) \;\big| 0 \bigr\rangle
    &\;=\;& \frac{3\,N_C}{2\pi}\;(2J+1)\;\bigl|R'(0)\bigr|^2,
    \\[4pt]
    \bigl\langle 0 \big|\; \mathcal{O}_1({}^3D_J) \;\big| 0 \bigr\rangle
    &\;=\;& \frac{15\,N_C}{4\pi}\;(2J+1)\;\bigl|R''(0)\bigr|^2,
  \end{eqnarray}
\end{subequations}
The LDMEs of colour-singlet quarkonia, $\bigl\langle \mathcal{O}_1({}^nL_J) \bigr\rangle$, are typically fitted from experimental data or derived from potential models, lattice QCD, or QCD sum rules~\cite{Colquhoun:2014ica,Shifman:1978by}. They play a pivotal role in determining quarkonium decay widths and production cross-sections. While the wavefunction at the origin $R(0)$ and its derivatives naturally arise from the Bethe-Salpeter formalism in potential models, they also serve as useful proxies in NRQCD for estimating LDMEs. To leading order in the non-relativistic expansion, the LDMEs can be expressed in terms of $|R^{(n)}(0)|^2$ via matching procedures that connect potential model inputs to effective field theory operators. The identification is not exact but provides a practical and widely adopted scheme for incorporating non-perturbative effects into quarkonium phenomenology.

Incorporating Eq.~\eqref{eq:prodAmplitude} into a parton shower framework necessitates careful matching between the hard MEs, the non-relativistic wavefunction factors, and the colour-flow structure. In \textsf{Herwig~7}, the final-state parton shower must handle the evolution of the heavy quark and antiquark prior to bound-state formation, ensuring that both singlet and octet channels are handled appropriately, with octet states undergoing further soft gluon emissions before transitioning into a singlet bound state. The resultant quarkonium can then undergo further decays, governed by NRQCD decay MEs, facilitating realistic final-state modelling. Additionally, feed-down from excited quarkonium states and the interplay of higher-order corrections can be incorporated by extending the set of MEs and wavefunction parameters.

\section{Quarkonium Splitting Functions}
\label{sec:splitting}

In this section, we present the splitting functions required for quarkonium production in the parton shower framework of \textsf{Herwig~7}. These functions describe the probability that an initial parton (quark or gluon) evolves into a final state containing a heavy-quark bound state ($\mathcal{O}_c({}^nL_J)$) alongside additional partons. We outline the key steps in implementing such branchings, including the kinematic setup, relevant Feynman diagrams, amplitude expressions, and resulting splitting probabilities. Additionally, we discuss the integration of these processes into the \textsf{Herwig} event-generation workflow, emphasising the factorisation of the quarkonium amplitude into short-distance and non-perturbative components while preserving spin and colour correlations in the shower evolution.

\subsection{Parton Shower Kinematics}
\label{subsec:kinematics}

In a parton shower framework such as \textsf{Herwig~7}, the kinematics of each branching process must be defined carefully to ensure a consistent and physically meaningful evolution of momenta. Following the original approach used in \textsf{Herwig++}~\cite{Bahr:2008pv}, we employ quasi-collinear kinematics, parameterising the momenta of the branching particle and its decay products in a light-cone Sudakov basis. This ensures that emissions remain consistent with both the parton shower evolution variable and the constraints imposed by the shower algorithm.

Specifically, we describe the momenta of the progenitor and its children using the reference vectors $p$, $n$, and $q_\perp$, which allow us to smoothly transition between different evolution scales while preserving energy-momentum conservation and angular-ordering. In this framework, the momenta of the parent parton and its two outgoing partons are parameterised as:
\begin{subequations}
  \begin{eqnarray}
    q_0 & = & z\,p \;+\;\beta_0\,n, \\
    q_1 & = & z\,p \;+\;\beta_1\,n \;+\;q_\perp, \\
    q_2 & = & (1-z)\,p \;+\;\beta_2\,n \;-\;q_\perp,
  \end{eqnarray}
\end{subequations}
where $q_0$, $q_1$, and $q_2$ are the four-momenta of the parent and daughter partons, respectively. The vectors $p$, $n$, and $q_\perp$ are chosen to ensure a light-cone decomposition suitable for shower evolution, satisfying:
\begin{equation}
  p^2 \;=\; m_A^2,\quad
  n^2 \;=\;0,\quad
  n\cdot p \;\neq\;0,\quad
  n\cdot q_\perp \;=\;0,\quad
  p\cdot q_\perp \;=\;0,\quad
  q_\perp^2 \;=\;-\,p_\perp^2,
\end{equation}
where $m_A$ is the mass of the branching particle. The light-like reference vector $n$ ensures that the evolution respects angular ordering, consistent with Sudakov suppression, allowing consistent recoil assignments.

For massive final-state particles, the on-shell conditions for the daughters ($m_B$, $m_C$) impose additional constraints, leading to:
\begin{subequations}
  \begin{eqnarray}
    \beta_1 &=& \frac{p_\perp^2 + m_B^2 - z^2\,m_A^2}{2\,z\,p\cdot n},\\
    \beta_2 &=& \frac{p_\perp^2 + m_C^2 - (1-z)^2\,m_A^2}{2\,(1-z)\,p\cdot n},\\
    \beta_0 &=& \beta_1 + \beta_2 \;=\; \frac{1}{2\,p\cdot n}\,\Bigl[\frac{p_\perp^2}{z\,(1-z)} + \frac{m_B^2}{z} + \frac{m_C^2}{1-z} - m_A^2\Bigr]
    \;=\;\frac{q_0^2 - m_A^2}{2\,p\cdot n}.
  \end{eqnarray}
\end{subequations}
From these, we obtain the squared virtuality of the branching particle:
\begin{equation}
  q_0^2 \;=\;\frac{p_\perp^2}{z\,(1-z)} \;+\;\frac{m_B^2}{z} \;+\;\frac{m_C^2}{1-z}.
\end{equation}

To facilitate shower evolution, it is useful to express the dot products of key momenta in terms of virtuality:
\begin{subequations}
\begin{align}
  n\cdot q_1 &= z\,n\cdot p, 
  & p\cdot q_1 &= \tfrac12\Bigl[(1-z)\,q_0^2 \;+\;z\,m_A^2 \;+\;m_B^2 \;-\;m_C^2\Bigr],\\
  n\cdot q_2 &= (1-z)\,n\cdot p, 
  & p\cdot q_2 &= \tfrac12\Bigl[z\,q_0^2 \;+\;(1-z)\,m_A^2 \;-\;m_B^2 \;+\;m_C^2\Bigr],\\
  q_1\cdot q_2 &= \tfrac12\Bigl[q_0^2 - m_B^2 - m_C^2\Bigr].
\end{align}\label{eqn:dot}
\end{subequations}
The additional phase space for an extra emission factorises as
\begin{equation}
  \mathrm{d}\Phi_{n+1} \;=\; \mathrm{d}\Phi_n \;\frac{1}{4\,(2\pi)^3}\;\mathrm{d}\phi\;\mathrm{d}q_0^2\;\mathrm{d}z,
\end{equation}
where $\mathrm{d}\Phi_{n+1}$ and $\mathrm{d}\Phi_n$ are the $(n+1)$- and $n$-body phase-space measures, respectively, and $\phi$ is the azimuthal angle of the emission. This phase-space factorisation follows from the standard Lorentz-invariant form, where the emitted parton’s phase space is integrated over its virtuality, energy fraction $z$, and azimuthal angle $\phi$. This formulation ensures consistency with parton shower evolution by preserving angular-ordering constraints and energy-momentum conservation. The use of quasi-collinear kinematics allows a smooth transition between different evolution scales while maintaining a Sudakov-reweighted emission probability. Since the splitting probability $P(z, q_0^2)$ is directly related to the squared amplitude and phase-space measure, we define:
\begin{equation}
  P(z, q_0^2) = \frac{1}{2} \sum_{\lambda} \big| \mathcal{M}_{\lambda}(z, q_0^2) \big|^2 \times \Phi(z, q_0^2),
\end{equation}
where $\mathcal{M}_{\lambda}(z, q_0^2)$ is the helicity-dependent production amplitude, summed over final-state polarisations and averaged over initial-state ones. Moreover, $\Phi(z, q_0^2)$ encodes the phase-space measure, ensuring that the probability is properly normalised in the parton shower evolution.

The processes $q \to q'\,\mathcal{O}_1({}^nL_J)$, $g \to g\,\mathcal{O}_1({}^nL_J)$, and $g \to \mathcal{O}_8({}^nL_J)$ encode quarkonium production across distinct spin, orbital, and colour configurations. A ${}^3S_1$ state (e.g.\ $J/\psi$, $\Upsilon$) is a spin-triplet, $S$-wave state with $J=1$ and a non-vanishing wavefunction at the origin. Its production is enhanced by the absence of an angular momentum barrier and favourable spin-colour alignment, with gluon fragmentation dominating in the high $p_T$ tails, primarily via colour-octet channels. A ${}^1P_1$ state (e.g.\ $h_c$, $h_b$) is a spin-singlet, $P$-wave configuration with vanishing wavefunction at the origin; production scales with $R'(0)$ and is suppressed by the angular momentum barrier, with feed-down from higher states often contributing significantly. A ${}^1S_0$ state (e.g.\ $\eta_c$, $\eta_b$) is also spin-singlet and $S$-wave, but its production is generally helicity-suppressed in processes such as $q \to q'\,\mathcal{O}_1({}^1S_0)$, due to spin constraints in the massless QCD limit. For $g \to g\,\mathcal{O}_1({}^1S_0)$, colour-octet fragmentation dominates at high $p_T$, where the gluon’s spin and colour structure preferentially couple to octet configurations. Spin-triplet $P$- and $D$-wave transitions (\mbox{${}^3P_{\,0,1,2}$}, \mbox{${}^3D_{\,1,2,3}$}) are also incorporated for both quark and gluon progenitors, in singlet and octet configurations. Their non-zero orbital angular momentum introduces derivative couplings and enables rich feed-down cascades. This comprehensive set of $S$-, $P$-, and $D$-wave modes ensures the parton shower spans all relevant quarkonium states up to $D$-wave, capturing both perturbative production and non-perturbative neutralisation within a consistent NRQCD framework.

Although colour-singlet transitions determine the observable bound-state quantum numbers, colour-octet channels are essential to capturing the complete production dynamics. In NRQCD, octet states such as $\mathcal{O}_8({}^3S_1)$ and $\mathcal{O}_8({}^3P_J)$ contribute at the same order in $v^2$ as singlet states and dominate in specific kinematic regimes, notably at large transverse momentum. The shower framework includes explicit $g \to \mathcal{O}_8$ transitions to ensure their contributions are correctly modelled and tunable via associated LDMEs.

\subsection{Colour-Singlet $q\to q'\,\mathcal{O}_1$ Splittings}
\label{sebsec:qqO}

\begin{figure}[h!]
  \begin{center}
    \includegraphics[width=0.4\textwidth]{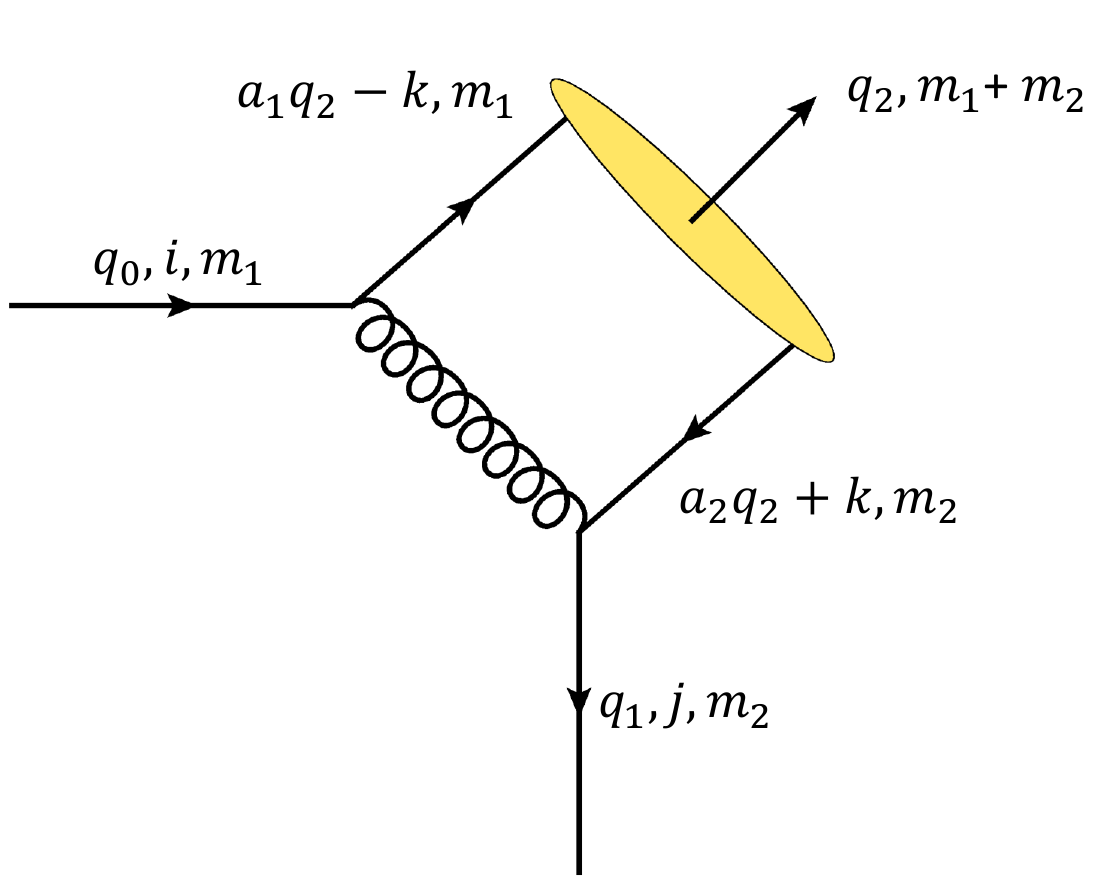}
  \end{center}
  \caption{\small Feynman diagram for colour-singlet $q\to q'\,\mathcal{O}_1$ splittings.}
  \label{fig:qtoqO}
\end{figure}

The Feynman diagram for the production of a colour-singlet quarkonium state from an initial quark is shown in Fig.~\ref{fig:qtoqO}, where a parent quark of mass $m_1$ branches into a daughter quark $q'$ with mass $m_2$ plus a heavy quark-antiquark pair that forms the quarkonium bound state with mass $M=m_1+m_2$. We can write the short-distance QCD amplitude for a generic $q\to q'\,\mathcal{O}_1$ process as
\begin{eqnarray}
  \mathcal{A}(q\to q'\,\mathcal{O}_1) &=& 
  \frac{g_S^2{\bf t}^a_{ki}{\bf t}^a_{jl}}{\left(q_0^2-m_1^2\right)\left(a_2q_2+k+q_1\right)^2} 
  \bar{u}_k(a_1q_2-k,m_1)\gamma^\mu u_i(q_0)
  \bar{u}_j(q_1)\gamma^\nu v_l(a_2q_2+k,m_2)
  \nonumber\\
  &\times& \left[ g_{\mu\nu} - \frac{(a_2q_2+k+q_1)^\alpha n^\beta   
  + [\mu\leftrightarrow\nu]}{n\cdot(a_2q_2+k+q_1)} \right],
  \label{AqS}
\end{eqnarray}
where $g_S$ is the strong coupling. Assuming $q_2$ as the momentum of the produced quarkonium, one can write the momenta of its constituent quarks as $a_1 q_2 - k$ and $a_2 q_2 + k$, with the momentum fractions $a_1+a_2 = 1$. In Eq.~\eqref{AqS}, ${\bf t}^a$ are the generators of the SU(3) colour group in the fundamental representation, satisfying the algebra $[{\bf t}^a, {\bf t}^b] = i f^{abc} {\bf t}^c$. These can be written in terms of the Gell-Mann matrices as ${\bf t}^a_{ij} = (\lambda^a)_{ij}/2$. The quarkonium state is projected onto the colour-singlet by inserting the operator $\delta_{ij}/\sqrt{N_C}$. Summing over the heavy-quark spins ensures that the short-distance NRQCD ME is properly factorised, isolating the non-perturbative contributions into the wavefunction-dependent term $R(0)$ for the $S$-wave states. 

The effective production amplitude of this branching process can be written in its most general form as:
\begin{eqnarray}
    \mathcal{M}(q &\to& q'\,\mathcal{O}_1({}^nL_J)) = 
    \frac{g_S^2 C_F \; \big\langle \mathcal{O}_1({}^nL_J) \big \rangle \; \delta_{ij}}{\sqrt{N_C} \left( q_0^2 - a_1^2 M^2 \right)}
    \bar{u}_j(q_1, m_2) \gamma^\nu \; \Pi_{{}^nS} \; \gamma^\mu u_i(p, m_1)
    \\
    &\times& \left[ g_{\mu\nu} - \frac{(q_1^\mu + a_2 q_2^\mu + k^\mu) (q_1^\nu + a_2 q_2^\nu + k^\nu) + (q_0^2 - a_1^2 M^2) g_{\mu\nu}}
    {\left( q_1 \cdot n + a_2 q_2 \cdot n + k \cdot n \right) 
    \left( k^2 + q_1^2 + 2 a_2 q_1 \cdot q_2 + 2 q_1 \cdot k + 2 a_2 q_2 \cdot k + 2 a_2 q_1 \cdot q_2 \right) } \right], \nonumber 
\end{eqnarray}
with $C_F$ as the colour factor and the projection operator $\Pi_{nS}$ being defined with:
\begin{equation}
\begin{pmatrix}
\Pi_{{}^1S} \\ \Pi_{{}^3S}
\end{pmatrix}  
= \frac{\sqrt{M}}{4 m_1 m_2} 
\left( a_2 \slashed{q_2} + \slashed{k} - m_2 \right) 
\begin{pmatrix}
\gamma^5 \\ \slashed{\epsilon}(q_2)
\end{pmatrix} 
\left( a_1 \slashed{q_2} - \slashed{k} + m_1 \right).
\end{equation}
Having the above information, one can calculate the differential branching probability of a given ${}^nL_J$ quarkonium state as:
\begin{equation}
  {\rm d}P_{q \to q' \mathcal{O}_1({}^nL_J)} = {\rm d}z {\rm d}q_0^2\frac{{\rm d}\phi}{2\pi} 
  F_{q \to q' \mathcal{O}_1({}^nL_J)}(z,q^2_0),
  \label{prob}
\end{equation}
where $F_{{}^nL_J}(z,q^2_0)$ is the spin-squared/averaged and colour-averaged splitting probability function via:
\begin{equation}
    F_{q \to q' \mathcal{O}_1({}^nL_J)}(z,q^2_0) = {1 \over 2} \sum_{\text{spins}} 
    {\left| \mathcal{M}({q \to q' \mathcal{O}_1({}^nL_J))} \right|^2 }.
    \label{Fgeneric}
\end{equation}

\subsubsection{$q\to q'\,\mathcal{O}_1({}^1S_0)$ Splittings}

For the spin-singlet $S$-wave case, ${}^1S_0$, the spinor projection is implemented by $\gamma_5$, ensuring the quark-antiquark pair is in the pseudoscalar channel. This yields an amplitude of the form:
\begin{eqnarray}
  \mathcal{M}(q \to q'\mathcal{O}_1({}^1S_0)) &=& 
  \frac{g_S^2\,R(0)\,C_F\,\delta_{ij}}{2\,\sqrt{N_C\,\pi\,M}\,\bigl(1-a_1\bigr)\,\bigl(q_0^2 - m_1^2\bigr)}
  \bigg[
   \frac{M}{\bigl(q_0^2 - m_1^2\bigr)}\;\bar{u}_j(q_1)\,\gamma_5\,u_i(p)
  \nonumber\\
    &+&   \frac{1+(1-a_1)(1-z)}{2(1 - a_1(1-z))\,n\cdot p}
  \bar{u}_j(q_1)\,\slashed{n}\,\gamma_5\,u_i(p)
    \bigg],
    \label{ME1S0}
\end{eqnarray}
where $R(0)$ has dimensions of GeV${}^{3/2}$. Combining Eqs.~\eqref{ME1S0} and~\eqref{Fgeneric}, results in the explicit expression,
\vspace{-0.1in}
\begin{eqnarray}
F_{q \to q'\mathcal{O}_1({}^1S_0)}(z,q_0^2) &=&
   \frac{g_S^4 |R(0)|^2 C_F^2}{4 \pi M N_C D_1}
   \Bigg[
   a_1^6 M^4 (z-1)^2 z
   + 2 a_1^5 M^4 (2 z^2 - 3 z + 1)
   \nonumber \\
   &+& a_1^4 \Big( M^4 (z^2 + 5z - 5) - 2 M^2 q_0^2 (z-1)^2 z \Big)
   \nonumber \\
   &-& 2 a_1^3 \Big( M^4 (z-3) z - M^2 q_0^2 (z^3 - 5z^2 + 5z - 1) \Big)
   \nonumber \\
   &+& a_1^2 \Big( M^4 (7 - 5z) - M^2 q_0^2 (z^3 - 7z^2 + 17z - 9)
   + q_0^4 (z-1)^2 z \Big)
   \nonumber \\
   &-& 2 a_1 \Big( 2 M^4 + M^2 q_0^2 (z^2 - 5z + 6) + q_0^4 z (z^2 - 3z + 2) \Big)
   \nonumber \\
   &+& q_0^2 \Big( M^2 (5 - 3z) + q_0^2 (z-2)^2 z \Big)
   \Bigg],
   \label{P1S0EX}
\end{eqnarray}
with $m_1 = a_1 M$, $m_2 = a_2 M$ and $a_2 = 1 - a_1$ and
$$
D_1 = (a_1-1)^2(a_1 (z-1)+1)^2 \left(q_0^2-a_1^2 M^2\right)^4.
$$
These processes have been implemented in \textsf{Herwig~7} via the splitting classes \textcolor{blue}{\texttt{QtoQP1S0SplitFn}} ($c \to b + B_c^+$ and $b \to c + B_c^-$) and \textcolor{blue}{\texttt{QtoQ1S0SplitFn}} ($c \to c + \eta_c$ and $b \to b + \eta_b$). In the latter case, since $m_1=m_2$ and $a_1 = 0.5$, one can further simplify the Eq.~\eqref{P1S0EX} to
\begin{eqnarray}
F_{q \to q\mathcal{O}_1({}^1S_0)}(z,q_0^2) &=&
\frac{16 g_S^4 |R(0)|^2 C_F^2}{\pi M N_C (z+1)^2 \left(M^2 - 4 q_0^2\right)^4}
\Bigg[
M^4 \left(z^3 - 6z^2 - 23z - 32\right)
\nonumber \\
&-& 8 M^2 q_0^2 \left(z^3 + 2z^2 + 9z - 8\right)
+ 16 q_0^4 (z-3)^2 z
\Bigg],
\end{eqnarray}
which is consistent with similar calculations for the fragmentation of charm quarks into quarkonia states, carried out in Ref.~\cite{Braaten:1993mp}. 

\subsubsection{$q\to q'\,\mathcal{O}_1({}^3S_1)$ Splittings}

For a spin-triplet $S$-wave state, ${}^3S_1$, the spinor projection employs a vector polarisation $\epsilon^\mu$. Then, the production amplitude takes the form
\begin{eqnarray}
  \mathcal{M}(q \to &q& \mathcal{O}_1({}^3S_1)) =
  \frac{g_S^2\,R(0)\,C_F}{a_2\,\sqrt{\pi M N_C}}
  \;\frac{1}{q_0^2 - m_1^2}
  \Biggl[
  \frac{(1-z)}{4\bigl(1-a_1(1-z)\bigr)\,n\cdot q_0}\;\bar{u}(q_1)\,\slashed{n}\,\slashed{\epsilon}(q_2)\,u(q_0)\nonumber\\
  &+& \;\frac{z\,n\cdot \epsilon(q_2)}{2\bigl(1 - a_1\,(1-z)\bigr)\,n\cdot q_0}\;\bar{u}(q_1)\,u(q_0)
  \;+\;\frac{M}{2\bigl(q_0^2 - m_1^2\bigr)}\;\bar{u}(q_1)\,\slashed{\epsilon}(q_2)\,u(q_0)
  \Biggr], 
\end{eqnarray}
and hence, the corresponding averaged splitting probability reads
\begin{eqnarray}
F_{q \to q' \mathcal{O}_1({}^3S_1)}(z,q^2_0) &=&
   \frac{g_S^4 |R(0)|^2 C_F^2}{4 \pi M N_C  D_1} 
   \Bigg[
   a_1^6 M^4 (z-1)^2 z
   - 2 a_1^5 M^4 \left(2 z^2-5 z+3\right) 
   \nonumber \\
   &+& a_1^4 \Big(M^4 \left(11 z^2-31 z+21\right)-2 M^2 q_0^2 (z-1)^2 z \Big) 
   \nonumber \\
   &-& 2 a_1^3 \Big(M^4 \left(3 z^2-19 z+18\right)+M^2 q_0^2 \left(-z^3+z^2+3 z-3\right) \Big) 
   \nonumber \\
   &+& a_1^2 \Big(M^4 (33-15 z)+M^2 q_0^2 \left(-3 z^3+9 z^2+z-9\right)+q_0^4 (z-1)^2 z \Big) 
   \nonumber \\
   &-& 2 a_1 \Big(6 M^4+M^2 q_0^2 z (3 z-5)+q_0^4 z \left(z^2-3 z+2\right) \Big) 
   \nonumber \\
   &+& q_0^2 \Big( M^2 (3-9 z)+q_0^2 z \left(3 z^2-8 z+6\right) \Big)
   \Bigg].
\end{eqnarray}
For non-degenerate mass configurations, the splitting $q \to q'\,\mathcal{O}_1({}^3S_1)$ encompasses two distinct processes: $c \to b + B_c^{\star +}$ and $b \to c + B_c^{\star -}$. In the equal-mass limit, this reduces to $c \to c + J/\psi$ and $b \to b + \Upsilon$. The expression above can therefore be simplified by restricting to the \textit{equal-mass case}:
\begin{eqnarray}
F_{q \to q\mathcal{O}_1({}^3S_1)}(z,q_0^2) &=&
   \frac{16 g_S^4 |R(0)|^2 C_F^2}{\pi  M N_C (z+1)^2 \left(M^2-4 q_0^2\right)^4}
   \Bigg[
   M^4 \left(z^3 - 14 z^2 - 39 z - 72\right) 
   \nonumber \\
   &-& 8 M^2 q_0^2 \left(5 z^3 + 6 z^2 + 37 z - 12\right) 
   + 16 q_0^4 z \left(9 z^2 - 22 z + 17\right)
   \Bigg].
\end{eqnarray}
The splitting classes \textcolor{blue}{\texttt{QtoQP3S1SplitFn}} and \textcolor{blue}{\texttt{QtoQ3S1SplitFn}} handle these types of splitting in \textsf{Herwig~7} quarkonium parton shower.

\subsubsection{$q\to q'\,\mathcal{O}_1({}^1P_1)$ Splittings}

The inclusion of $q\to q'\mathcal{O}_1({}^1P_1)$ splittings in the quarkonium parton shower extends heavy-flavour jet evolution by enabling the production of axial-vector quarkonium states, such as the spin-singlet $h_c$ and $h_b$. Unlike $S$-wave quarkonia, whose production is enhanced by a non-vanishing wavefunction at the origin, $P$-wave states have vanishing wavefunction values at the origin and instead require derivative contributions. This introduces a suppression in the splitting kernel, proportional to the square of the radial wavefunction derivative, $|R'(0)|^2$. Consequently, the production probability of $P$-wave quarkonia in the parton shower is reduced and features a different angular and energy dependence compared to $S$-wave modes.

In \textsf{Herwig~7}, the axial-vector singlet splittings $c \to c + h_c$ and $b \to b + h_b$ are implemented via the \textcolor{blue}{\texttt{QtoQ1P1SplitFn}} splitting class\footnote{In heavy-light systems such as $B_c$ mesons, the physical $1^+$ states can also arise from a mixing of ${}^1P_1$ and ${}^3P_1$ configurations, and are treated within the parton shower through the same axial-vector splitting class.}. For these states, we incorporate the first derivative of the radial wavefunction at the origin, $R'(0)$, due to the orbital angular momentum $L=1$. The resulting amplitude is
\begin{eqnarray}
  \mathcal{M}(q &&\to q'\mathcal{O}_1({}^1P_1)) =
  \mathrm{i}\,\sqrt{\frac{3\,M}{\pi\,N_C}}\;
  \frac{g_S^2\,R'(0)\,C_F}{a_2^2\,\bigl(q_0^2 - m_1^2\bigr)} \sum_{l=1}^6 \bar{u}_j(q_1) \mathcal{H}_l(z,q_0^2) u_i(p),
\end{eqnarray}
with $\mathcal{H}_l(z,q_0^2)$ parametrising the helicity terms of the produced quarkonia as:
\allowdisplaybreaks{
\begin{eqnarray}
    \mathcal{H}_1(z,q_0^2) &=& \frac{1 - 2\,a_1}{4\,a_1\,M\,\bigl(q_0^2 - m_1^2\bigr)} \slashed{\epsilon}(q_2)\,\gamma_5 ,
    \\
    \mathcal{H}_2(z,q_0^2) &=& \frac{a_1^2(1-z)^2 - a_1 \bigl(z^2 - 5z + 4\bigr) - 2z + 3}{4\,M\bigl(1 - a_1\,(1-z)\bigr)^2} 
      { n\cdot \epsilon(q_2) \over n\cdot p} \slashed{n}\,\gamma_5 ,
    \\
    \mathcal{H}_3(z,q_0^2) &=& \Bigl(\frac{1}{q_0^2 - m_1^2} + \frac{z}{2\,a_1\,M^2\,\bigl(1-a_1\,(1-z)\bigr)}\Bigr)\frac{n\cdot\epsilon(q_2)}{2\,n\cdot p} \gamma_5
    \\
    \mathcal{H}_4(z,q_0^2) &=& \frac{(1-z)}{8\,a_1\,M^2\,\bigl(1 - a_1\,(1-z)\bigr)} {\slashed{n}\,\slashed{\epsilon}(q_2) \over n\cdot p} \gamma_5,
    \\
    \mathcal{H}_5(z,q_0^2) &=& \frac{\bigl(2 - z - a_1\,(1-z)\bigr)}{2\,M\,\bigl(1-a_1\,(1-z)\bigr)\,\bigl(q_0^2 - m_1^2\bigr)}
    {p\cdot\epsilon(q_2) \over n\cdot p} \slashed{n}\,\gamma_5,
    \\
    \mathcal{H}_6(z,q_0^2) &=&\frac{1}{\bigl(q_0^2 - m_1^2\bigr)^2} p\cdot\epsilon(q_2) \; \gamma_5 .
\end{eqnarray}}
Using the above amplitude, one can write the averaged splitting probability for the relevant equal-mass $q\to q\,\mathcal{O}_1({}^1P_1)$ splittings as follows
\begin{eqnarray}
  F_{q\to q\,\mathcal{O}_1({}^1P_1)}(z,q^2_0) &=& \frac{192 g_S^4 |R'(0)|^2 C_F^2}{\pi  M^3 N_C (z+1)^4 \left(M^2-4 q_0^2\right)^5}
  \Big[
   M^6 \left(9 z^5-12 z^4-2 z^3-292 z^2-543 z-296\right)
   \nonumber \\
   &-&4 M^4 q_0^2 \left(27 z^5-12 z^4+442 z^3+644 z^2+163 z-192\right)
   \nonumber \\
   &+&16 M^2 q_0^4 \left(27 z^5-132 z^4-70 z^3+148 z^2+227 z-24\right)
   \nonumber \\
   &-&64 q_0^6 z \left(9 z^4-4 z^3-2 z^2-20 z+33\right)
  \Big]
\end{eqnarray}
This result is identical to the similar calculation carried out in Ref.~\cite{Chen:1993ii}.

\subsubsection{$q \to q'\, \mathcal{O}_1({}^3P_J)$ Splittings}

Quarkonium states with quantum numbers ${}^3P_J$ $(J=0,1,2)$ play a critical role in the structure of the angular-ordered parton shower due to their non-zero orbital angular momentum content. These states are spin-triplet ($S=1$), $P$-wave ($L=1$) configurations of a heavy quark-antiquark pair and contribute significantly to feed-down into lower-lying $S$-wave states. As such, their proper inclusion is essential for accurately modelling the production spectrum of quarkonia in Monte Carlo simulations such as \textsf{Herwig~7}.

The ${}^3P_J$ states are described by a derivative acting on the ${}^3S_1$ spin projector: 
\begin{equation}
\mathcal{M}(q \to q' \mathcal{O}_1({}^3P_J)) \;\propto\; \epsilon^{(\lambda)}_{\mu} \left. \frac{\partial}{\partial k_\mu} \mathcal{M}(q \to q' \mathcal{O}_1({}^3S_1))(k) \right|_{k=0},
\end{equation}
where $\epsilon^{(\lambda)}_\mu$ is the polarisation tensor corresponding to the total angular momentum $J$. This derivative structure emerges from the expansion of the Bethe-Salpeter wavefunction around the origin in relative momentum space, leading to a $P$-wave suppression proportional to the derivative of the radial wavefunction, $R'(0)$. The multiplicative prefactor includes $\sqrt{3/4\pi}$ to ensure proper angular normalisation in the NRQCD formalism.

In the context of \textsf{Herwig~7}, the $q \to q'\, \mathcal{O}_1({}^3P_J)$ splittings are implemented through dedicated splitting kernels, which encapsulate the helicity-dependent structure of the $P$-wave bound states and their evolution under angular ordering. These include:
\begin{itemize}
\item[--] $c \to c\,\chi_{c0}$ and $b \to b\,\chi_{b0}$ for $J = 0$, via the \textcolor{blue}{\texttt{QtoQ3P0SplitFn}} class,
\item[--] $c \to b\,B_{c0}^{\star +}$ and $b \to c\,B_{c0}^{\star -}$ for $J = 0$, via the \textcolor{blue}{\texttt{QtoQP3P0SplitFn}} class,
\item[--] $c \to c\,\chi_{c1}$ and $b \to b\,\chi_{b1}$ for $J = 1$, via the \textcolor{blue}{\texttt{QtoQ3P1SplitFn}} class,
\item[--] $c \to b\,B_{c1}^{+}$, $b \to c\,B_{c1}^{-}$, $c \to b\,B_{c1}^{\prime +}$ and $b \to c\,B_{c1}^{\prime  -}$, for $J = 1$, via the \textcolor{blue}{\texttt{QtoQPP1SplitFn}} class,
\item[--] $c \to c\,\chi_{c2}$ and $b \to b\,\chi_{b2}$ for $J = 2$, via the \textcolor{blue}{\texttt{QtoQ3P2SplitFn}} class,
\item[--] $b \to c\,B_{c2}^-$ and $c \to b\,B_{c2}^+$ for $J = 2$, via the \textcolor{blue}{\texttt{QtoQP3P2SplitFn}} class.
\end{itemize}
Each of these splittings accounts for the specific Lorentz structure of the corresponding polarisation tensor and its contraction with the derivative of the short-distance amplitude. The corresponding averaged splitting probability and splitting functions in the same-flavour limit admit compact analytic expressions, which we present below. The general cases with unequal masses lead to significantly more cumbersome formulae and are therefore omitted.
\allowdisplaybreaks{
\begin{eqnarray}
  F_{q\to q\,\mathcal{O}_1({}^3P_0)}(z,q^2_0) &=& \frac{64 g_S^4 |R'(0)|^2 C_F^2 }{\pi  M^3 N_C (z+1)^4 \left(M^2-4 q_0^2\right)^5}
  \Big[
   - 64 q_0^6 z \left(z^2-2 z+5\right)^2
   \nonumber \\
   &+&16 M^2 q_0^4 \left(19 z^5+76 z^4+266 z^3-76 z^2-357 z-264\right)
   \nonumber \\
   &-&4 M^4 q_0^2 \left(67 z^5+308 z^4+362 z^3-1020 z^2-1845 z-896\right)
   \nonumber \\
   &+&M^6 \left(49 z^5+100 z^4-402 z^3-1732 z^2-1975 z-760\right)
   \Big],
\\
  F_{q\to q\,\mathcal{O}_1({}^3P_1)}(z,q^2_0) &=& \frac{384 g_S^4 |R'(0)|^2 C_F^2 }{\pi  M^3 N_C (z+1)^4 \left(M^2-4 q_0^2\right)^5}
  \Big[
   -64 q_0^6 z \left(z^4-4 z^3+6 z^2-4 z+17\right)
   \nonumber \\
   &+&16 M^2 q_0^4 \left(5 z^5-22 z^4+102 z^3+160 z^2+93 z-34\right)
   \nonumber \\
   &-&4 M^4 q_0^2 \left(15 z^5+96 z^4+426 z^3+364 z^2-49 z-164\right)
   \nonumber \\
   &+&M^6 \left(11 z^5+50 z^4+74 z^3-184 z^2-381 z-194\right)
   \Big],
\\
  F_{q\to q\,\mathcal{O}_1({}^3P_2)}(z,q^2_0) &=& \frac{128 g_S^4 |R'(0)|^2 C_F^2 }{\pi  M^3 N_C (z+1)^4 \left(M^2-4 q_0^2\right)^5}
  \Big[
   -64 q_0^6 z \left(z^4-4 z^3+38 z^2-68 z+49\right)
   \nonumber \\
   &+&16 M^2 q_0^4 \left(121 z^5-98 z^4-82 z^3+104 z^2+417 z-30\right)
   \nonumber \\
   &-&4 M^4 q_0^2 \left(151 z^5+200 z^4+1178 z^3+1596 z^2+663 z-284\right)
   \nonumber \\
   &+&M^6 \left(31 z^5-26 z^4+18 z^3-496 z^2-985 z-574\right)
  \Big].
\end{eqnarray}}
The above expressions are in complete agreement with similar derivations in~\cite{Yuan:1994hn,Chen:1993ii}. 

In heavy-quarkonium systems, mixing between spin-singlet and spin-triplet states can play a non-negligible role in observables, particularly for the physical $h_c$ and $\chi_{c1}$, $h_b$ and $\chi_{b1}$, or $B_{c1}$ and $B_{c1}^\prime$ mesons, where the mass splittings are small and the quantum numbers allow such mixing. This phenomenon arises from spin-dependent interactions that weakly break heavy-quark spin symmetry, especially in the presence of relativistic corrections or loop-induced effects. Here, we consider a mixed physical state defined as
\begin{equation}
    \mathcal{O}_1^\text{mixed}(P_1) = \cos\theta_{P_1}\, \mathcal{O}_1({}^1P_1) + \sin\theta_{P_1}\, \mathcal{O}_1({}^3P_1),
    \nonumber
\end{equation}
where $\theta_{P_1}$ is the mixing angle. The amplitude for the corresponding mixed splitting is given by
\begin{equation}
    \mathcal{M}_{[\mathcal{O}_1({}^1P_1)+\mathcal{O}_1({}^3P_1)]} = \cos\theta_{P_1}\, \mathcal{M}_{{}^1P_1} + \sin\theta_{P_1}\, \mathcal{M}_{{}^3P_1}.
    \nonumber
\end{equation}
The squared amplitude, after averaging over initial-state spin and summing over final-state polarisation, is then
\begin{equation}
    |\mathcal{M}_{[\mathcal{O}_1({}^1P_1)+\mathcal{O}_1({}^3P_1)]}|^2 = \cos^2\theta_{P_1}\, |\mathcal{M}_{{}^1P_1}|^2 + \sin^2\theta_{P_1}\, |\mathcal{M}_{{}^3P_1}|^2 + 2\cos\theta_{P_1}\sin\theta_{P_1}\, \text{Re}\left(\mathcal{M}_{{}^1P_1}^* \mathcal{M}_{{}^3P_1}\right),
    \nonumber
\end{equation}
where the interference term accounts for the quantum coherence between the two components. The splitting kernel is computed as a function of $z$, $q_0^2$, and the mixing angle $\theta_{P_1}$.

In our setup, the $q \to q'\, \mathcal{O}_1({}^1P_1)$ and $q \to q'\, \mathcal{O}_1({}^3P_1)$ amplitudes are calculated independently, each with their own spinor structures and projection tensors. The mixed amplitude is constructed by linearly combining these components with mixing weights, using the default mixing angle $\theta_{P_1}=25^\circ$, and the resulting probability is obtained by contracting with the appropriate external polarisation tensors. The explicit expression for this contraction is implemented numerically as
\begin{eqnarray}
F_{q\to q'\,[\mathcal{O}_1({}^1P_1)+\mathcal{O}_1({}^3P_1)]}(z,q^2_0) &=& 
    \frac{1}{2} \sum_{\text{spins}} |\mathcal{M}_{[\mathcal{O}_1({}^1P_1)+\mathcal{O}_1({}^3P_1)]}|^2 
    \nonumber \\
    &=& \cos^2\theta_{P_1}\, F_{q\to q'\,\mathcal{O}_1({}^1P_1)}(z,q^2_0) + \sin^2\theta_{P_1}\, F_{q\to q'\,\mathcal{O}_1({}^3P_1)}(z,q^2_0).
    \nonumber
\end{eqnarray}
This method ensures that interference effects are retained consistently and automatically within the framework. The implementation of this mixing within the parton shower framework requires modifying the standard splitting functions to accommodate the interference contributions, thereby improving the modelling of $P$-wave quarkonia in angular-ordered evolution. This approach ensures consistency with NRQCD expectations and allows for more accurate phenomenological descriptions of spin-sensitive observables.

\subsubsection{$q\to q'\,\mathcal{O}_1({}^1D_2)$ Splitting}

The ${}^1D_2$ heavy-quarkonium state represents a spin-singlet configuration, with total angular momentum $J=2$ and orbital angular momentum $L=2$. Although phenomenologically less dominant compared to lower orbital states, these splittings contribute meaningfully to the Sudakov region and improve the theoretical control over subleading corrections in heavy-quarkonium evolution. In particular, the $b \to b\,\eta_{b2}$ transition is implemented via \textcolor{blue}{\texttt{QtoQ1D2SplitFn}}, following the same conventions adopted for the ${}^1P_1$ and ${}^3P_J$ configurations.

The production amplitude is derived by projecting the short-distance amplitude onto the spin-singlet $L=2$ configuration, with the orbital component carried by a rank-2 symmetric traceless polarisation tensor. In our implementation, this is computed using a tensorial projection constructed from the second derivative of the Bethe–Salpeter radial wavefunction evaluated at the origin. The analytic form of the projected helicity amplitude reads
\begin{equation}
\mathcal{M}(q\to q'\,\mathcal{O}_1({}^1D_2)) = \sqrt{\frac{15}{8\pi}}\, \frac{R''(0)}{2} \cdot \frac{1}{2} \bar{u}(q_1) \left[ \gamma^\mu \left( \slashed{q}_0 - \slashed{q}_1 \right) \gamma^\nu \right] u(q_0) \, \epsilon^{*}_{\mu\nu}(q_2),
\end{equation}
where $\epsilon^{*}_{\mu\nu}$ is the symmetric traceless polarisation tensor of the ${}^1D_2$ state and $R''(0)$ is the second derivative of the radial wavefunction evaluated at the origin. The Lorentz structure is fully consistent with the $L=2$ angular momentum decomposition and matches the result in Ref.~\cite{Cheung:1995ir}.

The spin-averaged and colour-averaged splitting kernel is given by (the corresponding expression for the flavour-changing case, $F_{q\to q'\,\mathcal{O}_1({}^1D_2)}(z,q_0^2)$, is omitted here due to its length):
\begin{eqnarray}
F_{q\to q\,\mathcal{O}_1({}^1D_2)}(&z&,q^2_0) = \frac{80\, g_S^4\, |R''(0)|^2\, C_F^2}{\pi^3 M^5 N_C\, (z+1)^6 (M^2 - 4 q_0^2)^6}
  \Big[
      M^8 \big(97 z^7+118 z^6+1239 z^5+2708 z^4
      \nonumber \\
      &-& 1281 z^3-8938 z^2-8599 z-2688\big)
      - 16 M^6 q_0^2 \big(77 z^7+6 z^6+811 z^5+3996 z^4
      \nonumber \\
      &+& 6403 z^3+3462 z^2-91 z-520\big)
      + 32 M^4 q_0^4 \big(163 z^7-414 z^6-2043 z^5-1540 z^4
      \nonumber \\
      &+& 3229 z^3+4194 z^2+1435 z-224\big)
      - 256 M^2 q_0^6 \big(29 z^7+70 z^6+27 z^5-260 z^4 + 19 z^3
      \nonumber \\
      &+& 230 z^2+213 z-8\big)
      + 256 q_0^8 z \big(z^6-10 z^5+55 z^4-44 z^3+31 z^2-42 z+73\big)
  \Big].
\end{eqnarray}
This formulation is numerically consistent with the splitting functions computed in~\cite{Cheung:1995ir}, and its inclusion enhances the fidelity of parton shower evolution in heavy-flavour event generation, especially in observables sensitive to higher orbital excitations.

\subsubsection{$q \to q'\,\mathcal{O}_1({}^3D_J)$ Splittings}

In the context of parton shower evolution involving heavy-quarkonium states, the inclusion of $q \to q'\,\mathcal{O}_1({}^3D_J)$ splittings is necessary for accurate modelling of subleading components in the NRQCD expansion, particularly in the charm and bottom sectors. The $D$-wave contributions, while suppressed by higher powers of the heavy-quark velocity $v$, are phenomenologically relevant for mesons such as $\psi(3770)$, $\psi_2(1D)$, and $\Upsilon_J(1D)$, whose production channels are experimentally resolved and kinematically accessible at collider energies. The incorporation of these ${}^3D_J$ states ensures the completeness of the angular-momentum basis used in quarkonium production, allowing for improved precision in angular observables, polarisation measurements, and non-trivial interference effects in NRQCD matrix-element matching. In particular, for states such as $\Upsilon(1D)$ and $\psi(1D)$, the evolution of colour-singlet $D$-wave components requires dedicated treatment due to their $L=2$ orbital configuration and non-trivial spin-alignment patterns.

The amplitude for the $q \to q'\, \mathcal{O}_1({}^3D_J)$ transition is defined through the second derivative of the quark-antiquark wavefunction and the corresponding multipole expansion of the spin-triplet current. It reads
\begin{equation}
\mathcal{M}(q \to q'\, \mathcal{O}_1({}^3D_J)) = 
\sqrt{\frac{15}{8\pi}}\, \frac{R''(0)}{2} \, 
\bar{u}(q_1)\, \Gamma^{\mu\nu} \, u(q_0) \; 
\epsilon^*_{\mu\nu}(q_2),
\end{equation}
where $\Gamma^{\mu\nu}$ denotes the rank-2 tensor current derived from the divergence of the spin-triplet amplitude with respect to the relative momentum, and $\epsilon^*_{\mu\nu}$ is the polarisation tensor associated with the ${}^3D_J$ state (with $J=1,2,3$). In the NRQCD framework, this corresponds to projecting the amplitude onto the $D$-wave sector using the symmetrised, traceless polarisation tensor $\epsilon^{\mu\nu}$ of the ${}^3D_J$ state. Explicitly, $\Gamma^{\mu\nu}$ can be defined as
\begin{equation}
    \Gamma^{\mu\nu} = \left. \frac{\partial^2}{\partial k_\mu \partial k_\nu} 
    \left[ \bar{u}(q_1)\, \slashed{\epsilon}(q_2)\, u(q_0) \right] \right|_{k=0},
\end{equation}
where $\slashed{\epsilon}(q_2)$ corresponds to the spin-triplet $S=1$ polarisation structure, and $k$ is the relative momentum within the bound state. In practice, the structure contributing to the amplitude can be written as
\begin{equation}
    \mathcal{M}(q \to q'\, \mathcal{O}_1({}^3D_J)) \sim \bar{u}(q_1)\, 
    \left[ \gamma^\mu (q_1 - q_0)^\nu + \gamma^\nu (q_1 - q_0)^\mu \right] 
    u(q_0)\, \epsilon^*_{\mu\nu}(q_2),
\end{equation}
with the symmetric, traceless rank-2 tensor $\epsilon^*_{\mu\nu}$ encoding the $J=1,2,3$ polarisations. In \textsf{Herwig}'s quarkonium parton shower, these splittings are implemented through the following classes:
\begin{itemize}
  \item[--] \textcolor{blue}{\texttt{QtoQ3D1SplitFn}} class for $J=1$: $c \to c\,\psi(3770)$, $b \to b\,\Upsilon_1(1D)$,
  \item[--] \textcolor{blue}{\texttt{QtoQP3D1SplitFn}} class for $J=1$: $c \to b\,B_c(1D)^{\star +}$, $b \to c\,B_c(1D)^{\star -}$,
  \item[--] \textcolor{blue}{\texttt{QtoQ3D2SplitFn}} class for $J=2$: $c \to c\,\psi_2(1D)$, $b \to b\,\Upsilon_2(1D)$,
  \item[--] \textcolor{blue}{\texttt{QtoQD2SplitFn}} class for $J=2$: $c \to b\,B_{c2}(L)^+$, $b \to c\,B_{c2}(L)^-$, $c \to b\,B_{c2}(H)^+$, and $b \to c\,B_{c2}(H)^-$,
  \item[--] \textcolor{blue}{\texttt{QtoQ3D3SplitFn}} class for $J=3$: $c \to c\,\psi_3(1D)$, $b \to b\,\Upsilon_3(1D)$,
  \item[--] \textcolor{blue}{\texttt{QtoQP3D3SplitFn}} class for $J=3$: $c \to b\,B_{c3}(1D)^{\star +}$, $b \to c\,B_{c3}(1D)^{\star -}$.
\end{itemize}
Here, $L$ and $H$ denote the low- and high-mass eigenstates arising from spin-orbit mixing in the $B_c$ system. 

Having calculated the production amplitudes, we now derive the splitting probabilities for the $J = 1, 2, 3$ polarisations. As the general case is again lengthy, we present only the significantly simpler same-flavour cases below.
\allowdisplaybreaks{
\begin{eqnarray}
  F_{q\to q\,\mathcal{O}_1({}^3D_1)}(z,q^2_0) &=& \frac{8 g_S^4 |R''(0)|^2 C_F^2 }{\pi ^3 M^5 N_C (z+1)^6 \left(M^2-4 q_0^2\right)^6}
  \Big[
   M^8 \big( 1539 z^7+10824 z^6+31609 z^5+16318 z^4
   \nonumber \\
   &-&81355 z^3-163092 z^2-116113 z-29906\big)
   -8 M^6 q_0^2 \big(685 z^7+7035 z^6+38961 z^5
   \nonumber \\
   &+&78823 z^4+52135 z^3-31247 z^2-51781 z-17683\big)
   -32 M^4 q_0^4 \big(2 z^7+1039 z^6
   \nonumber \\
   &-& 6168 z^5-27203 z^4-36598 z^3-5751 z^2+12716 z+6475\big)
   -128 M^2 q_0^6 \big(91 z^7
   \nonumber \\
   &+&417 z^6+1391 z^5+2413 z^4+5761 z^3+1491 z^2-523 z-673\big)
   +256 q_0^8 z \big(17 z^6
   \nonumber \\
   &+&14 z^5-105 z^4-412 z^3+1071 z^2-82 z+73\big)
   \Big],
\\
F_{q \to q\, \mathcal{O}_1({}^3D_2)}(z, q_0^2) &=&
\frac{40 g_S^4 |R''(0)|^2 C_F^2}{\pi^3 M^5 N_C (z+1)^6 (M^2 - 4 q_0^2)^6}
\Big[
M^8 (365 z^7 + 2396 z^6 + 9551 z^5 + 14318 z^4
\nonumber \\
&-& 277 z^3 - 22512 z^2 - 21319 z - 6394)
- 8 M^6 q_0^2 (329 z^7 + 1855 z^6 + 11101 z^5 
\nonumber \\
&+& 28379 z^4 + 32827 z^3 + 12669 z^2 - 3233 z - 2903)
+ 32 M^4 q_0^4 (148 z^7 - 439 z^6 
\nonumber \\
&+& 138 z^5 + 5639 z^4 +13488 z^3 + 9831 z^2 + 2002 z - 791)
- 128 M^2 q_0^6 (7 z^7
\nonumber \\
&-& 83 z^6 - 5 z^5 - 263 z^4 + 1061 z^3 + 1239 z^2 + 793 z - 61)
+ 256 q_0^8 z (11 z^6 
\nonumber \\
&-& 22 z^5 + 77 z^4 - 84 z^3 + 117 z^2 - 54 z + 147)
\Big],
\\
F_{q \to q\, \mathcal{O}_1({}^3D_3)}(z, q_0^2) &=&
\frac{16 g_S^4 |R''(0)|^2 C_F^2}{\pi^3 M^5 N_C (z+1)^6 (M^2 - 4 q_0^2)^6}
\Big[
M^8 (1123 z^7 + 3058 z^6 + 11053 z^5 + 19996 z^4
\nonumber \\
&-& 1235 z^3 - 45294 z^2 - 46301 z - 14912)
- 16 M^6 q_0^2 (1115 z^7 + 3730 z^6 + 12241 z^5 
\nonumber \\
&+& 31088 z^4 + 43645 z^3 + 25098 z^2 + 2359 z - 2348)
+ 32 M^4 q_0^4 (1601 z^7 + 2782 z^6
\nonumber \\
&-& 1209 z^5 - 2924 z^4 + 16191 z^3 + 20382 z^2 + 8153 z - 880)
+ 256 M^2 q_0^6 (29 z^7
\nonumber \\
&-& 202 z^6 - 721 z^5 + 752 z^4 - 101 z^3 - 546 z^2 - 967 z + 28)
+ 256 q_0^8 z (19 z^6 - 62 z^5 
\nonumber \\
&+& 205 z^4 - 324 z^3 + 477 z^2 - 414 z + 291)
\Big],
\end{eqnarray}}
which are identical to the results obtained in~\cite{Cheung:1995ir}. 

Similar to the ${}^1P_1$-${}^3P_1$ case, the mixing between $^1D_2$ and $^3D_2$ states becomes phenomenologically relevant due to their shared $J^{PC}=2^{-+}$ quantum numbers. Such mixing, driven by spin-dependent corrections and violations of heavy-quark spin symmetry, notably affects transitions involving $\eta_{c2}$ and $\psi_2(1D)$ or $\eta_{b2}$ and $\Upsilon_2(1D)$. We define the physical eigenstate as
\begin{equation}
\mathcal{O}_1^\text{mixed}(D_2) = \cos\theta_{D_2}\, \mathcal{O}_1(^1D_2) + \sin\theta_{D_2}\, \mathcal{O}_1(^3D_2),
\nonumber
\end{equation}
with the default mixing angle $\theta_{D_2} = 34.4^\circ$. The corresponding splitting amplitude is given by
\begin{equation}
\mathcal{M}_{[\mathcal{O}(^1D_2)+\mathcal{O}(^3D_2)]} = \cos\theta_{D_2}\, \mathcal{M}_{^1D_2} + \sin\theta_{D_2}\, 
\mathcal{M}_{^3D_2},
\nonumber
\end{equation}
where $\mathcal{M}_{^1D_2}$ and $\mathcal{M}_{^3D_2}$ are computed independently using their respective Dirac structures and projection tensors. The helicity-averaged splitting probability is then
\begin{eqnarray}
F_{q\to q'\,[\mathcal{O}_1(^1D_2)+\mathcal{O}_1(^3D_2)]}(z,q^2_0) &=&
\frac{1}{2} \sum_{\text{spins}} \left|\mathcal{M}_{[\mathcal{O}_1(^1D_2)+\mathcal{O}_1(^3D_2)]}\right|^2
\nonumber \\
&=& \cos^2\theta_{D_2} F_{q\to q'\,\mathcal{O}_1(^1D_2)}(z,q^2_0) + \sin^2\theta_{D_2} F_{q\to q'\,\mathcal{O}_1(^3D_2)}(z,q^2_0), 
\nonumber
\end{eqnarray}
where the interference vanishes due to orthogonality in the non-relativistic limit. This prescription is embedded in the shower algorithm, preserving the spin structure of the amplitude throughout evolution.

\subsection{Colour-Singlet $g \to g\,\mathcal{O}_1$ Splittings}
\label{sebsec:ggO}

\begin{figure}[h!]
  \begin{center}
    \includegraphics[width=0.8\textwidth]{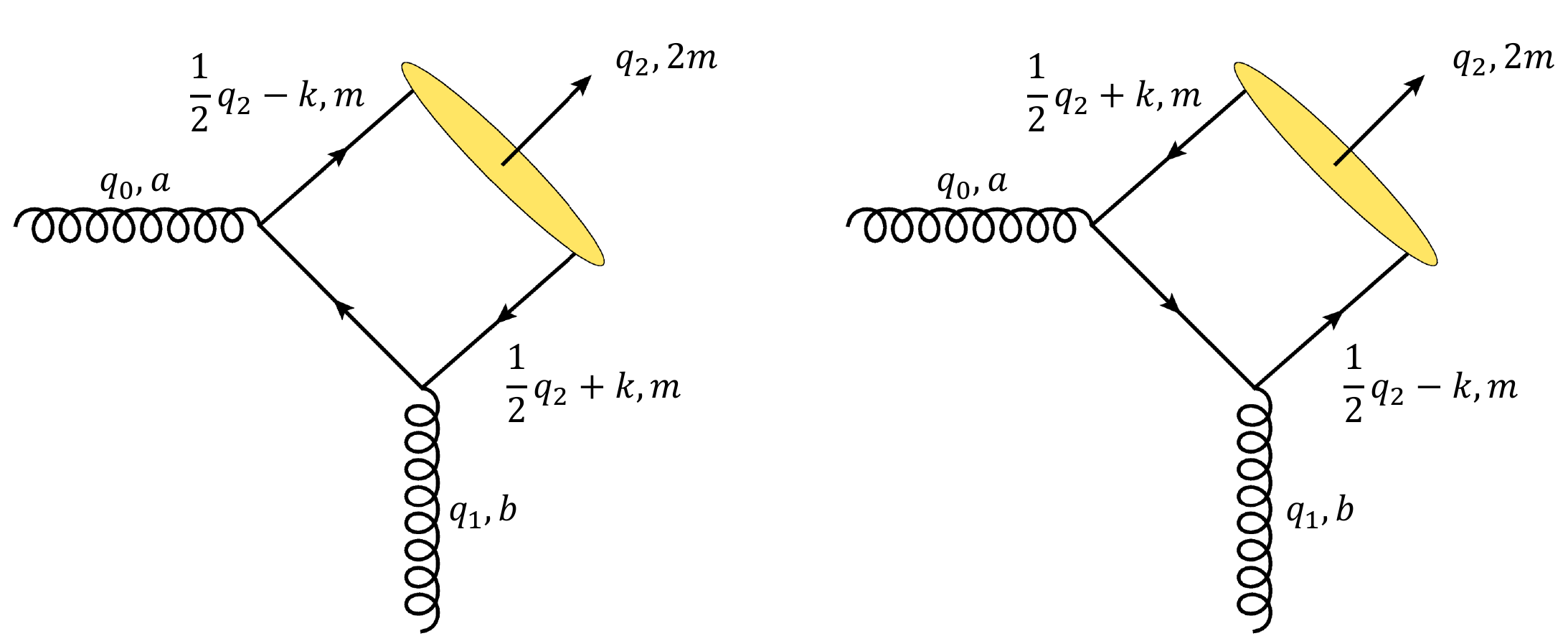}
  \end{center}
  \caption{\small Feynman diagrams for colour-singlet $g\to g\,\mathcal{O}_1$ splittings.}
  \label{fig:gtogO}
\end{figure}

Fig.~\ref{fig:gtogO} shows the Feynman diagrams for the production of a quarkonium state from a gluon in a colour-singlet configuration. In this scenario, a gluon of virtuality $q_0^2$ branches into a final-state gluon plus a heavy quark-antiquark pair that forms the quarkonium bound state.
Here, the $gQ\bar{Q}$ interactions follow the QCD Feynman rules, where the strong coupling and the virtuality of the initial gluon enter explicitly as prefactors. Here, the amplitude involves two distinct topologies in which the heavy quark and antiquark can couple to the gluon lines. A typical structure has
\begin{eqnarray}
  \mathcal{A}(g \to &g&\,\mathcal{O}_1) = \frac{g_S^2}{q_0^2}\;\bar{u}\Bigl(\tfrac12\,q_2 + k\Bigr)
  \Biggl[
    -\;\frac{{\bf t}^a_{ik}\,{\bf t}^b_{kj}}{\bigl(q_1 + \tfrac12\,q_2 + k\bigr)^2 - m^2}
    \;\slashed{\epsilon}(q_0)\;\bigl(\slashed{q}_1 + \tfrac12\,\slashed{q}_2 + \slashed{k} - m\bigr)\;\slashed{\epsilon}(q_1)
  \nonumber\\
    &&\qquad +\;\frac{{\bf t}^b_{ik}\,{\bf t}^a_{kj}}{\bigl(q_1 + \tfrac12\,q_2 - k\bigr)^2 - m^2}
    \;\slashed{\epsilon}(q_1)\;\bigl(\slashed{q}_1 + \tfrac12\,\slashed{q}_2 - \slashed{k} + m\bigr)\;\slashed{\epsilon}(q_0)
  \Biggr]
  v\Bigl(\tfrac12\,q_2 + k\Bigr).
\end{eqnarray}
The projection onto the colour-singlet state involves the sum over exchanged gluon colours, ensuring that only the singlet component contributes. This simplifies the ME to a form that isolates the short-distance QCD coefficient.

Hence, the production amplitude for a colour-singlet quarkonium state from initial-state gluon splitting can be written as
\begin{eqnarray}
    \mathcal{M}(g \to &g&\,\mathcal{O}_1({}^nL_J)) =
    \frac{g_S^2 C_F \; \big\langle \mathcal{O}_1({}^nL_J) \big \rangle \; \delta^{ab}}{4 m \sqrt{2 N_C m} \, q_0^2}
    \, \text{Tr} \left[
        \left( \frac{1}{2} \slashed{q}_2 + \slashed{k} + m \right)
        \begin{pmatrix}
            \gamma^5 \\
            -\slashed{\epsilon}(q_2)
        \end{pmatrix}
        \left( \frac{1}{2} \slashed{q}_2 - \slashed{k} + m \right)
    \right. \nonumber \\
    & \times & \left.
    \left(
        -\frac{
            \slashed{\epsilon}(q_0)^a \left( \slashed{q}_1 + \frac{1}{2} \slashed{q}_2 + \slashed{k} - m \right) \slashed{\epsilon}(q_1)^b
        }{
            \left(q_1 + \frac{1}{2} q_2 + k \right)^2 - m^2
        }
        + \frac{
            \slashed{\epsilon}(q_1)^b \left( \slashed{q}_1 + \frac{1}{2} \slashed{q}_2 - \slashed{k} + m \right) \slashed{\epsilon}(q_0)^a
        }{
            \left(q_1 + \frac{1}{2} q_2 - k \right)^2 - m^2
        }
    \right)
    \right],
\end{eqnarray}
where $m$ is the heavy-quark mass, $q_1$ is the momentum of the initial gluon, $q_2$ is the quarkonium momentum, and $k$ is the relative momentum of the $Q\bar{Q}$ pair in the quarkonium rest frame. The quarkonium projection operator depends on the spin configuration: $\gamma^5$ for ${}^1S_0$, and $-\slashed{\epsilon}(q_2)$ for ${}^3S_1$. The colour-singlet projection is implemented via $\delta^{ab}/\sqrt{N_C}$, and helicity dependence enters via $\epsilon(q_0)$ and $\epsilon(q_1)$, the polarisation vectors of the initial and final gluons.

\subsubsection{$g\to g \mathcal{O}_1({}^1S_0)$ Splittings}

A concrete example of a $g \to g\,\mathcal{O}_1$ splitting is the pseudoscalar channel ${}^1S_0$. The antisymmetric tensor structure, 
\begin{eqnarray}
  \mathcal{M}(g\to g \mathcal{O}_1({}^1S_0)) &=&
  -\,\frac{\mathrm{i}\,g_S^2 \,\sqrt{2}\,R(0)}{\sqrt{m\,N_C\,\pi}\,q_0^2\,\bigl(q_0^2 - 4\,m^2\bigr)}
  \;\epsilon_{\alpha\beta \mu \nu}\;\epsilon(q_0)^\alpha\,\epsilon(q_1)^\beta\,q_1^\mu\,q_2^\nu,
  \label{Mg1S0}
\end{eqnarray}
manifests the pseudoscalar nature of the bound state through $\gamma_5$ insertion\footnote{Here, the appearance of the antisymmetric Levi-Civita tensor $\epsilon_{\alpha\beta\mu\nu}$ ensures the correct parity-odd behaviour of the amplitude, effectively encoding the pseudoscalar nature of the state. This structure is commonly encountered in gluon-induced pseudoscalar production, where the dual field strength $G_{\mu\nu} \tilde{G}^{\mu\nu}$ similarly captures negative parity.} and leads to the appropriate splitting probability for a gluon to produce a ${}^1S_0$ quarkonium. Similar constructions exist for other angular-momentum and spin assignments, involving the non-perturbative radial wavefunction $R(0)$ or its derivatives for higher orbital excitations. From the amplitude Eq.~\eqref{Mg1S0}, one can derive the spin- and colour-averaged splitting function as
\begin{eqnarray}
F_{g\to g \mathcal{O}_1({}^1S_0)}(z,q_0^2) = \frac{2 \pi |R(0)|^2}{3 m \left(q_0^2-4 m^2\right)^2} \left[16 m^4+8 m^2 q_0^2 (z-1)+q_0^4 (2 (z-1) z+1)\right],
\end{eqnarray}
that is in agreement with the similar calculations carried out in Ref.~\cite{Braaten:1993rw}. These types of splittings are implemented in \textsf{Herwig~7} within the class \textcolor{blue}{\texttt{GtoG1S0SplitFn}}, which governs the $g \to g \eta_c$ and $g \to g \eta_b$ splittings in the parton shower. 

\subsubsection{$g\to g \mathcal{O}_1({}^3S_1)$ Splittings}

The amplitude for the production of the triplet-bound states through gluon fragmentation can be given as
\begin{eqnarray}
\mathcal{M}(g\to g \mathcal{O}_1({}^3S_1)) &=& \frac{
    2 \sqrt{2} g_S^2 
}{
    m^{3/2} q_0^2 D
}
\Big[
        -8 q_2^\mu k^\nu q_1^\rho m^2 
        -8 q_2^\mu k^\nu q_2^\rho m^2 
        -8 g^{\mu \nu} k^\rho k m^2 
        +16 q_1^\mu g^{\nu \rho} (k \cdot q_1) m^2 
        \nonumber \\
        &+&8 q_2^\mu g^{\nu \rho} (k \cdot q_1) m^2 
        -16 g^{\mu \nu} q_1^\rho (k \cdot q_1) m^2 
        -8 g^{\mu \nu} q_2^\rho (k \cdot q_1) m^2 
        +4 g^{\mu \nu} k^\rho k 
        \nonumber \\
        &-&16 g^{\mu \nu} k^\rho (k \cdot q_1) 
        -4 g^{\mu \nu} k^\rho (k \cdot q_2) 
        +2 q_0^2 q_2^\mu k^\nu q_1^\rho 
        +2 q_0^2 q_2^\mu k^\nu q_2^\rho 
        +2 q_0^2 g^{\mu \nu} k^\rho k 
        \nonumber \\
        &+&4 q_2^\mu k^\nu q_1^\rho k 
        +4 q_2^\mu k^\nu q_2^\rho k 
        -2 q_0^2 q_2^\mu g^{\nu \rho} (k \cdot q_1) 
        +16 q_1^\mu k^\nu k^\rho (k \cdot q_1) 
        +2 q_0^2 g^{\mu \nu} q_2^\rho (k \cdot q_1) 
        \nonumber \\
        &-&4 q_2^\mu q_1^\nu q_2^\rho (k \cdot q_1) 
        -4 q_1^\mu q_2^\nu q_2^\rho (k \cdot q_1) 
        -4 q_2^\mu q_2^\nu q_2^\rho (k \cdot q_1) 
        -8 q_1^\mu g^{\nu \rho} k (k \cdot q_1) 
        \nonumber \\
        &+&8 g^{\mu \nu} q_1^\rho k (k \cdot q_1) 
        +4 g^{\mu \nu} q_2^\rho k (k \cdot q_1) 
    \Big],
\end{eqnarray}
with 
$$
D =  (-4 m^2 + q_0^2 + 2 k - 4 (k \cdot q_1) - 2 (k \cdot q_2)) 
        (-4 m^2 + q_0^2 + 2 k + 4 (k \cdot q_1) + 2 (k \cdot q_2)).
$$
This results in a ME that vanishes in the $k \to 0$ limit due to the Landau-Yang theorem~\cite{Landau:1948kw,Yang:1950rg}, which forbids on-shell $g \to g\,\mathcal{O}_1({}^3S_1)$ transitions. Thus, in principle, the correct leading-order fragmentation process would be $g \to g g \,\mathcal{O}_1({}^3S_1)$, but implementing such a $1 \to 3$ splitting would require major modifications to the parton shower kinematics and recoil strategy. Given that this channel is both suppressed and subleading, we instead employ the $g \to g\,\mathcal{O}_1({}^3S_1)$ process as a technically convenient surrogate, with its ME fixed to unity. This may be viewed as standing in for the integrated effect of an unresolved extra gluon in $g \to g g \,\mathcal{O}_1({}^3S_1)$.

This simplification allows efficient tuning of colour-octet contributions within the shower without compromising numerical stability. It aligns with the broader NRQCD framework, where high-$p_T$ quarkonium production proceeds predominantly via gluon fragmentation into intermediate colour-octet ${}^3S_1$ states~\cite{Braaten:1993rw,Cho:1995ce,Cacciari:1994dr}, followed by non-perturbative evolution into physical quarkonia. While dedicated $g \to \mathcal{O}_8({}^3S_1)$ octet splittings are included explicitly in the parton shower, this effective $g \to g\,\mathcal{O}_1({}^3S_1)$ channel provides a minimal and IR-safe implementation path and is numerically negligible in all results presented here. The implementation in \textsf{Herwig~7} is handled by \textcolor{blue}{\texttt{GtoG3S1SplitFn}}, which governs the $g \to g\,J/\psi$ and $g \to g\,\Upsilon$ splittings as part of the spin- and colour-correlated evolution framework; users who prefer to disable the surrogate may remove it at run time via the \texttt{delete} command (see App.~\ref{sec:AppB} for an example).

\subsubsection{$g\to g \mathcal{O}_1({}^3P_J)$ Splittings}

The production amplitude for the $g\to g \mathcal{O}_1({}^1P_1)$ splitting is forbidden at the Landau-Yang limit, as the amplitude for the transition of a gluon into a ${}^1P_1$ quarkonium state, 
\begin{equation}
    \mathcal{M}(g \to g \mathcal{O}_1({}^1P_1)) = -i \sqrt{\frac{3}{4\pi}} \, R'(0) \, \left[ \nabla^\mu \mathcal{A}(g\to g \mathcal{O}_1) \right] \Big|_{\mathbf{k} = 0} = 0.
\end{equation}
More generally, the four-divergence of the singlet amplitude, $ \nabla^\mu \mathcal{M}_{\text{singlet}} $, vanishes in the absence of an axial anomaly or an external current. Consequently, no contribution arises for the ${}^1P_1$ channel in the parton shower evolution. Instead, these states are expected to form predominantly via two-gluon fusion or through electric dipole transitions from higher excitations.

Therefore, we shift our focus over to gluon fragmentation to $P$-wave ${}^3P_{0,1,2}$ quarkonia states, i.e. $\chi_{cJ}, \chi_{bJ}$ and explicitly, the splittings
$$
g \to g\,\chi_{q0} \, ({}^3P_0), \quad
g \to g\,\chi_{q1} \, ({}^3P_1), \quad
g \to g\,\chi_{q2} \, ({}^3P_2), \quad q = c,b,
$$
coded into splitting kernels \textcolor{blue}{\texttt{GtoG3P0SplitFn}}, \textcolor{blue}{\texttt{GtoG3P1SplitFn}} and \textcolor{blue}{\texttt{GtoG3P2SplitFn}}, respectively. For these states, one can write
\begin{eqnarray}
\mathcal{M}^{\mu \nu}(g \to g \mathcal{O}_1({}^3P_J)) &=& 
-\frac{i \sqrt{6} g_S^2 \, R'(0)}{m^{3/2} \sqrt{\pi} q_0^2 (q_0^2 - 4m^2)^2} 
\Bigg[
    (4m^2 - q_0^2) g^{\alpha \mu} \left( (4m^2 - q_0^2) g^{\nu \rho} + 2 q_2^{\nu} q_1^{\rho} \right) 
    \nonumber \\
    &+& (4m^2 - q_0^2) g^{\alpha \nu} \left( (4m^2 + q_0^2) g^{\mu \rho} - 2 q_2^{\mu} (q_1^{\rho} + q_2^{\rho}) \right) 
    + 8m^2 q_2^{\alpha} q_1^{\mu} g^{\nu \rho} 
    \nonumber \\
    &+& 8m^2 q_1^{\alpha} q_2^{\mu} g^{\nu \rho} 
    + 8m^2 q_2^{\alpha} q_1^{\nu} g^{\mu \rho} 
    + 8m^2 q_1^{\alpha} q_2^{\nu} g^{\mu \rho} 
    - 8m^2 q_2^{\alpha} q_1^{\rho} g^{\mu \nu} 
    \nonumber \\
    &-& 8m^2 q_1^{\alpha} q_2^{\rho} g^{\mu \nu} 
    + 16m^2 q_1^{\alpha} q_1^{\mu} g^{\nu \rho} 
    + 16m^2 q_1^{\alpha} q_1^{\nu} g^{\mu \rho} 
    - 16m^2 q_1^{\alpha} q_1^{\rho} g^{\mu \nu} 
    \nonumber \\
    &+& 4m^2 q_2^{\alpha} q_2^{\mu} g^{\nu \rho} 
    + 4m^2 q_2^{\alpha} q_2^{\nu} g^{\mu \rho} 
    - 2q_0^2 q_1^{\alpha} q_2^{\mu} g^{\nu \rho} 
    + 2q_0^2 q_1^{\alpha} q_2^{\nu} g^{\mu \rho} 
    + 2q_0^2 q_1^{\alpha} q_2^{\rho} g^{\mu \nu} 
    \nonumber \\
    &-& q_0^2 q_2^{\alpha} q_2^{\mu} g^{\nu \rho} 
    + q_0^2 q_2^{\alpha} q_2^{\nu} g^{\mu \rho} 
    - 4 q_1^{\alpha} q_2^{\mu} q_1^{\nu} q_2^{\rho} 
    - 2 q_2^{\alpha} q_2^{\mu} q_1^{\nu} q_2^{\rho} 
    - 4 q_1^{\alpha} q_1^{\mu} q_2^{\nu} q_2^{\rho} 
    \nonumber \\
    &-& 2 q_2^{\alpha} q_1^{\mu} q_2^{\nu} q_2^{\rho} 
    - 4 q_1^{\alpha} q_2^{\mu} q_2^{\nu} q_2^{\rho} 
    - 2 q_2^{\alpha} q_2^{\mu} q_2^{\nu} q_2^{\rho}
\Bigg].
\end{eqnarray}
And from this, one can derive the production amplitude for the scalar, the axial-vector and the tensor states as 
\begin{equation}
\begin{pmatrix}
\mathcal{M}(g \to g \mathcal{O}_1({}^3P_0)) \\
\mathcal{M}(g \to g \mathcal{O}_1({}^3P_1)) \\
\mathcal{M}(g \to g \mathcal{O}_1({}^3P_2))
\end{pmatrix}
=
\begin{pmatrix}
\frac{1}{\sqrt{3}} g_{\mu\nu} \\
\frac{i}{\sqrt{2}} \varepsilon_{\mu\nu\alpha\beta} q^\alpha P^\beta \\
q_{\mu} q_{\nu} - \frac{1}{3} g_{\mu\nu} q^2
\end{pmatrix}
\mathcal{M}^{\mu \nu}(g \to g \mathcal{O}_1({}^3P_J)),
\end{equation}
which results in the following splitting functions:
\begin{eqnarray}
F_{g \to g \mathcal{O}_1({}^3P_0)}(z,q_0^2) &=& \frac{2 \pi  |R'(0)|^2 \left(q_0^2-12 m^2\right)^2}{3 m^3 
\left( q_0^2 -4 m^2\right)^4}
\Big[16 m^4+8 m^2 q_0^2 (z-1)+ q_0^4 \left(2 z^2-2z+1\right)\Big],
\\
F_{g \to g \mathcal{O}_1({}^3P_1)}(z,q_0^2) &=& \frac{4 \pi  q_0^4 |R'(0)|^2 }{m^3 \left(q_0^2-4 m^2\right)^4} 
\Big[ m^4 (16-64 z)-8 m^2 q_0^2 \left(2 z^2-3 z+1\right)
\nonumber \\
&+& q_0^4 \left(2 z^2-2 z+1\right) \Big],
\\
F_{g \to g \mathcal{O}_1({}^3P_2)}(z,q_0^2) &=& \frac{4 \pi  |R'(0)|^2 }{3 m^3 \left(q_0^2-4 m^2\right)^4}
\Big[ 1536 m^8+768 m^6 q_0^2 (z-1)+16 m^4 q_0^4 \left(12 z^2-24 z+7\right)
\nonumber \\
&-&8 m^2 q_0^6 \left(6 z^2-7 z+1\right)+q_0^8 \left(2 z^2-2 z+1\right) \Big],
\end{eqnarray}
that are numerically identical to similar calculations made in Ref.~\cite{Braaten:1994kd}. Trivially, the non-perturbative parameter $R'(0)$ could be different in any of these cases. However, for simplicity, we are using a unique notation for it, instead of adding complex indices to differentiate them for different states at different quantum numbers. We will discuss the tuning tactics for these parameters in Sec .~\ref{sec:matrixelementTuning}.

\subsection{Colour-Octet $g \to \mathcal{O}_8$ Splittings}
\label{sebsec:gO}

The inclusion of colour-octet states is essential for a theoretically consistent and phenomenologically accurate description of quarkonium production. Within the NRQCD factorisation framework~\cite{Bodwin:1994jh}, the production cross-section is expressed as a sum over intermediate $\mathcal{O}_c({}^nL_J)$ states. Both singlet and octet channels contribute at leading power in the $v^2$ expansion, with short-distance coefficients calculable in perturbative QCD and LDMEs $\langle \mathcal{O}_c({}^nL_J) \rangle$ fitted to data. Phenomenologically, colour-singlet contributions alone fail to reproduce observed production rates of $J/\psi$, $\psi(2S)$, and $\Upsilon(nS)$ at high transverse momentum. The dominant contribution in this regime arises from gluon fragmentation to colour-octet $\mathcal{O}_8({}^3S_1)$ states, followed by non-perturbative evolution into physical quarkonium via soft gluon emission. Additional octet channels, including $\mathcal{O}_8({}^1S_0)$ and $\mathcal{O}_8({}^3P_{0,1,2})$, play a subleading role in matching differential cross-sections and polarisation observables. Nevertheless, excluding these channels would not only violate the completeness of the NRQCD expansion but also lead to significant disagreement with experimental measurements. The \textsf{Herwig} implementation therefore incorporates dedicated splitting kernels for octet states, enabling systematic tuning of the non-perturbative parameters and ensuring compatibility with the theoretical structure and experimental constraints of modern quarkonium phenomenology.

\begin{figure}[h!]
  \begin{center}
    \includegraphics[width=0.4\textwidth]{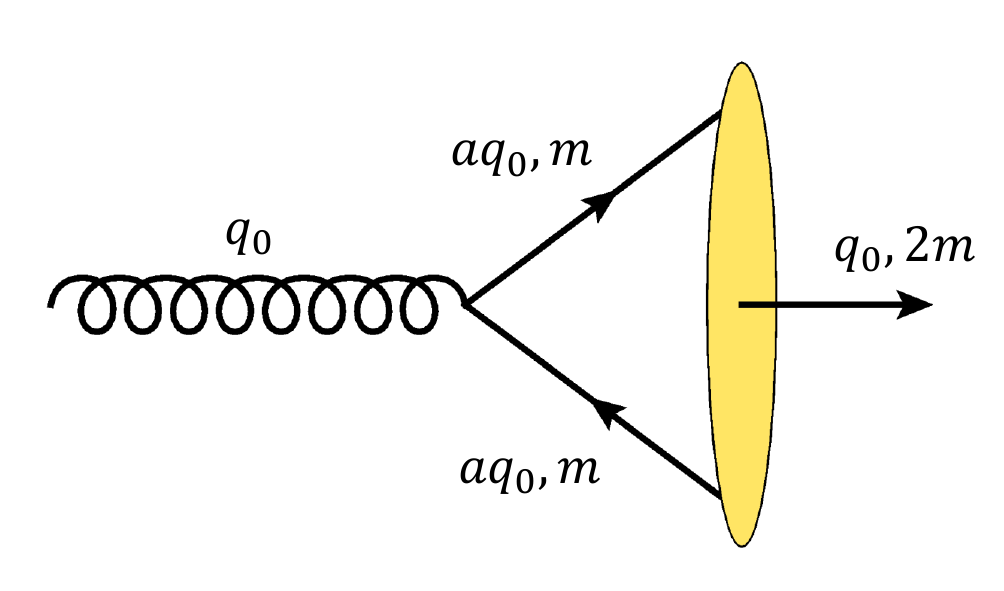}
  \end{center}
  \caption{\small Feynman diagram for colour-octet $g\to \mathcal{O}_8$ splittings.}
  \label{fig:gtoO}
\end{figure}

As illustrated in Fig.~\ref{fig:gtoO}, the gluon fragments into a heavy quark-antiquark pair in a colour-octet configuration, which subsequently hadronises into a quarkonium state. At leading order in $\alpha_S$, the dominant mechanism for producing colour-octet heavy quarkonium is through the fragmentation process $g \to \mathcal{O}_8({}^nL_J)$. Since the transition from the perturbative $Q\bar{Q}$ construct to the physical quarkonium proceeds via non-perturbative soft-gluon radiation, the shower only needs to initialise the $g \to \mathcal{O}_8$ vertex. Subsequent evolution into a singlet bound state is absorbed into the LDME $\langle \mathcal{O}_8({}^nL_J) \rangle$, obviating the need for intermediate gluon emissions within the perturbative expansion. This fragmentation limit is theoretically justified by factorisation theorems in NRQCD~\cite{Bodwin:1994jh, Nayak:2005rw}, where the perturbative kernel is independent of the non-perturbative hadronisation dynamics. As such, the $g \to \mathcal{O}_8$ transition is sufficient and complete at leading power, and no explicit inclusion of $g \to g \mathcal{O}_8$ or $g \to gg \mathcal{O}_8$ vertices is required in the shower.

Rather than computing the full production amplitudes and subsequently the splitting functions for octet production, a simplified approach is adopted in which the $g \to \mathcal{O}_8$ process is modelled via a fixed probability per unit evolution time. This probability is governed by the non-perturbative LDME associated with the colour-octet state and is given by:
\begin{equation}
    P_{g \to \mathcal{O}_8({}^nL_J)} = \frac{\pi \, \alpha_S(4m^2)}{24 m^3} \langle 0|\mathcal{O}_8({}^nL_J)|0 \rangle,
\end{equation}
where $m$ is the heavy quark mass. Under this scheme, the structural differences between $S$-wave and $P$-wave octet channels are absorbed into the non-perturbative MEs. This enables a uniform treatment of all octet states, simplifying tuning and maintaining compatibility with the NRQCD power counting. Such implementations are motivated by the practical dominance of colour-octet contributions in high-$p_T$ quarkonium production~\cite{Braaten:1993rw, Cho:1995ce} and are supported by previous parton shower models that treat octet fragmentation inclusively to retain computational efficiency and phenomenological realism~\cite{Kniehl:2006sk}.

The implementation in \textsf{Herwig~7} includes several dedicated classes to simulate colour-octet quarkonium production through gluon fragmentation. These classes handle effective $g \to \mathcal{O}_8$ transitions and are tuned via non-perturbative parameters. The class \textcolor{blue}{\texttt{Gto3S1OctetSplitFn}} governs the $g \to J/\psi$, $\psi(2S)$, and $\Upsilon$ splittings into vector quarkonia. Scalar $P$-wave octet splittings such as $g \to \chi_{c0}$ and $\chi_{b0}$ are handled by \textcolor{blue}{\texttt{Gto3P0OctetSplitFn}}, while axial-vector $P$-wave transitions like $g \to \chi_{c1}$ and $\chi_{b1}$ are described by \textcolor{blue}{\texttt{Gto3P1OctetSplitFn}}. The tensor $P$-wave states, $g \to \chi_{c2}$ and $\chi_{b2}$, are treated using \textcolor{blue}{\texttt{Gto3P2OctetSplitFn}}. 

\subsection{Diquark Splittings}
\label{subsec:diquark}

Diquark splittings enable the simulation of doubly heavy baryons such as $\Xi_{cc}$, $\Xi_{bc}$, and $\Xi_{bb}$ within the parton shower by modelling the formation of compact colour-antitriplet $QQ'$ pairs. Analogous to quarkonium production, these transitions capture the dominant kinematic configurations leading to baryon formation, particularly at high transverse momentum. In \textsf{Herwig~~7}, the implemented splitting kernels provide effective $1 \to 2$ transitions corresponding to the emission of spin-0 or spin-1 heavy diquark clusters. These splittings are especially relevant for current and upcoming searches for doubly heavy baryons at hadron colliders~\cite{Falk:1993gb,Ma:2003zk,Chen:2017sbg}, where accurate predictions of production rates and kinematic distributions depend sensitively on the modelling of diquark formation. Their inclusion enhances the realism of simulated final states in both inclusive and exclusive measurements.

The implementation of diquark splittings in \textsf{Herwig~7} includes several dedicated classes for simulating the production of heavy diquark systems via quark-to-diquark transitions. These fall into three structurally distinct configurations:
\begin{itemize}
    \item \textcolor{blue}{\texttt{QtoQBarQQ1SplitFn}}: $q \to \bar{q} + (qq)_1$, corresponding to same-flavour antiquark emission and formation of a spin-1 (axial-vector) diquark, e.g.\ $c \to \bar{c}\,cc_1$ or $b \to \bar{b}\,bb_1$. This class is structurally analogous to \textcolor{blue}{\texttt{QtoQ3S1SplitFn}}, used for spin-triplet quarkonium;
    \item \textcolor{blue}{\texttt{QtoQPBarQQP0SplitFn}}: $q \to \bar{q}' + (qq')_0$, involving the emission of a flavour-distinct antiquark and formation of a spin-0 (scalar) diquark, such as $c \to \bar{b}\,bc_0$ or $b \to \bar{c}\,bc_0$. This class corresponds to the quarkonium pseudoscalar splitting \textcolor{blue}{\texttt{QtoQP1S0SplitFn}};
    \item \textcolor{blue}{\texttt{QtoQPBarQQP1SplitFn}}: $q \to \bar{q}' + (qq')_1$, forming a spin-1 diquark, e.g.\ $c \to \bar{b}\,bc_1$ or $b \to \bar{c}\,bc_1$, analogous to the vector quarkonium splitting \textcolor{blue}{\texttt{QtoQP3S1SplitFn}}.
\end{itemize}
These splittings model baryon-like colour-triplet correlations within the parton shower, thereby extending coverage of non-perturbative spin-flavour structures. Scalar diquarks, typically more tightly bound, contribute dominantly to low-lying baryon states, while axial-vector diquarks populate higher-spin multiplets. The spin assignments follow SU(3) colour and spin-symmetry constraints, with room for future phenomenological tuning.

The diquark and quarkonium splitting classes in \textsf{Herwig~7} share a unified structural template for Sudakov evolution. Each diquark splitting function has a one-to-one correspondence with a quarkonium analogue, differing only in the prefactor used to compute the branching probability. These normalisation factors can be generically written as
\begin{equation}
F_{q \to q' \mathcal{O}_1({}^nL_J)}(z, q_0^2) \propto \frac{\alpha_S^2\, C_F\, |R_{{}^nL}(0)|^2}{4\pi\, a_2^2\, M},
\end{equation}
where $a_2$ and $M$ represent the phase-space and mass-dependent terms. For spin-1 transitions, the diquark splitting function is suppressed relative to its quarkonium analogue:
\begin{align}
F_{q \to q\, \mathcal{O}_1({}^3S_1)}(z, q_0^2) = \frac{1}{27} F_{q \to q\, (\bar{q}q)}(z, q_0^2),
\end{align}
For spin-0 $bc$-type diquarks versus the pseudoscalar quarkonium channel:
\begin{align}
F_{q \to q'\, \mathcal{O}_1({}^1S_0)}(z, q_0^2) = \frac{2}{9} F_{q \to \bar{q}'\, (qq')}(z, q_0^2),
\end{align}
And for spin-1 $bc$-type diquarks compared to vector quarkonia:
\begin{align}
F_{q \to q'\, \mathcal{O}_1({}^3S_1)}(z, q_0^2) = \frac{2}{27} F_{q \to \bar{q}'\, (qq')}(z, q_0^2).
\end{align}
These relations reflect the relative colour and spin algebra factors between the diquark and quarkonium production mechanisms and are consistent with the scaling expected from non-relativistic bound-state formation.

\subsection{Impact}
\label{sec:impact}

\begin{figure}[htbp]
\centering
\includegraphics[width=.32\textwidth]{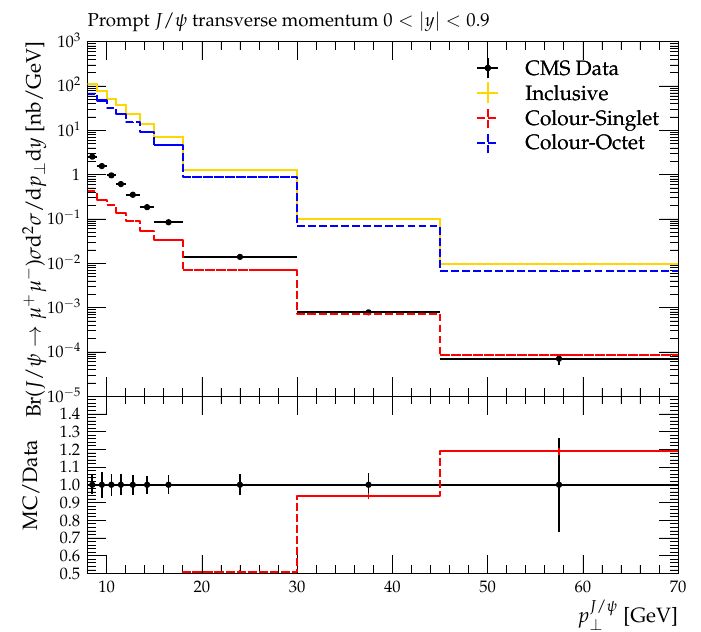}
\includegraphics[width=.32\textwidth]{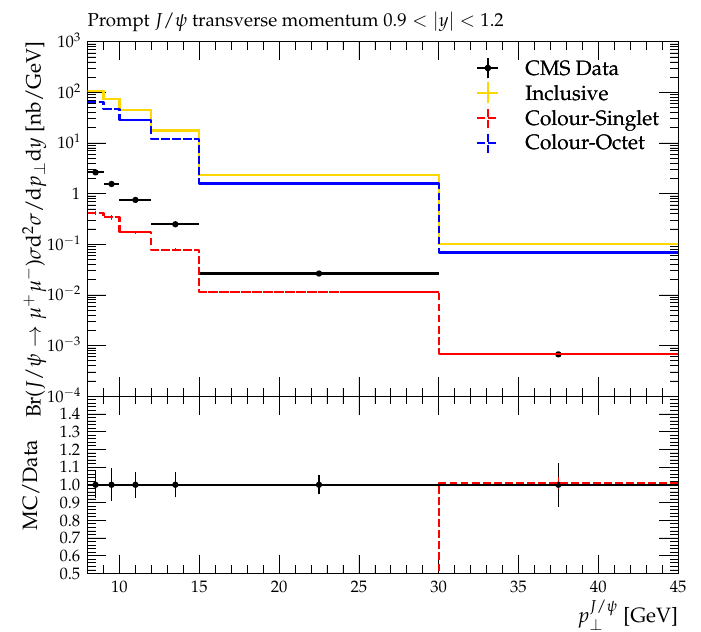}
\includegraphics[width=.32\textwidth]{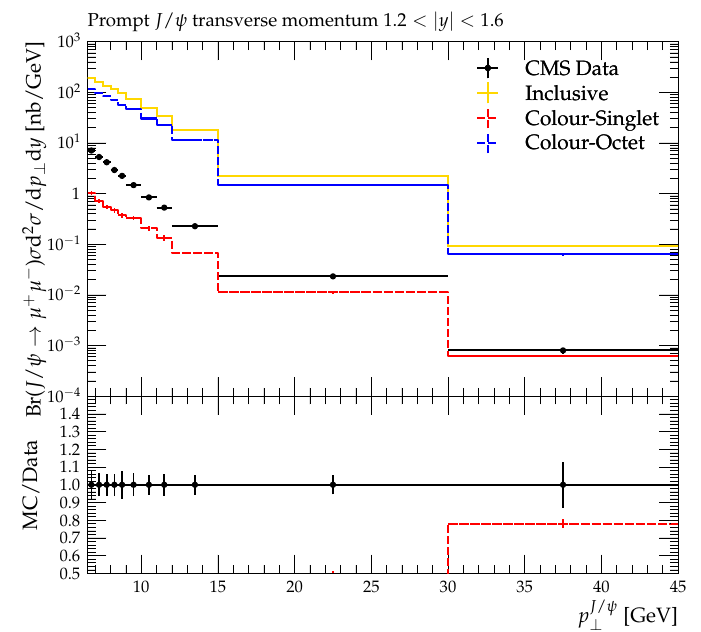}
\includegraphics[width=.32\textwidth]{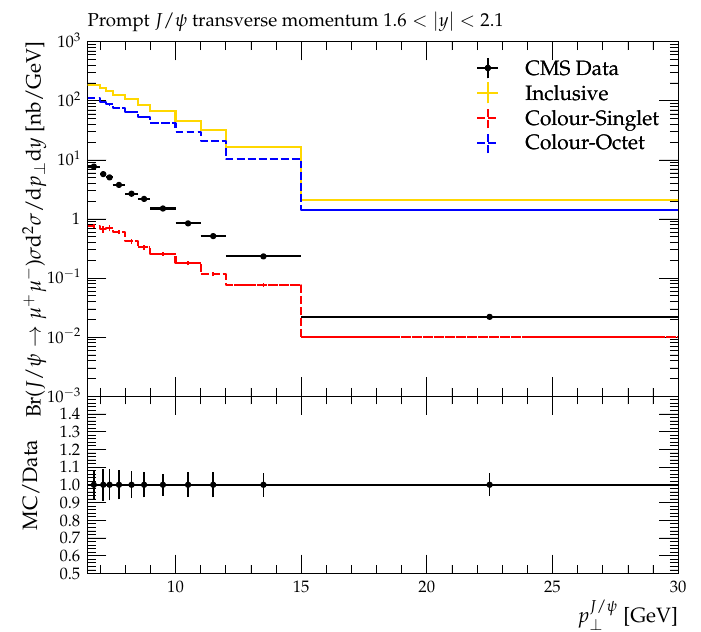}
\includegraphics[width=.32\textwidth]{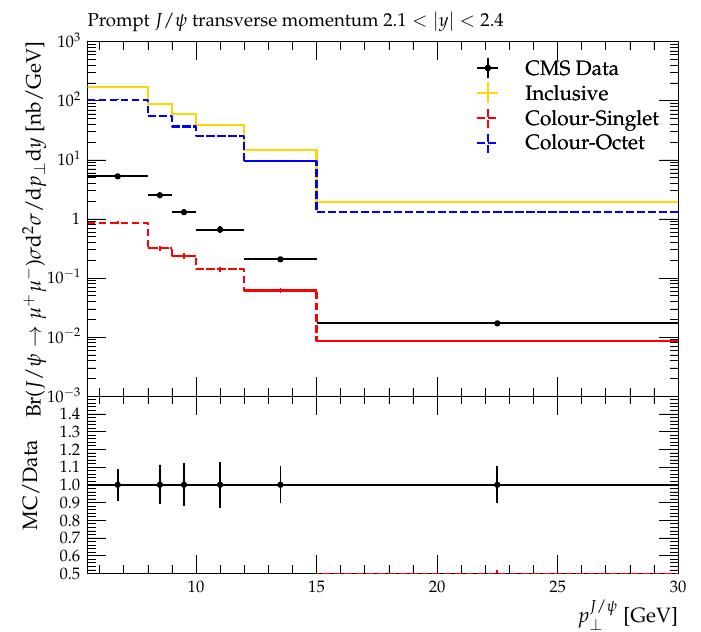}
\caption{\small
Differential cross-section $\mathrm{d}^2\sigma/\mathrm{d}p_T\,\mathrm{d}y$ as a function of $p_T$ for prompt $J/\psi$ production at $\sqrt{s} = 7\,\mathrm{TeV}$, shown in various rapidity bins and compared with CMS data~\cite{CMS:2011rxs}. The dashed lines indicate the individual contributions from colour-singlet (red) and colour-octet (blue) channels. The solid yellow curves denote the inclusive predictions generated using the \textsf{Herwig~7} quarkonium parton shower, including feed-down from higher quarkonium states.}
\label{fig:impactJpsi}
\end{figure}

\begin{figure}[htbp]
\centering
\includegraphics[width=.32\textwidth]{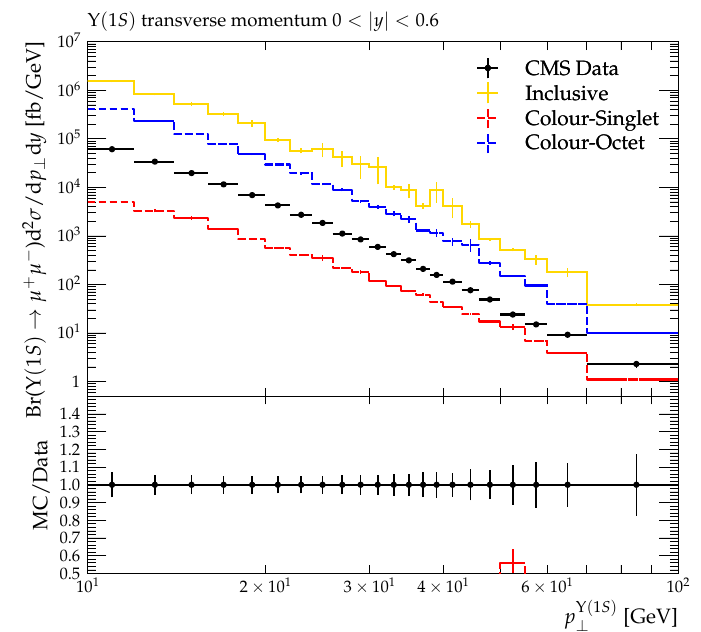}
\includegraphics[width=.32\textwidth]{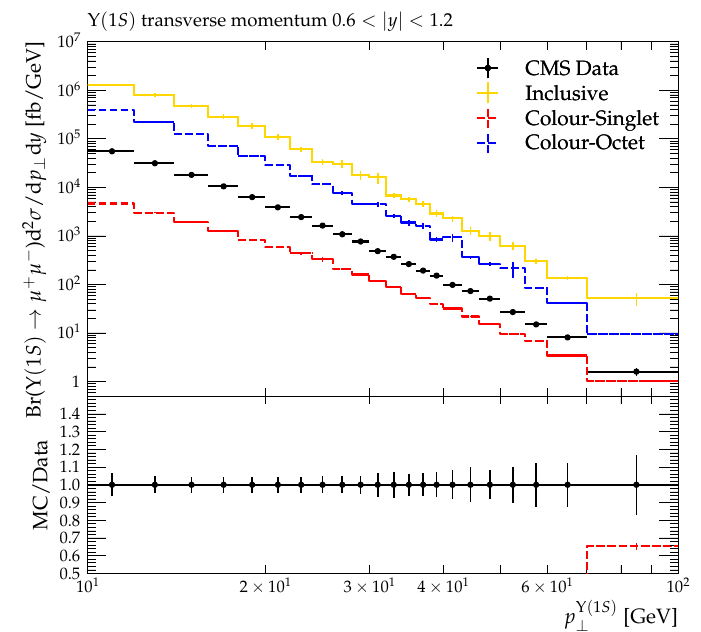}
\includegraphics[width=.32\textwidth]{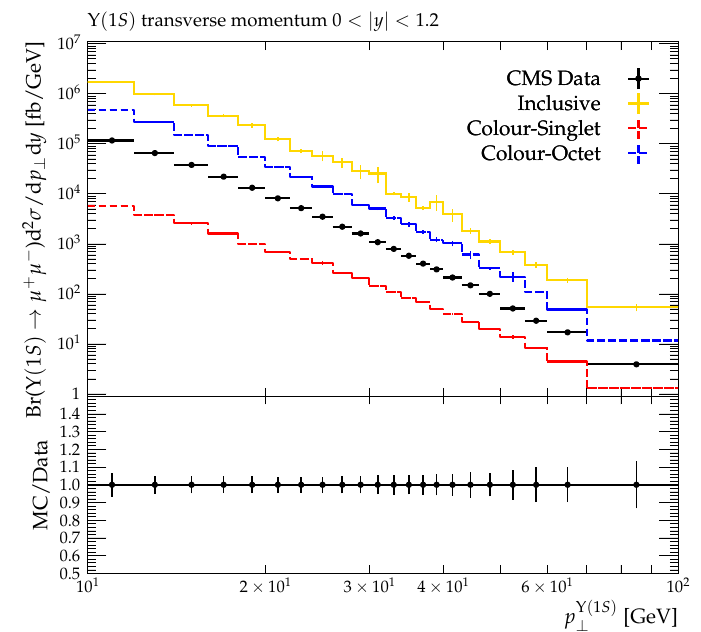}
\caption{\small
Differential cross-section $\mathrm{d}^2\sigma/\mathrm{d}p_T\,\mathrm{d}y$ as a function of $p_T$ for prompt $\Upsilon(1S)$ production at $\sqrt{s} = 7\,\mathrm{TeV}$, shown in various rapidity bins and compared with CMS data~\cite{CMS:2015xqv}. The notation follows that of Fig.~\ref{fig:impactJpsi}.}
\label{fig:impactUpsilon}
\end{figure}

To highlight the effect of the quarkonium parton shower implementation introduced in this work, we present comparisons between \textsf{Herwig~7} predictions and experimental data for prompt quarkonium production at the LHC. Figures~\ref{fig:impactJpsi} and~\ref{fig:impactUpsilon} show differential cross-sections $\mathrm{d}^2\sigma/\mathrm{d}p_T\,\mathrm{d}y$ as functions of the transverse momentum $p_T$ at $\sqrt{s} = 7\,\mathrm{TeV}$ for $J/\psi$ and $\Upsilon(1S)$ production, respectively, across multiple rapidity bins. These results are compared to CMS measurements from Refs.~\cite{CMS:2011rxs,CMS:2015xqv}.

In these plots, the dashed lines in each plot represent the resolved contributions from individual NRQCD channels: the colour-singlet terms are shown in red, while the colour-octet terms are displayed in blue. These correspond to parton-level splittings governed by the short-distance matrix elements of NRQCD, followed by parton shower evolution. Their shapes reflect the expected kinematic behaviour, with singlet contributions dominating at low-$p_T$ and octet terms becoming increasingly relevant in the high-$p_T$ tail due to their enhanced scaling and fragmentation-like structure. The solid yellow histograms represent what we label as the \emph{inclusive} predictions. This nomenclature refers to the fully showered and hadronised quarkonium signal, which includes all prompt production mechanisms: direct production via singlet and octet channels, feed-down from heavier quarkonium states (such as $\chi_{cJ}$ and $\psi(2S)$), and contributions from parton-level and hadronisation-stage dynamics. Non-prompt production via decays of long-lived heavy-flavour hadrons is explicitly excluded to match the experimental selection criteria. These inclusive predictions therefore reflect the complete \emph{prompt} quarkonium signal as reconstructed by the simulation and serve as the baseline observable output of the \textsf{Herwig~7} event generator.

\begin{figure}[htbp]
\centering
\includegraphics[width=.32\textwidth]{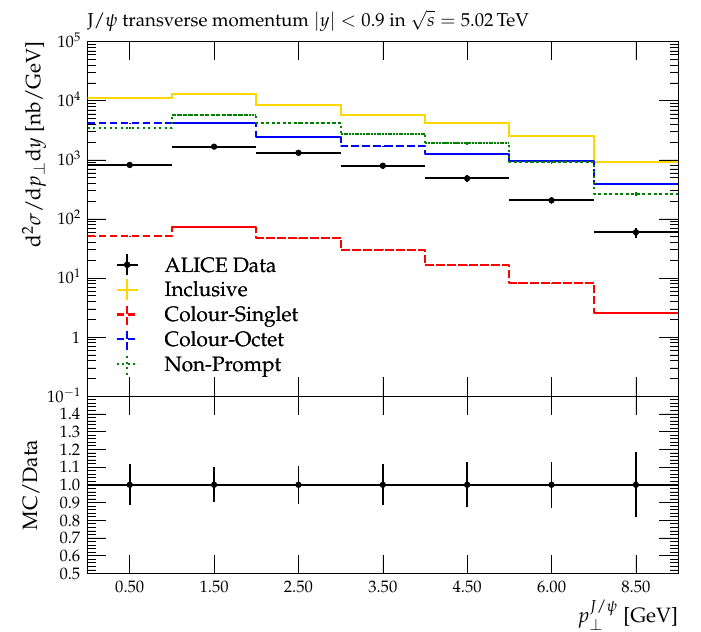}
\includegraphics[width=.32\textwidth]{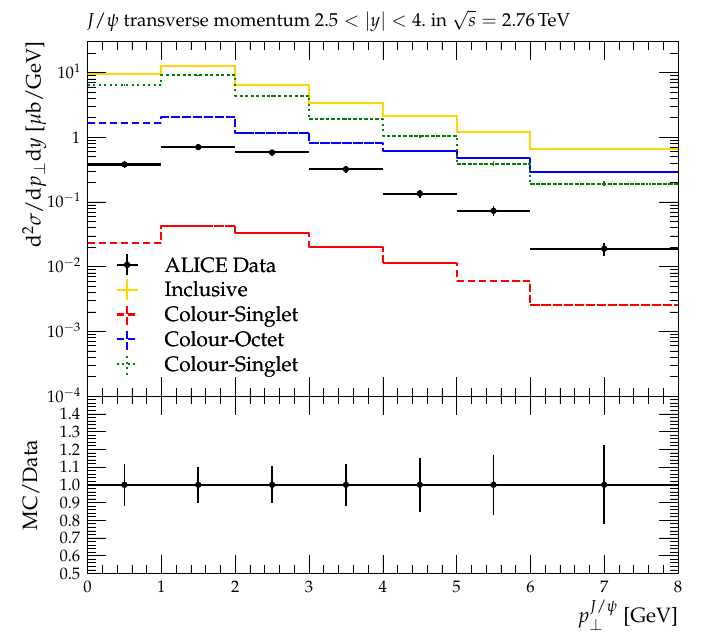}
\includegraphics[width=.32\textwidth]{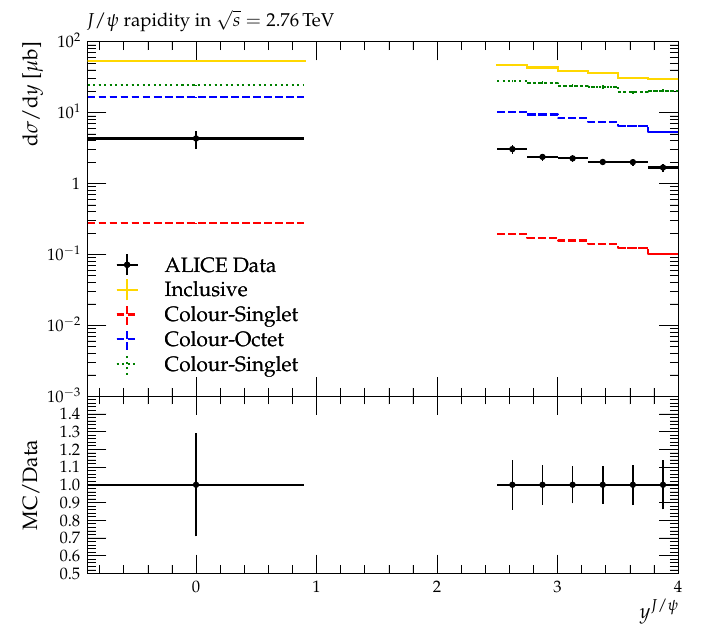}
\caption{\small
Differential cross-sections $\mathrm{d}^2\sigma/\mathrm{d}p_T\,\mathrm{d}y$ (left and middle panels) and $\mathrm{d}^2\sigma/\mathrm{d}y$ (right panel) as functions of $p_T$ and rapidity $y$ for forward $J/\psi$ production at $\sqrt{s} = 5.02\,\mathrm{TeV}$ (left panel) and $\sqrt{s} = 2.76\,\mathrm{TeV}$ (middle and right panels), compared with ALICE data~\cite{ALICE:2012vup,ALICE:2019pid}. The green dotted lines represent non-prompt $J/\psi$ contributions arising from feed-down, hadronisation, and decays of unstable hadrons. The rest of the notation follows that of Fig.~\ref{fig:impactJpsi}.}
\label{fig:impactNP}
\end{figure}

To illustrate the composition of the inclusive quarkonium signal, Fig.~\ref{fig:impactNP} displays differential cross-sections for forward $J/\psi$ production measured by ALICE at $\sqrt{s} = 5.02$ and $2.76$\,TeV~\cite{ALICE:2012vup,ALICE:2019pid}. These measurements are explicitly inclusive, encompassing both prompt and non-prompt $J/\psi$ production. The green dotted curves isolate the non-prompt component, which arises primarily from decays of $b$-hadrons and hadronic feed-down from higher charmonium states. As the figure shows, non-prompt contributions dominate the total cross-section across much of the phase space, especially at low-$p_T$ and large rapidity. Nonetheless, the direct colour-singlet and octet contributions remain non-negligible, particularly at higher transverse momenta. Accurate modelling of these perturbative channels remains essential for precision comparisons and for isolating new-physics effects in heavy-flavour production.

It is important to emphasise that the results shown here are untuned. Generic values have been used for non-perturbative parameters and LDMEs, with no adjustments made to improve agreement with data. The aim of these comparisons is to isolate and highlight the structural impact of the quarkonium parton shower within the event generation framework. A systematic discussion of parameter tuning and the resulting improvements in phenomenological agreement is deferred to the next section.

\section{Long-Distance Matrix Elements, Decay Widths and Tuning}
\label{sec:matrixelementTuning}

Heavy quarkonium phenomenology relies crucially on the determination of non-perturbative parameters that encode the bound-state dynamics beyond leading-order perturbation theory. In this work, these parameters are identified with the radial wavefunctions at the origin, $\bigl|R_{{}^nL}(0)\bigr|^2$ (and their derivatives), as well as the colour-octet MEs, $\langle \mathcal{O}_c({}^nL_J) \rangle$, that enter the NRQCD factorisation formalism. Both singlet and octet contributions are implemented in \textsf{Herwig} to describe quarkonium production and decay.

The decay of a heavy quarkonium state depends on its spin and colour configuration, proceeding through electromagnetic or gluonic channels. For colour-singlet states, processes such as $\mathcal{O}_1({}^3S_1) \to \ell^+\ell^-$, $\mathcal{O}_1({}^1S_0) \to \gamma\gamma$, or gluonic decays like $\mathcal{O}_1({}^1S_0) \to gg$ and $\mathcal{O}_1({}^3S_1) \to 3g$ dominate at leading order. These are direct decays of physical, colour-singlet quarkonium. Colour-octet contributions, by contrast, do not correspond to physical hadrons and do not decay directly into observable final states. Instead, they contribute to the decay rate of the physical singlet states through higher-order partonic subprocesses, which are matched onto the observable decay via non-perturbative soft-gluon transitions. Therefore, comparing theoretical predictions with experimental decay rates enables the extraction and tuning of $\bigl|R_{{}^nL}(0)\bigr|^2$ and, more indirectly, the non-perturbative MEs associated with colour-octet configurations.

The following equations, which are representative of standard potential-model and NRQCD-based approaches, illustrate how the decay width of a $\mathcal{O}_1({}^nL_J)$ state depends on the relevant non-perturbative inputs. They are not exhaustive but serve to highlight the key dependencies on $\bigl|R_{{}^nL}(0)\bigr|^2$ and $\alpha_S$. The theoretical decay widths for the most significant decay channels of heavy quarkonium can be written as:
\allowdisplaybreaks{
\begin{eqnarray}
\Gamma\bigl(\mathcal{O}_1({}^{3}S_1) \to \ell \ell\bigr) &=& \frac{16 \pi \alpha_{\text{em}}^2}{9}\,\frac{\bigl|R_{{}^nL}(0)\bigr|^2}{4 \pi (2m)^2}\,\left(1 - \frac{16}{3}\,\frac{\alpha_S(m^2)}{\pi}\right),
\\[0.05in]
\Gamma\bigl(\mathcal{O}_1({}^1S_0) \to \gamma\gamma\bigr) &=& \frac{48\,\pi\,\alpha_{\text{em}}^2}{81} \frac{\bigl|R_{{}^nL}(0)\bigr|^2}{4 \pi (2m)^2} \,\left(1 - 3.4\,\frac{\alpha_S(m^2)}{\pi}\right),
\\[0.05in]
\Gamma\bigl(\mathcal{O}_1({}^1S_0) \to gg\bigr) &=& \frac{32\,\pi\,\alpha_S^2(m^2)}{3} \frac{\bigl|R_{{}^nL}(0)\bigr|^2}{4 \pi\,(2m)^2}\,\left(1 + 4.4\,\frac{\alpha_S(m^2)}{\pi}\right),
\\[0.05in]
\Gamma\bigl(\mathcal{O}_1({}^3S_1) \to 3g\bigr) &=& \frac{160\,(\pi^2 - 9)\,\alpha_S^3(m^2)}{81} \frac{\bigl|R_{{}^nL}(0)\bigr|^2}{4\pi\,(2m)^2}\,\left(1 - 4.9\,\frac{\alpha_S(m^2)}{\pi}\right),
\\[0.05in]
\Gamma\bigl(\mathcal{O}_1({}^3S_1) \to 3\gamma\bigr) &=& \frac{64\,(\pi^2 - 9)\,\alpha_{\text{em}}^3}{2187} \frac{\bigl|R_{{}^nL}(0)\bigr|^2}{4 \pi \,(2m)^2}\,\left(1 - 12.6\,\frac{\alpha_S(m^2)}{\pi}\right),
\\[0.05in]
\Gamma\bigl(\mathcal{O}_1({}^3S_1) \to gg\gamma\bigr) &=& \frac{128\,(\pi^2 - 9)\,\alpha_{\text{em}}\,\alpha_S^2(m^2)}{81} \frac{\bigl|R_{{}^nL}(0)\bigr|^2}{4 \pi \,(2m)^2}\,\left(1 - 1.7\,\frac{\alpha_S(m^2)}{\pi}\right).
\end{eqnarray}}
These formulae show that colour-singlet states often decay via electromagnetic channels (photon exchange or photon pairs) or via a small number of gluons, whereas colour-octet states necessarily involve higher-order gluonic processes. In practice, one can isolate $\bigl|R_{{}^nL}(0)\bigr|^2$ from the measured total or partial decay widths. This is done by calculating each channel’s width as above, identifying the dominant mechanism for a given state, and fitting $\bigl|R_{{}^nL}(0)\bigr|^2$ (and higher-derivative terms for $P$- and $D$-waves) against experimental data. This procedure provides the baseline colour-singlet wavefunction values.

Once these colour-singlet parameters are established for each charmonium and bottomonium state, production data at high-energy colliders can be used to tune the colour-octet MEs that appear in the NRQCD framework. Hadron-collider cross-sections at moderate and high $p_\perp$ often reveal that purely colour-singlet descriptions underestimate the observed rates of states like $J/\psi$ and $\Upsilon(nS)$. By introducing colour-octet channels with appropriate MEs and comparing to the differential cross-sections measured by ATLAS and CMS, it is possible to adjust these values to obtain improved agreement. In addition, feed-down from higher excited states must be taken into account, particularly in bottomonium. A sizeable fraction of $\Upsilon(1S)$ emerges through radiative transitions from $\chi_{bJ}(nP)$ states. Large branching fractions, such as 33.9\% for $\chi_{b1}(1P)\to \Upsilon(1S)$, imply that even a moderate shift in the production cross-section for $\chi_{bJ}(1P)$ can cause a large variation in the $\Upsilon(1S)$ yield. Hence, the tuning of octet MEs for $P$-wave bottomonia must be carefully moderated so as not to overshoot the eventual $\Upsilon(1S)$ flux.

Tables~\ref{tab:TuneSinglet},~\ref{tab:TuneOctet}, and~\ref{tab:TuneDiquark} summarise the non-perturbative wavefunction parameters for quarkonium and diquark states, as well as the tuned colour-octet matrix elements, extracted from the latest experimental data and theoretical inputs. All values are expressed in units of $\text{GeV}^3$ and are essential for accurately modelling the production and decay of heavy-quark systems in high-energy collisions. Only the values shown in bold font are directly extracted using our tuning framework. For Tables~\ref{tab:TuneSinglet} and~\ref{tab:TuneDiquark}, the remaining (non-bold) colour-singlet parameters are taken from Ref.~\cite{Eichten:2019hbb} for quarkonium and from Ref.~\cite{Falk:1993gb} for diquark states. This is because reliable experimental data on total and partial decay widths are not available for many observed quarkonium states, a limitation reflected in the most recent PDG compilations~\cite{ParticleDataGroup:2022pth, ParticleDataGroup:2024cfk}.

\begin{table}[h!]
\centering
\begin{minipage}[t]{0.30\textwidth}
\centering
\resizebox{0.9\textwidth}{!}{%
\begin{tabular}{c c c}
Type & Structure & $|R_{{}^nL}(0)|^2$ \\
\hline \hline
$c\bar{c}$ & ${}^1S$ & \textbf{1.0285} \\ 
$c\bar{c}$ & ${}^1P$ & \textbf{0.0013} \\
$c\bar{c}$ & ${}^1D$ & 0.0329 \\
$c\bar{c}$ & ${}^2S$ & \textbf{0.4262} \\
$c\bar{c}$ & ${}^2P$ & 0.1767 \\
$c\bar{c}$ & ${}^2D$ & 0.0692 \\
$c\bar{c}$ & ${}^3S$ & 0.5951 \\
$c\bar{c}$ & ${}^3P$ & 0.2106 \\
$c\bar{c}$ & ${}^3D$ & 0.1074 \\
$c\bar{c}$ & ${}^4S$ & 0.5461 \\
$c\bar{c}$ & ${}^4P$ & 0.2389 \\
$c\bar{c}$ & ${}^5S$ & 0.5160 \\
\hline
\end{tabular}}
\end{minipage}
\begin{minipage}[t]{0.30\textwidth}
\centering
\resizebox{0.9\textwidth}{!}{%
\begin{tabular}{c c c}
Type & Structure & $|R_{{}^nL}(0)|^2$ \\
\hline \hline
$b\bar{b}$ & ${}^1S$ & \textbf{0.6364} \\
$b\bar{b}$ & ${}^1P$ & 1.6057 \\
$b\bar{b}$ & ${}^1D$ & 0.8394 \\
$b\bar{b}$ & ${}^2S$ & \textbf{0.2300} \\
$b\bar{b}$ & ${}^2P$ & 1.8240 \\
$b\bar{b}$ & ${}^2D$ & 1.5572 \\
$b\bar{b}$ & ${}^3S$ & \textbf{0.5548} \\
$b\bar{b}$ & ${}^3P$ & 1.9804 \\
$b\bar{b}$ & ${}^3D$ & 2.2324 \\
$b\bar{b}$ & ${}^4S$ & 1.2863 \\
$b\bar{b}$ & ${}^4P$ & 2.1175 \\
$b\bar{b}$ & ${}^4D$ & 2.8903 \\
$b\bar{b}$ & ${}^5S$ & 1.7990 \\
$b\bar{b}$ & ${}^5P$ & 2.2430 \\
$b\bar{b}$ & ${}^5D$ & 3.5411 \\
$b\bar{b}$ & ${}^6S$ & 1.6885 \\
$b\bar{b}$ & ${}^6P$ & 2.3600 \\
$b\bar{b}$ & ${}^7S$ & 1.6080 \\
\hline
\end{tabular}}
\end{minipage}
\begin{minipage}[t]{0.30\textwidth}
\centering
\resizebox{0.9\textwidth}{!}{%
\begin{tabular}{c c c}
Type & Structure & $|R_{{}^nL}(0)|^2$ \\
\hline \hline
$b\bar{c}$ & ${}^1S$ & 1.9943 \\
$b\bar{c}$ & ${}^1P$ & 0.3083 \\
$b\bar{c}$ & ${}^1D$ & 0.0986 \\
$b\bar{c}$ & ${}^2S$ & 1.1443 \\
$b\bar{c}$ & ${}^2P$ & 0.3939 \\
$b\bar{c}$ & ${}^2D$ & 0.1989 \\
$b\bar{c}$ & ${}^3S$ & 0.9440 \\
$b\bar{c}$ & ${}^3P$ & 0.4540 \\
$b\bar{c}$ & ${}^4S$ & 0.8504 \\
\hline
\end{tabular}}
\end{minipage}
\caption{\small Non-perturbative Singlet wavefunctions, representing $|R(0)|^2$, $|R'(0)|^2$ and $|R''(0)|^2$ for $S$, $P$ and $D$-state quarkonia, given in $GeV^3$ units. The values highlighted in bold fonts are calculated using the method of Section~\ref{sec:matrixelementTuning}. The rest of these parameters are extracted from~\cite{Eichten:2019hbb}.}
\label{tab:TuneSinglet}
\end{table}

\begin{table}[h!]
\centering
\resizebox{0.42\textwidth}{!}{%
\begin{tabular}{l c c c}
Process & Type & Structure & $\langle \mathcal{O}_8({}^nL_J) \rangle$ \\
\hline \hline
$g \to J/\psi$        & $c\bar{c}$ & ${}^1S_3$ & $1.09\times10^{-4}$ \\ 
$g \to \psi(2S)$      & $c\bar{c}$ & ${}^1S_3$ & $6.23\times10^{-5}$ \\

$g \to \chi_{c0}$     & $c\bar{c}$ & ${}^1S_3$ & $5.99\times10^{-5}$ \\
$g \to \chi_{c1}$     & $c\bar{c}$ & ${}^1S_3$ & $1.80\times10^{-4}$ \\
$g \to \chi_{c2}$     & $c\bar{c}$ & ${}^1S_3$ & $2.99\times10^{-4}$ \\
$g \to \Upsilon(1S)$  & $b\bar{b}$ & ${}^1S_3$ & $2.17\times10^{-4}$ \\
$g \to \Upsilon(2S)$  & $b\bar{b}$ & ${}^1S_3$ & $1.14\times10^{-4}$ \\
$g \to \Upsilon(3S)$  & $b\bar{b}$ & ${}^1S_3$ & $6.88\times10^{-5}$ \\

$g \to \chi_{b0}$     & $b\bar{b}$ & ${}^1S_3$ & $1.55\times10^{-4}$ \\ 
$g \to \chi_{b1}$     & $b\bar{b}$ & ${}^1S_3$ & $4.65\times10^{-4}$ \\   
$g \to \chi_{b2}$     & $b\bar{b}$ & ${}^1S_3$ & $7.75\times10^{-4}$ \\   

$g \to \chi_{b0}(2P)$ & $b\bar{b}$ & ${}^1S_3$ & $1.55\times10^{-4}$ \\ 
$g \to \chi_{b1}(2P)$ & $b\bar{b}$ & ${}^1S_3$ & $4.65\times10^{-4}$ \\   
$g \to \chi_{b2}(2P)$ & $b\bar{b}$ & ${}^1S_3$ & $7.75\times10^{-4}$ \\

$g \to \chi_{b0}(3P)$ & $b\bar{b}$ & ${}^1S_3$ & $1.55\times10^{-4}$ \\ 
$g \to \chi_{b1}(3P)$ & $b\bar{b}$ & ${}^1S_3$ & $4.65\times10^{-4}$ \\   
$g \to \chi_{b2}(3P)$ & $b\bar{b}$ & ${}^1S_3$ & $7.75\times10^{-4}$ \\
\hline
\end{tabular}}
\caption{\small Tuned octet parameters for selected states using all available experimental data. The values are given in $GeV^3$ units.}
\label{tab:TuneOctet}
\end{table}

\begin{table}[h!]
\centering
\resizebox{0.35\textwidth}{!}{%
\begin{tabular}{c c c}
Diquark Type & Structure & $|R_{{}^nL}(0)|^2$ \\
\hline \hline
$cc$ & ${}^1S$ & 1.9943 \\
$bb$ & ${}^1S$ & 0.3083 \\
$bc$ & ${}^1S$ & 0.0986 \\
\hline
\end{tabular}}
\caption{\small Non-perturbative Singlet wavefunctions $|R(0)|^2$ for $S$-state diquarks, given in $GeV^3$ units. The values of these parameters are extracted from~\cite{Falk:1993gb}.}
\label{tab:TuneDiquark}
\end{table}

After the wavefunction parameters were tuned using decay widths, the colour-octet MEs were adjusted by comparing theoretical predictions of shower prompt production cross-sections to LHC data for states such as $J/\psi$, $\psi(2S)$, $\chi_{cJ}$, $\Upsilon(1S)$, $\Upsilon(2S)$, and $\Upsilon(3S)$. The modifications to the octet MEs were often performed through multiplicative rescalings, with relative ratios guided by spin-counting or earlier theoretical arguments. In cases like $\chi_{bJ}(nP)$, where feed-down to $\Upsilon(1S)$ is crucial, the $P$-wave octet parameters were significantly lowered to prevent an overestimate of the final $\Upsilon(1S)$ yield. This tuning procedure mitigated the tendency of the fit to drive the $\Upsilon(1S)$ octet component to unphysical values in order to compensate for too-large $P$-wave feed-down.

While tuning the colour-octet non-perturbative MEs, $p_T$ bins below 10 GeV are neglected due to the dominance of non-perturbative effects such as soft gluon emissions, multiple parton interactions, and feed-down contributions from higher excited states. The breakdown of collinear factorisation, increased sensitivity to underlying event activity, and larger theoretical uncertainties in scale dependence further limit the reliability of perturbative QCD-based models in this region. Additionally, experimental resolution and acceptance issues complicate the extraction of clean signals at low $p_T$, potentially biasing the tuning process. By focusing on $p_T > 10$ GeV, where hard scattering dynamics prevail, the parton shower can be tuned more accurately to perturbative QCD predictions.

\section{Comparing Results and Discussions}
\label{sec:results}

In this section, we compare our predictions with experimental measurements for both charmonium and bottomonium production. In all figures, we show the ATLAS data points together with three theoretical curves: (i) the default \textit{Generic Quarkonium Shower} in \textsf{Herwig~7} (blue lines), which are the predictions produced with our newly implemented shower before tuning, (ii) the \textit{Default Pythia8 Quarkonium Shower} produced using \textsf{Pythia8.313} (green lines), and (iii) the \textit{Tuned Quarkonium Shower} model, which will form part of the complete implementation in \textsf{Herwig~7.4.0} (red lines).  

For the \textsf{Herwig~7} predictions, we use internal MEs to simulate $2\to2$ QCD processes, including either explicit heavy quark pair production ($g g \to Q\bar{Q}$, $q\bar{q} \to Q\bar{Q}$, with $Q=b,c$) for colour-singlet states, or inclusive light-flavour processes for colour-octet contributions. Event generation is partitioned into five transverse momentum intervals: $0$-$10$ GeV, $10$-$20$ GeV, $20$-$40$ GeV, $40$-$80$ GeV, and $80$ GeV to the luminosity limit. The resulting samples are analysed with \textsf{Rivet-4.0.1}~\cite{Buckley:2010ar,Bierlich:2024vqo} and \textsf{YODA-2.0.1}~\cite{Buckley:2023xqh}. Further tuning details are provided in Section~\ref{sec:matrixelementTuning}. Calculations have been performed at centre-of-mass energies of $5.02$ TeV and $7$ TeV, and compared to ATLAS data from Refs.~\cite{ATLAS:2017prf, ATLAS:2014ala}. For the \textsf{Pythia8} predictions, we adopt the default quarkonium shower settings as provided in their standard example templates, following the official online tutorial guidance. Both the \textsf{Herwig~7} and \textsf{Pythia8} simulations are performed with comparable statistics and similar $p_T$ cuts. 

\subsection{Charmonia}

\begin{figure} [h]
\centering
\includegraphics[width=.32\textwidth]{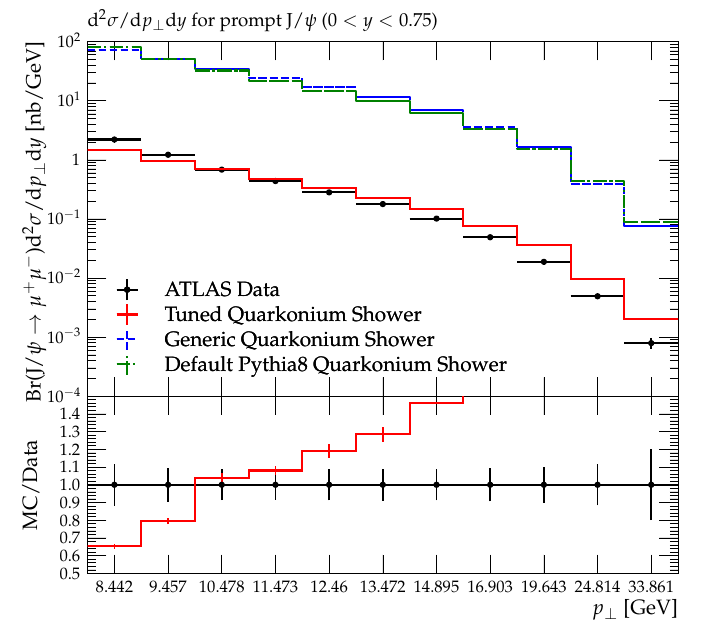}
\includegraphics[width=.32\textwidth]{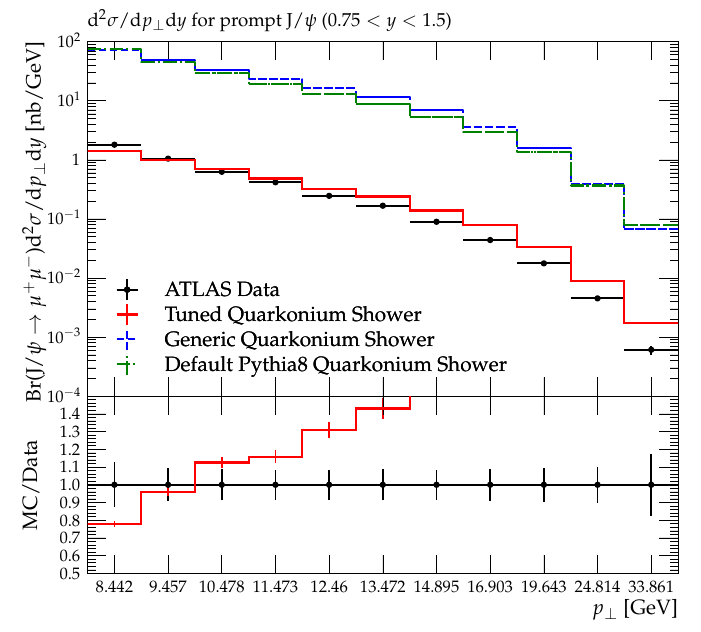}
\includegraphics[width=.32\textwidth]{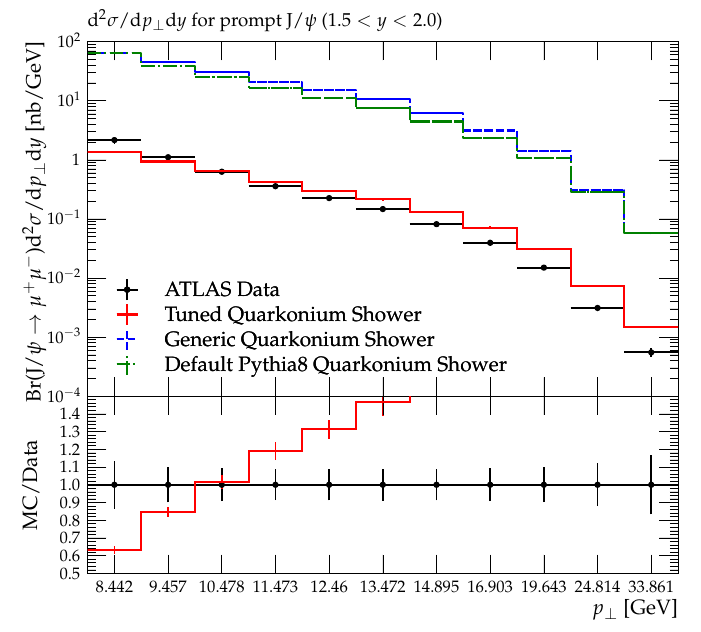}
\caption{\small Differential cross-section $d^2\sigma/(dp_T dy)$ as a function of $p_T$ for prompt $J/\psi$ production in $\sqrt{s}=5.02$ TeV within various rapidity bins, compared with ATLAS data \cite{ATLAS:2017prf}. The tuned quarkonium parton shower (red lines) in \textsf{Herwig} shows improved agreement compared to the generic model (blue lines). Disabling the quarkonium parton shower results in no signal.}
\label{fig:Jpsi}
\end{figure}

\begin{figure} [h]
\centering
\includegraphics[width=.32\textwidth]{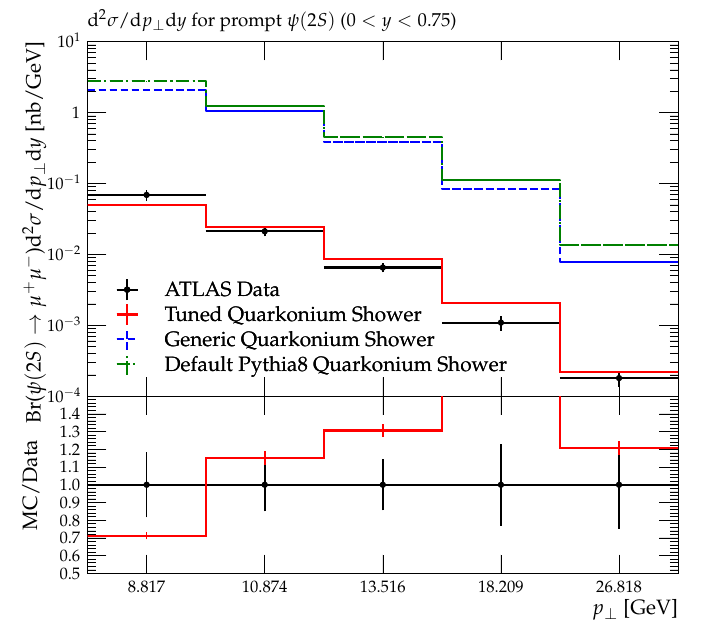}
\includegraphics[width=.32\textwidth]{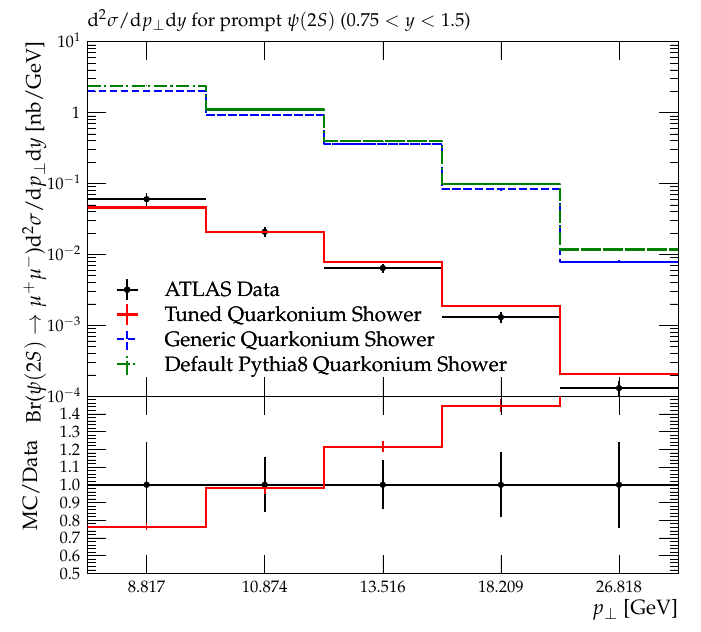}
\includegraphics[width=.32\textwidth]{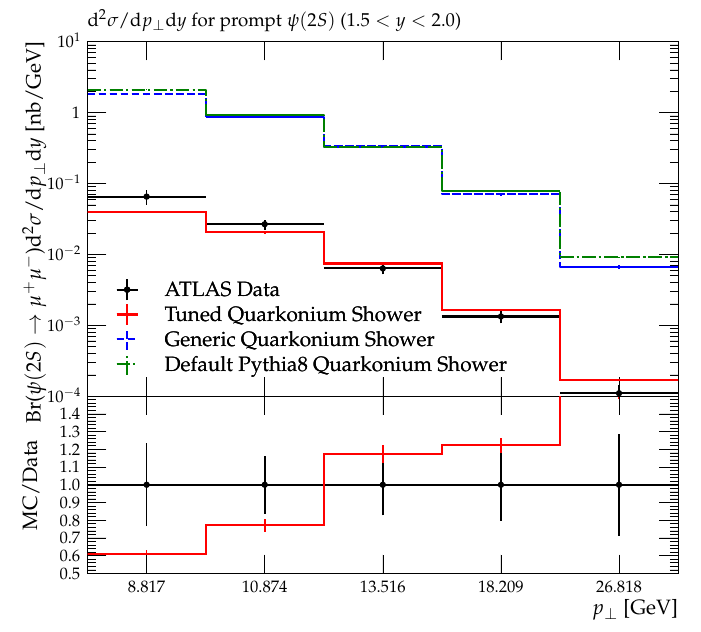}
\caption{\small Differential cross-section $d^2\sigma/(dp_T dy)$ as a function of $p_T$ for prompt $\psi(2S)$ production in $\sqrt{s}=5.02$ TeV within various rapidity bins, compared with ATLAS data \cite{ATLAS:2017prf}. The notation follows the same convention as in Figure~\ref{fig:Jpsi}.}
\label{fig:psi2S}
\end{figure}

\begin{figure} [h]
\centering
\includegraphics[width=.32\textwidth]{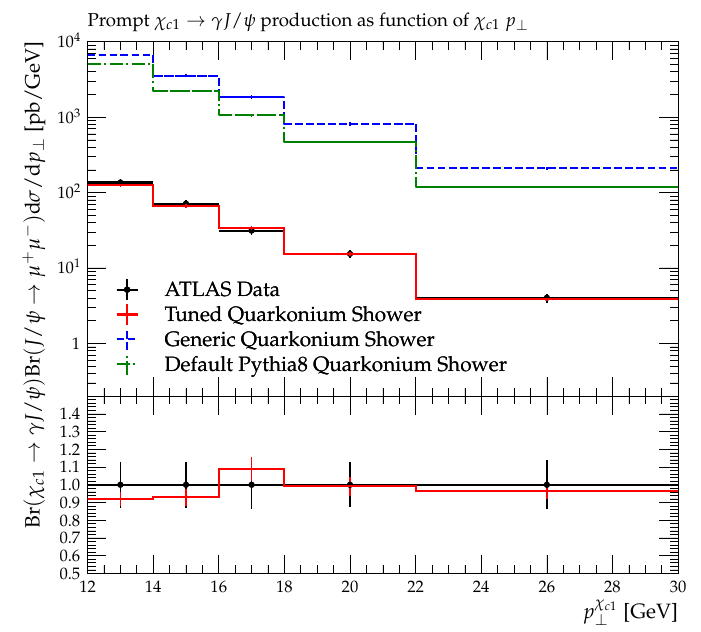}
\includegraphics[width=.32\textwidth]{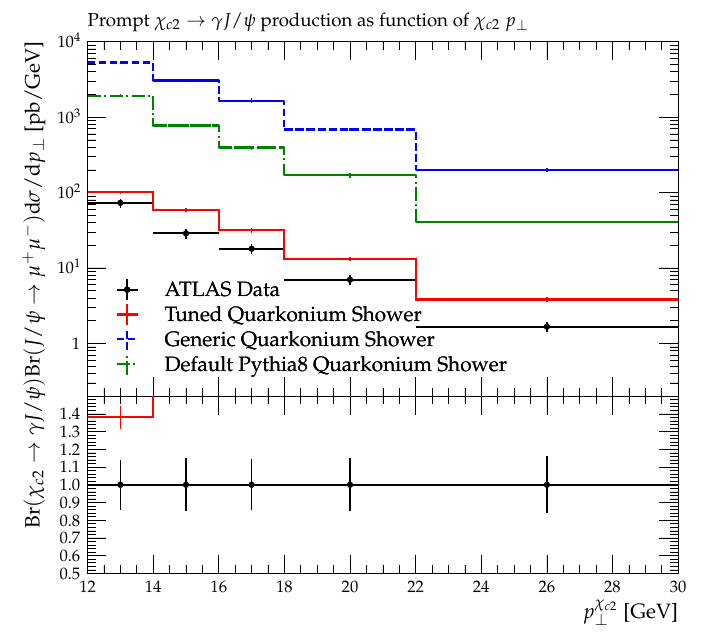}
\caption{\small Differential cross-section $d^2\sigma/dp_T$ as a function of $p_T$ for prompt $\chi_{c1}$ (left) and $\chi_{c2}$ (right) production in $\sqrt{s}=7$ TeV, compared with ATLAS data \cite{ATLAS:2014ala}. The notation follows the same convention as in Figure~\ref{fig:Jpsi}.}
\label{fig:chi_c}
\end{figure}

Figures~\ref{fig:Jpsi}, \ref{fig:psi2S}, and \ref{fig:chi_c} present the differential cross-sections 
${d^2 \sigma}/({d p_T \, d y})$
as functions of $p_T$ for prompt $J/\psi$, $\psi(2S)$, and $\chi_{c1}, \chi_{c2}$ production, respectively. The new quarkonium shower exhibits a marked improvement in the overall normalisation and shape, particularly at intermediate and high $p_T$. In contrast, the generic shower often underestimates the cross-sections and displays a flatter slope in the tail region. This underlines the importance of both colour-singlet and colour-octet channels, as well as a proper treatment of feed-down from excited states. Although agreement is generally good, some tension remains at very low $p_T$, where non-perturbative effects such as multiple parton interactions, non-factorisable soft gluon exchanges, and hadronisation details can become dominant. Overall, however, the tuned quarkonium shower captures the main features of charmonium production more accurately than the default model.

\subsection{Bottomonia}

\begin{figure} [h]
\centering
\includegraphics[width=.32\textwidth]{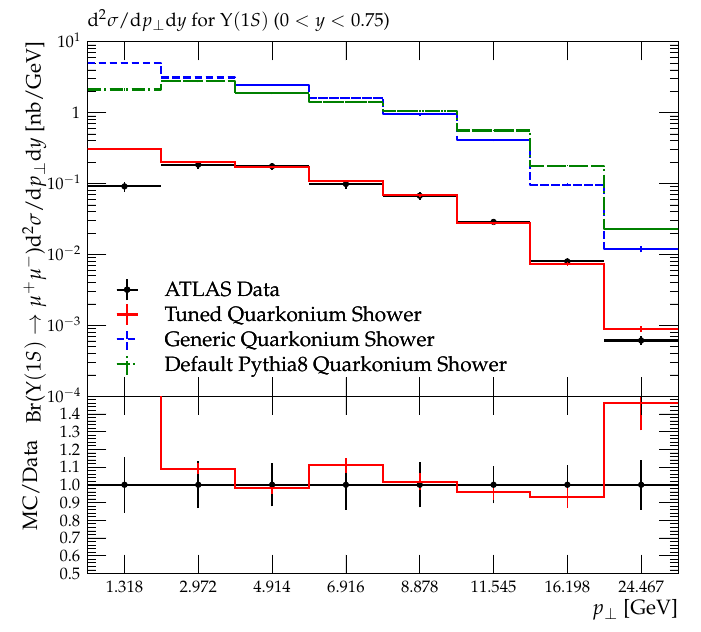}
\includegraphics[width=.32\textwidth]{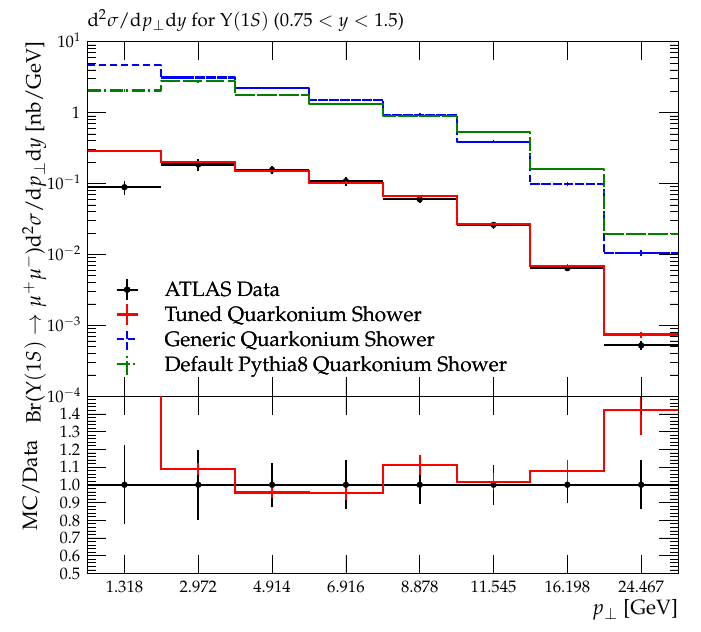}
\includegraphics[width=.32\textwidth]{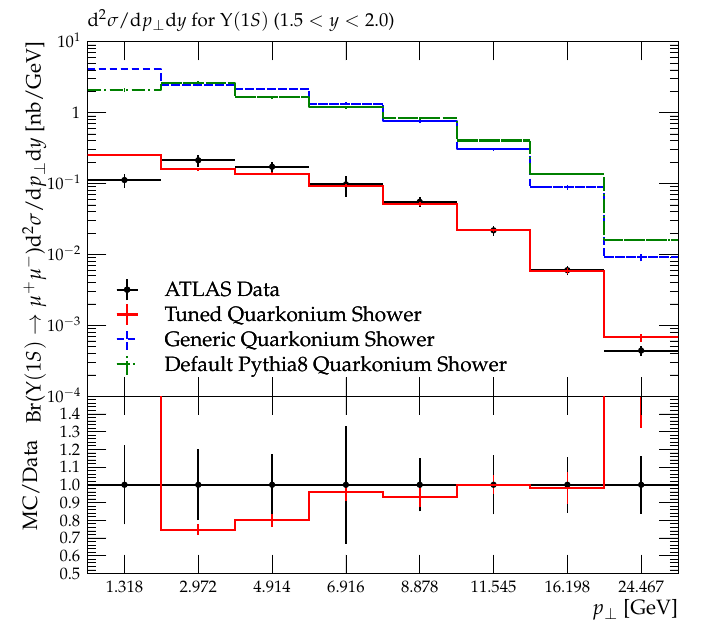}
\caption{\small Differential cross-section $d^2\sigma/(dp_T dy)$ as a function of $p_T$ for prompt $\Upsilon(1S)$ production in $\sqrt{s}=5.02$ TeV within various rapidity bins, compared with ATLAS data \cite{ATLAS:2017prf}. The notation follows the same convention as in Figure~\ref{fig:Jpsi}.}
\label{fig:Upsilon1S}
\end{figure}

\begin{figure} [h]
\centering
\includegraphics[width=.32\textwidth]{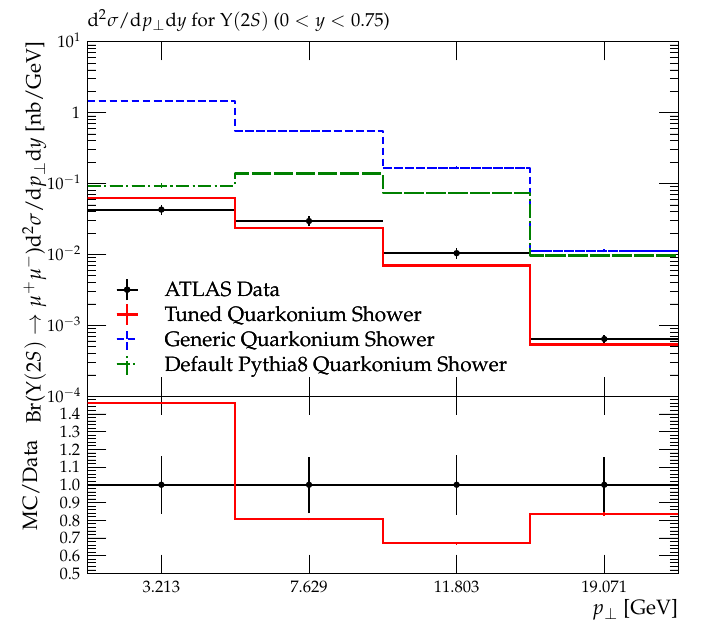}
\includegraphics[width=.32\textwidth]{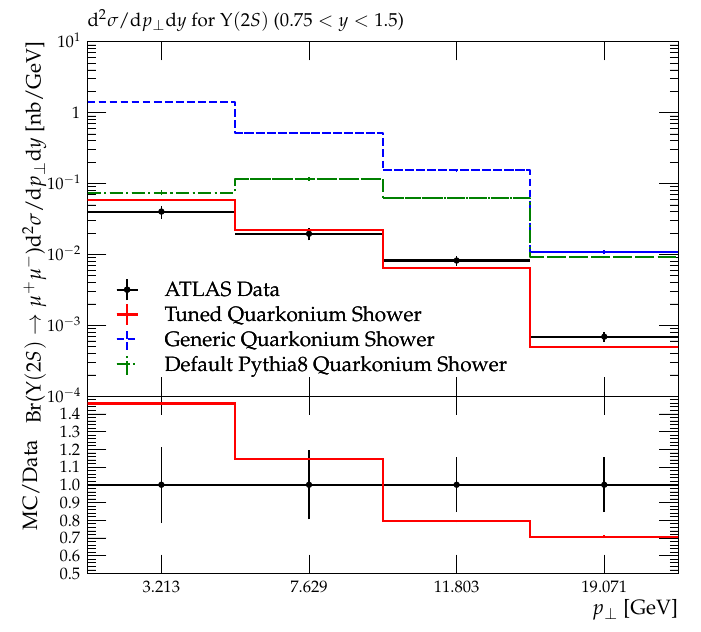}
\includegraphics[width=.32\textwidth]{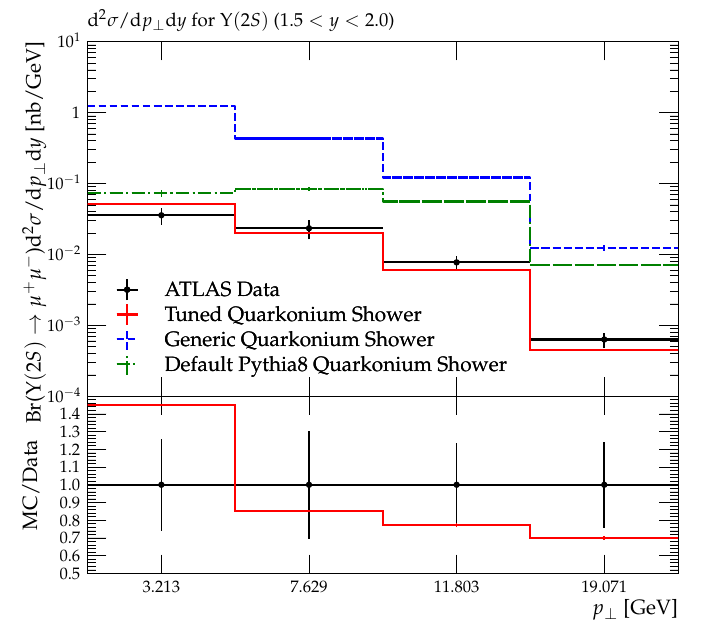}
\caption{\small Differential cross-section $d^2\sigma/(dp_T dy)$ as a function of $p_T$ for prompt $\Upsilon(2S)$ production in $\sqrt{s}=5.02$ TeV within various rapidity bins, compared with ATLAS data \cite{ATLAS:2017prf}. The notation follows the same convention as in Figure~\ref{fig:Jpsi}.}
\label{fig:Upsilon2S}
\end{figure}

\begin{figure} [h]
\centering
\includegraphics[width=.32\textwidth]{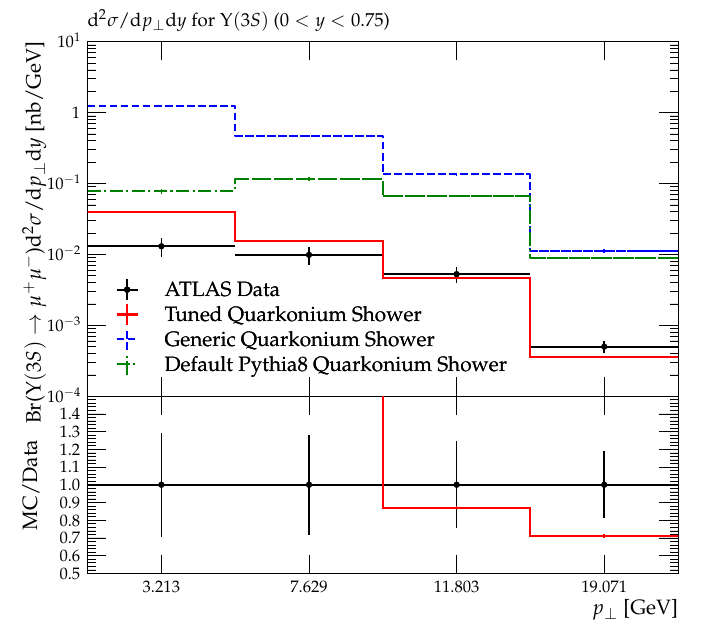}
\includegraphics[width=.32\textwidth]{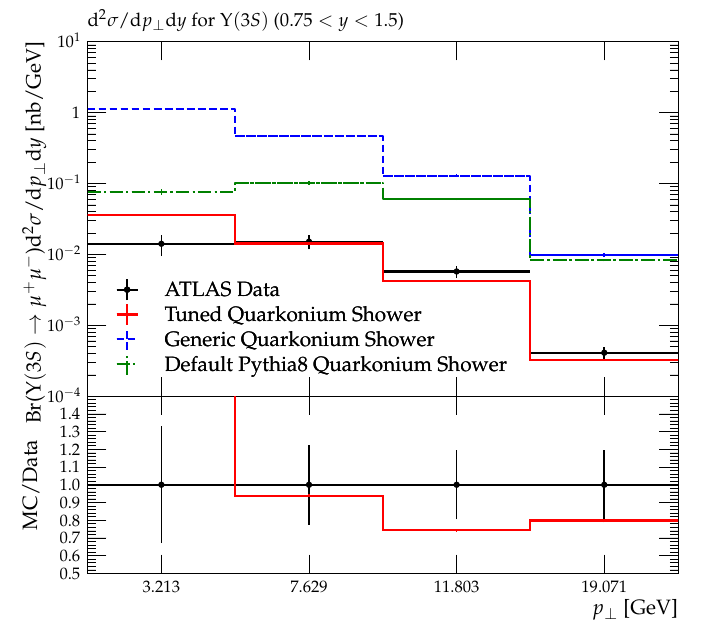}
\includegraphics[width=.32\textwidth]{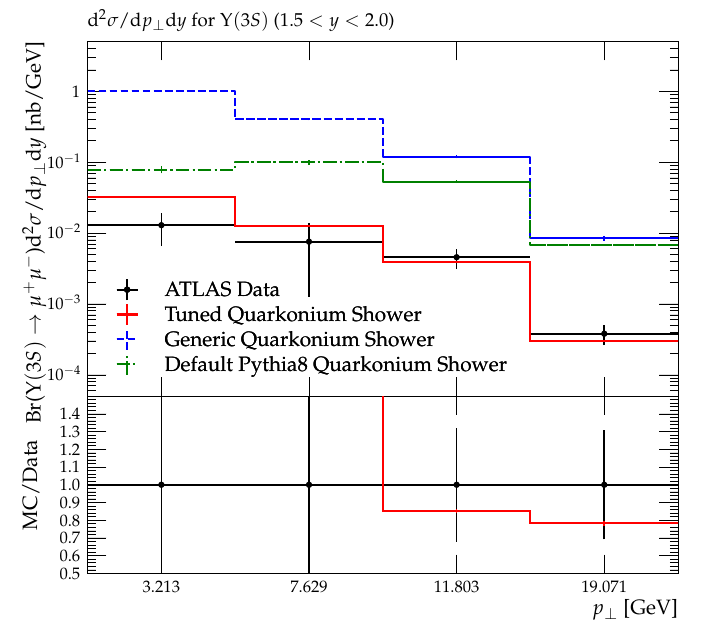}
\caption{\small Differential cross-section $d^2\sigma/(dp_T dy)$ as a function of $p_T$ for prompt $\Upsilon(3S)$ production in $\sqrt{s}=5.02$ TeV within various rapidity bins, compared with ATLAS data \cite{ATLAS:2017prf}. The notation follows the same convention as in Figure~\ref{fig:Jpsi}.}
\label{fig:Upsilon3S}
\end{figure}

Turning to bottomonia, Figures~\ref{fig:Upsilon1S}, \ref{fig:Upsilon2S}, and \ref{fig:Upsilon3S} show analogous differential cross-sections for prompt $\Upsilon(1S)$, $\Upsilon(2S)$, and $\Upsilon(3S)$ production. Again, we observe that the tuned quarkonium parton shower tracks the measured shapes better than the generic shower. For $\Upsilon(2S)$ and $\Upsilon(3S)$, the improvements are especially evident in the mid-to-high $p_T$ regime. After tuning the non-perturbative bounding energies, the new shower model achieves a substantially improved description of bottomonium spectra. The fact that it simultaneously describes $\Upsilon(1S)$, $\Upsilon(2S)$ and $\Upsilon(3S)$ data with consistent quarkonium parameters underscores its robustness. Overall, these comparisons confirm that the newly implemented quarkonium parton shower in \textsf{Herwig} successfully captures the main kinematic trends in both charmonium and bottomonium production, providing a more realistic simulation framework for heavy quarkonia at the LHC.

\section{Conclusions}
\label{sec:conc}

In this paper, we have presented the implementation of a quarkonium parton shower within the \textsf{Herwig~7} framework, incorporating both colour-singlet and colour-octet channels based on the NRQCD factorisation approach. Specifically, we include colour-singlet channels such as $q \to q' \mathcal{O}_1$ and $g \to g \mathcal{O}_1$, colour-octet channels like $g \to \mathcal{O}_8$, and also diquark production processes. Our implementation systematically accounts for $S$-, $P$-, and $D$-wave quarkonium states, along with feed-down contributions from higher excitations.

We detail the underlying splitting kinematics, MEs, splitting functions, and associated probabilities, all of which are integrated into a consistent parton shower framework. The shower comprises 30 newly implemented splitting classes, designed to accommodate 103 registered splittings. These are fully customisable and can be expanded as new experimental data becomes available, allowing the model to remain flexible and future-proof. To accurately simulate heavy-quarkonium production, we have tuned both the non-perturbative MEs (relevant for colour-octet states) and wavefunctions (critical for colour-singlet and diquark cases) using a hybrid methodology. This incorporates available experimental results, established theoretical predictions, and our own computations to constrain the relevant parameters effectively.

Our implementation has been evaluated in terms of its impact on both inclusive and prompt production of quarkonia states. Comparisons with ATLAS data for charmonium and bottomonium production, alongside the \textsf{Pythia8} quarkonium parton shower, demonstrate that our fully tuned \textsf{Herwig~7} quarkonium shower noticeably improves the description of observables, particularly in the intermediate to high $p_T$ regimes. The model captures the key phenomenological features of both charmonium and bottomonium spectra, outperforming the generic default shower in terms of shape fidelity and cross-section normalisation. Some residual discrepancies at low $p_T$ suggest further refinements in modelling non-perturbative effects such as multiple parton interactions and hadronisation. Nonetheless, the overall agreement with experimental observations affirms the robustness of our approach.

This development considerably improves the predictive capabilities of \textsf{Herwig~7} in the domain of heavy-flavour physics, offering a comprehensive and reliable tool for simulating quarkonium production at the LHC. Future efforts will aim to refine the non-perturbative tuning via global fits and to include higher-order QCD corrections, enhancing the accuracy and scope of the simulation. The quarkonium parton shower presented here will be publicly available with the release of \textsf{Herwig~7.4.0}.

\section*{Acknowledgements}
\noindent We thank our fellow \textsf{Herwig} authors for useful discussions. This work has received funding from the European Union’s Horizon 2020 research and innovation programme as part of the Marie Sk\l{}odowska-Curie Innovative Training Network MCnetITN3 (grant agreement no.~722104). \textit{MRM} is also supported by the UK Science and Technology Facilities Council (grant numbers~ST/P001246/1).

\appendix

\section{Available Experimental Data on Quarkonium Production}
\label{sec:AppA}

Tables~\ref{tab:A1} and \ref{tab:A2} provide a comprehensive list of experimental measurements relevant to the study of charmonium and bottomonium production, including cross-section and polarisation observables across various centre-of-mass energies. The data are sourced from LHC experiments (ATLAS, CMS, LHCb, and ALICE) and are presented alongside their corresponding Rivet plugin identifiers. These data sets are used in tuning the non-perturbative parts of the octet states MEs in this study. 

\setcounter{table}{0}
\renewcommand{\thetable}{A\arabic{table}} 

\allowdisplaybreaks{}
\begin{table}[]
  \begin{center}
  \resizebox{0.8\textwidth}{!}{%
    \begin{tabular}{|c|c|c|c|c|}
      \hline
      Observable & Energy & Experiment & Rivet Plugin & Notes \\
      \hline
      \multicolumn{5}{|c|}{$\eta_c(1S)$}\\
      \hline
      $\frac{{\rm d}\sigma}{{\rm d}p_\perp}$ & 7,8\,TeV     & LHCb~\cite{Aaij:2014bga} & \href{https://rivet.hepforge.org/analyses/LHCB_2015_I1316329.html}{\texttt{LHCB\_2015\_I1316329}}     & $2<y<4.5$, $6.5<p_\perp<14$\,GeV \\[1mm]
      $\frac{{\rm d}\sigma}{{\rm d}p_\perp}$ & 13\,TeV & LHCb~\cite{Aaij:2019gsn} & \href{https://rivet.hepforge.org/analyses/LHCB_2020_I1763898.html}{\texttt{LHCB\_2020\_I1763898}}     & $2<y<4.5$, $6.5<p_\perp<14$\,GeV \\[1mm]
      \hline
      \multicolumn{5}{|c|}{$J/\psi$}\\
      \hline
      $\frac{{\rm d}\sigma}{{\rm d}p_\perp}$, $\frac{{\rm d}\sigma}{{\rm d}y}$ & 2.76\,TeV &  ALICE~\cite{ALICE:2012vup} & \href{https://rivet.hepforge.org/analyses/ALICE_2012_I1094079.html}{\texttt{ALICE\_2012\_I1094079}} & $2.5<y<4$, $p_\perp<8$\,GeV \\[1mm]
      $\frac{{\rm d}\sigma}{{\rm d}p_\perp}$ & 2.76\,TeV & LHCb~\cite{LHCb:2012kaz} & \href{https://rivet.hepforge.org/analyses/LHCB_2013_I1205646.html}{\texttt{LHCB\_2013\_I1205646}} & $2<y<4.5$, $p_\perp<12$\,GeV\\[1mm]
      $\frac{{\rm d}\sigma}{{\rm d}p_\perp}$ & 5.02\,TeV & ALICE~\cite{ALICE:2019pid} & \href{https://rivet.hepforge.org/analyses/ALICE_2019_I1735351.html}{\texttt{ALICE\_2019\_I1735351}} &  $|y|<0.9$, $0<p_\perp<10$\,GeV \\[1mm]
      $\frac{{\rm d}^2\sigma}{{\rm d}p_\perp{\rm d}y}$ & 5.02\,TeV & ATLAS~\cite{ATLAS:2017prf} & \href{https://rivet.hepforge.org/analyses/ATLAS_2018_I1622737.html}{\texttt{ATLAS\_2018\_I1622737}} & $|y|<2$, $8<p_\perp<40$\,GeV \\[1mm]
      $\frac{{\rm d}\sigma}{{\rm d}p_\perp}$ & 5.02\,TeV & CMS~\cite{CMS:2017exb} & \href{https://rivet.hepforge.org/analyses/CMS_2017_I1512296.html}{\texttt{CMS\_2017\_I1512296}} & $|y|<2.4$, $2p_\perp<30$\,GeV\\[1mm]
      $\frac{{\rm d}^2\sigma}{{\rm d}p_\perp{\rm d}y}$,  $\frac{{\rm d}\sigma}{{\rm d}p_\perp}$,  $\frac{{\rm d}\sigma}{{\rm d}y}$ & 5.02\,TeV & LHCb~\cite{LHCb:2021pyk} & \href{https://rivet.hepforge.org/analyses/LHCB_2021_I1915030.html}{\texttt{LHCB\_2021\_I1915030}} & $2.5<y<4.5$, $p_\perp<15$\,GeV \\[1mm]
      $\frac{{\rm d}^2\sigma}{{\rm d}p_\perp{\rm d}y}$ & 7\,TeV & ATLAS~\cite{Aad:2011sp} & \href{https://rivet.hepforge.org/analyses/ATLAS_2011_I896268.html}{\texttt{ATLAS\_2011\_I896268}} & $|y|<2.4$, $5<p_\perp<70$\,GeV \\[1mm]
      $\frac{{\rm d}^2\sigma}{{\rm d}p_\perp{\rm d}y}$ & 7\,TeV & ATLAS~\cite{ATLAS:2015zdw} & \href{https://rivet.hepforge.org/analyses/ATLAS_2016_I1409298.html}{\texttt{ATLAS\_2016\_I1409298}} & $|y|<2.4$, $8<p_\perp<100$\,GeV \\[1mm]
      $\frac{{\rm d}^2\sigma}{{\rm d}p_\perp{\rm d}y}$ & 7\,TeV & LHCb~\cite{Aaij:2013nlm} & \href{https://rivet.hepforge.org/analyses/LHCB_2013_I1244315.html}{\texttt{LHCB\_2013\_I1244315}} & $2<y<4.5$, $2<p_\perp<14$\,GeV \\[1mm]
      $J/\psi$ polarization & 7\,TeV & LHCb~\cite{Aaij:2013nlm} & \href{https://rivet.hepforge.org/analyses/LHCB_2013_I1244315.html}{\texttt{LHCB\_2013\_I1244315}} & $2<y<4.5$, $2<p_\perp<15$\,GeV \\[1mm]
      $\frac{{\rm d}^2\sigma}{{\rm d}p_\perp{\rm d}y}$ & 7\,TeV & LHCb~\cite{Aaij:2011jh} & \href{https://rivet.hepforge.org/analyses/LHCB_2011_I891233.html}{\texttt{LHCB\_2011\_I891233}} & $2<y<4.5$, $0<p_\perp<14$\,GeV \\[1mm]
      $\frac{{\rm d}^2\sigma}{{\rm d}p_\perp{\rm d}y}$ & 7\,TeV & CMS~\cite{CMS:2011rxs} & \href{https://rivet.hepforge.org/analyses/CMS_2012_I944755.html}{\texttt{CMS\_2012\_I944755}} & $|y|<2.4$, $1<p_\perp<30$\,GeV \\[1mm]
      $\frac{{\rm d}^2\sigma}{{\rm d}p_\perp{\rm d}y}$ & 8\,TeV & ATLAS~\cite{ATLAS:2015zdw} & \href{https://rivet.hepforge.org/analyses/ATLAS_2016_I1409298.html}{\texttt{ATLAS\_2016\_I1409298}} & $|y|<2.4$, $8<p_\perp<100$\,GeV \\[1mm]
      $\frac{{\rm d}^2\sigma}{{\rm d}p_\perp{\rm d}y}$ & 8\,TeV & LHCb~\cite{Aaij:2013yaa} & \href{https://rivet.hepforge.org/analyses/LHCB_2013_I1230344.html}{\texttt{LHCB\_2013\_I1230344}} & $2<y<4.5$, $0<p_\perp<14$\,GeV \\[1mm]
      $\frac{{\rm d}\sigma}{{\rm d}p_\perp}$, $\frac{{\rm d}\sigma}{{\rm d}y}$  & 13\,TeV & ALICE~\cite{ALICE:2021dtt} & \href{https://rivet.hepforge.org/analyses/ALICE_2021_I1898832.html}{\texttt{ALICE\_2021\_I1898832}} & $|y|<0.9$, $0<p_\perp<40$\,GeV \\[1mm]
      $\frac{{\rm d}\sigma}{{\rm d}p_\perp}$, $\frac{{\rm d}\sigma}{{\rm d}y}$ & 13\,TeV & ALICE~\cite{ALICE:2017leg} & \href{https://rivet.hepforge.org/analyses/ALICE_2017_I1511865.html}{\texttt{ALICE\_2017\_I1511865}} & $2.5<y<4$, $p_\perp<30$\,GeV \\[1mm]
      $\frac{{\rm d}^2\sigma}{{\rm d}p_\perp{\rm d}y}$,  $\frac{{\rm d}\sigma}{{\rm d}p_\perp}$ & 13\,TeV & CMS~\cite{CMS:2017dju} & \href{https://rivet.hepforge.org/analyses/CMS_2018_I1633431.html}{\texttt{CMS\_2018\_I1633431}} & $|y|<1.2$, $20<p_\perp<150$\,GeV\\[1mm]
      \hline
      \multicolumn{5}{|c|}{$\psi(2S)$}\\
      \hline
      $\frac{{\rm d}^2\sigma}{{\rm d}p_\perp{\rm d}y}$     & 5.02\,TeV & ATLAS~\cite{ATLAS:2017prf} & \href{https://rivet.hepforge.org/analyses/ATLAS_2018_I1622737.html}{\texttt{ATLAS\_2018\_I1622737}} & $|y|<2$, $8<p_\perp<40$\,GeV \\[1mm]
      $\frac{{\rm d}^2\sigma}{{\rm d}p_\perp{\rm d}y}$ & 5.02\,TeV & CMS~\cite{CMS:2018gbb} & \href{https://rivet.hepforge.org/analyses/CMS_2019_I1672011.html}{\texttt{CMS\_2019\_I1672011}} & $|y|<2.4$, $4<p_\perp<30$\,GeV\\[1mm]
      $\frac{{\rm d}^2\sigma}{{\rm d}p_\perp{\rm d}y}$ & 7\,TeV & ATLAS~\cite{ATLAS:2015zdw} & \href{https://rivet.hepforge.org/analyses/ATLAS_2016_I1409298.html}{\texttt{ATLAS\_2016\_I1409298}} & $|y|<2.4$, $8<p_\perp<60$\,GeV \\[1mm]
      $\frac{{\rm d}^2\sigma}{{\rm d}p_\perp{\rm d}y}$ & 8\,TeV & ATLAS~\cite{ATLAS:2015zdw} & \href{https://rivet.hepforge.org/analyses/ATLAS_2016_I1409298.html}{\texttt{ATLAS\_2016\_I1409298}} & $|y|<2.4$, $8<p_\perp<60$\,GeV \\[1mm]
      $\frac{{\rm d}\sigma}{{\rm d}p_\perp}$, $\frac{{\rm d}\sigma}{{\rm d}y}$  & 13\,TeV & ALICE~\cite{ALICE:2017leg} & \href{https://rivet.hepforge.org/analyses/ALICE_2017_I1511865.html}{\texttt{ALICE\_2017\_I1511865}} & $2.5<y<4$, $p_\perp<16$\,GeV \\[1mm]
      $\frac{{\rm d}^2\sigma}{{\rm d}p_\perp{\rm d}y}$,  $\frac{{\rm d}\sigma}{{\rm d}p_\perp}$ & 13\,TeV & CMS~\cite{CMS:2017dju} & \href{https://rivet.hepforge.org/analyses/CMS_2018_I1633431.html}{\texttt{CMS\_2018\_I1633431}} & $|y|<1.2$, $20<p_\perp<130$\,GeV\\[1mm]
      \hline
      \multicolumn{5}{|c|}{$\chi_{c1}$, $\chi_{c2}$}\\
      \hline
      $\frac{{\rm d}\sigma}{{\rm d}p_\perp}$ & 7 TeV & ATLAS~\cite{ATLAS:2014ala} & \href{https://rivet.hepforge.org/analyses/ATLAS_2014_I1292798.html}{\texttt{ATLAS\_2014\_I1292798}} &  $12<p_\perp<30$\,GeV\\[1mm]
      \hline
    \end{tabular}}
  \end{center}
  \caption{Charmonium production cross-sections and polarisation measurements from LHC experiments. The data are binned in transverse momentum and rapidity. Rivet plugin references provide access to the corresponding experimental analyses.}
  \label{tab:A1}
\end{table}

\begin{table}[]
  \begin{center}
  \resizebox{0.8\textwidth}{!}{%
    \begin{tabular}{|c|c|c|c|c|}
      \hline
      Observable & Energy & Experiment & Rivet Plugin & Notes \\
      \hline
      \multicolumn{5}{|c|}{$\Upsilon(1S)$}\\
      \hline
      $\frac{{\rm d}^2\sigma}{{\rm d}p_\perp{\rm d}y}$     & 5.02\,TeV & ATLAS~\cite{ATLAS:2017prf} & \href{https://rivet.hepforge.org/analyses/ATLAS_2018_I1622737.html}{\texttt{ATLAS\_2018\_I1622737}} & $|y|<2$, $p_\perp<40$\,GeV \\[1mm]
      $\frac{{\rm d}^2\sigma}{{\rm d}p_\perp{\rm d}y}$     & 7\,TeV & ATLAS~\cite{Aad:2012dlq} & \href{https://rivet.hepforge.org/analyses/ATLAS_2013_I1204994.html}{\texttt{ATLAS\_2013\_I1204994}} & $|y|<2.25$, $0<p_\perp<70$\,GeV \\[1mm]
      $\frac{{\rm d}^2\sigma}{{\rm d}p_\perp{\rm d}y}$     & 7\,TeV & CMS~\cite{CMS:2013qur} & \href{https://rivet.hepforge.org/analyses/CMS_2013_I1225274.html}{\texttt{CMS\_2013\_I1225274}} & $|y|<2.4$, $p_\perp<50 $\,GeV \\[1mm]
      $\frac{{\rm d}^2\sigma}{{\rm d}p_\perp{\rm d}y}$     & 7\,TeV & CMS~\cite{CMS:2015xqv}  & \href{https://rivet.hepforge.org/analyses/CMS_2015_I1342266.html}{\texttt{CMS\_2015\_I1342266}} & $|y|<1.2$, $10<p_\perp<100$\,GeV \\[1mm]
      Polarization                                        & 7\,TeV & CMS~\cite{CMS:2012bpf}  & \href{https://rivet.hepforge.org/analyses/CMS_2013_I1185414.html}{\texttt{CMS\_2013\_I1185414}} & $|y|<1.2$, $10<p_\perp<50$\,GeV \\[1mm]
      $\frac{{\rm d}^2\sigma}{{\rm d}p_\perp{\rm d}y}$     & 7\,TeV & LHCb~\cite{LHCb:2012aa}  & \href{https://rivet.hepforge.org/analyses/LHCB_2012_I1091071.html}{\texttt{LHCB\_2012\_I1091071}} & $2<y<4.5$, $0<p_\perp<15$\,GeV \\[1mm]
      $\frac{{\rm d}^2\sigma}{{\rm d}p_\perp{\rm d}y}$    & 8\,TeV & LHCb~\cite{Aaij:2013yaa} & \href{https://rivet.hepforge.org/analyses/LHCB_2013_I1230344.html}{\texttt{LHCB\_2013\_I1230344}} & $2<y<4.5$, $0<p_\perp<14$\,GeV \\[1mm]
      $\frac{{\rm d}^2\sigma}{{\rm d}p_\perp{\rm d}y}$,  $\frac{{\rm d}\sigma}{{\rm d}p_\perp}$ & 13\,TeV & CMS~\cite{CMS:2017dju} & \href{https://rivet.hepforge.org/analyses/CMS_2018_I1633431.html}{\texttt{CMS\_2018\_I1633431}} & $|y|<1.2$, $20<p_\perp<130$\,GeV\\[1mm]
      \hline
      \multicolumn{5}{|c|}{$\Upsilon(2S)$}\\
      \hline
      $\frac{{\rm d}^2\sigma}{{\rm d}p_\perp{\rm d}y}$     & 5.02\,TeV & ATLAS~\cite{ATLAS:2017prf} & \href{https://rivet.hepforge.org/analyses/ATLAS_2018_I1622737.html}{\texttt{ATLAS\_2018\_I1622737}} & $|y|<2$, $p_\perp<40$\,GeV \\[1mm]
      $\frac{{\rm d}^2\sigma}{{\rm d}p_\perp{\rm d}y}$     & 7\,TeV & ATLAS~\cite{Aad:2012dlq} & \href{https://rivet.hepforge.org/analyses/ATLAS_2013_I1204994.html}{\texttt{ATLAS\_2013\_I1204994}} & $|y|<2.25$, $0<p_\perp<70$\,GeV \\[1mm]
      $\frac{{\rm d}^2\sigma}{{\rm d}p_\perp{\rm d}y}$     & 7\,TeV & CMS~\cite{CMS:2013qur} & \href{https://rivet.hepforge.org/analyses/CMS_2013_I1225274.html}{\texttt{CMS\_2013\_I1225274}} & $|y|<2.4$, $p_\perp<50$\,GeV \\[1mm]
      $\frac{{\rm d}^2\sigma}{{\rm d}p_\perp{\rm d}y}$     & 7\,TeV & CMS~\cite{CMS:2015xqv}  & \href{https://rivet.hepforge.org/analyses/CMS_2015_I1342266.html}{\texttt{CMS\_2015\_I1342266}} & $|y|<1.2$, $10<p_\perp<100$\,GeV \\[1mm]
      Polarization                                        & 7\,TeV & CMS~\cite{CMS:2012bpf}  & \href{https://rivet.hepforge.org/analyses/CMS_2013_I1185414.html}{\texttt{CMS\_2013\_I1185414}} & $|y|<1.2$, $10<p_\perp<50$\,GeV \\[1mm]
      $\frac{{\rm d}^2\sigma}{{\rm d}p_\perp{\rm d}y}$     & 7\,TeV & LHCb~\cite{LHCb:2012aa}  & \href{https://rivet.hepforge.org/analyses/LHCB_2012_I1091071.html}{\texttt{LHCB\_2012\_I1091071}} & $2<y<4.5$, $0<p_\perp<15$\,GeV \\[1mm]
      $\frac{{\rm d}^2\sigma}{{\rm d}p_\perp{\rm d}y}$    & 8\,TeV & LHCb~\cite{Aaij:2013yaa} & \href{https://rivet.hepforge.org/analyses/LHCB_2013_I1230344.html}{\texttt{LHCB\_2013\_I1230344}} & $2<y<4.5$, $0<p_\perp<14$\,GeV \\[1mm]
      $\frac{{\rm d}^2\sigma}{{\rm d}p_\perp{\rm d}y}$,  $\frac{{\rm d}\sigma}{{\rm d}p_\perp}$ & 13\,TeV & CMS~\cite{CMS:2017dju} & \href{https://rivet.hepforge.org/analyses/CMS_2018_I1633431.html}{\texttt{CMS\_2018\_I1633431}} & $|y|<1.2$, $20<p_\perp<130$\,GeV\\[1mm]
      \hline
      \multicolumn{5}{|c|}{$\Upsilon(3S)$}\\
      \hline
      $\frac{{\rm d}^2\sigma}{{\rm d}p_\perp{\rm d}y}$     & 5.02\,TeV & ATLAS~\cite{ATLAS:2017prf} & \href{https://rivet.hepforge.org/analyses/ATLAS_2018_I1622737.html}{\texttt{ATLAS\_2018\_I1622737}} & $|y|<2$, $p_\perp<40$\,GeV \\[1mm]
      $\frac{{\rm d}^2\sigma}{{\rm d}p_\perp{\rm d}y}$     & 7\,TeV & ATLAS~\cite{Aad:2012dlq} & \href{https://rivet.hepforge.org/analyses/ATLAS_2013_I1204994.html}{\texttt{ATLAS\_2013\_I1204994}} & $|y|<2.25$, $0<p_\perp<70$\,GeV \\[1mm]
      $\frac{{\rm d}^2\sigma}{{\rm d}p_\perp{\rm d}y}$     & 7\,TeV & CMS~\cite{CMS:2013qur} & \href{https://rivet.hepforge.org/analyses/CMS_2013_I1225274.html}{\texttt{CMS\_2013\_I1225274}} & $|y|<2.4$, $p_\perp<50$\,GeV \\[1mm]
      $\frac{{\rm d}^2\sigma}{{\rm d}p_\perp{\rm d}y}$     & 7\,TeV & CMS~\cite{CMS:2015xqv}  & \href{https://rivet.hepforge.org/analyses/CMS_2015_I1342266.html}{\texttt{CMS\_2015\_I1342266}} & $|y|<1.2$, $10<p_\perp<100$\,GeV \\[1mm]
      Polarization                                        & 7\,TeV & CMS~\cite{CMS:2012bpf}  & \href{https://rivet.hepforge.org/analyses/CMS_2013_I1185414.html}{\texttt{CMS\_2013\_I1185414}} & $|y|<1.2$, $10<p_\perp<50$\,GeV \\[1mm]
      $\frac{{\rm d}^2\sigma}{{\rm d}p_\perp{\rm d}y}$     & 7\,TeV & LHCb~\cite{LHCb:2012aa}  & \href{https://rivet.hepforge.org/analyses/LHCB_2012_I1091071.html}{\texttt{LHCB\_2012\_I1091071}} & $2<y<4.5$, $0<p_\perp<15$\,GeV \\[1mm]
      $\frac{{\rm d}^2\sigma}{{\rm d}p_\perp{\rm d}y}$    & 8\,TeV & LHCb~\cite{Aaij:2013yaa} & \href{https://rivet.hepforge.org/analyses/LHCB_2013_I1230344.html}{\texttt{LHCB\_2013\_I1230344}} & $2<y<4.5$, $0<p_\perp<14$\,GeV \\[1mm]
      $\frac{{\rm d}^2\sigma}{{\rm d}p_\perp{\rm d}y}$,  $\frac{{\rm d}\sigma}{{\rm d}p_\perp}$ & 13\,TeV & CMS~\cite{CMS:2017dju} & \href{https://rivet.hepforge.org/analyses/CMS_2018_I1633431.html}{\texttt{CMS\_2018\_I1633431}} & $|y|<1.2$, $20<p_\perp<130$\,GeV\\[1mm]
      \hline
    \end{tabular}}
  \end{center}
  \caption{Bottomonium production cross-sections and polarisation measurements from LHC experiments. The dataset spans multiple centre-of-mass energies, with differential distributions in transverse momentum and rapidity. Rivet plugin references link to the encoded experimental analyses.}
  \label{tab:A2}
\end{table}

\section{Herwig~7 User Interface for the Quarkonium Parton Shower}
\label{sec:AppB}

The quarkonium parton shower in \textsf{Herwig~7} is configured using the standard run-card syntax through the \texttt{SplittingGenerator} framework. Quarkonium-producing branchings are implemented as Sudakov form factors and require explicit registration within the angular-ordered parton shower. No quarkonium splittings are enabled by default, ensuring complete user control over the production content and quantum state coverage. The default setting can be activated using
\begin{verbatim}
read snippets/OniumShower.in
\end{verbatim}
Each splitting function is implemented in a plugin library (e.g.\ \texttt{HwOniumShower.so}) and instantiated as a named object, which must be appropriately configured before being registered via:
\begin{verbatim}
do SplittingGenerator:AddFinalSplitting a->b,H; <SplittingObject>
\end{verbatim}
Here, \texttt{a} and \texttt{b} denote partons, and \texttt{H} a heavy quarkonium state. Splittings are defined independently for each quantum configuration. A generic configuration sequence is given by:
\begin{verbatim}
# Create the splitting object
create Herwig::BaseSplitting MySplitting HwOniumShower.so
set MySplitting:AngularOrdered Yes
set MySplitting:StrictAO Yes
set MySplitting:InteractionType QCD

# set the principal quantum number n (default value is 1)
set MySplitting:PrincipalQuantumNumber <n>

# set QCD coupling constant for final state radiations
set MySplitting:Alpha AlphaQCDFSR

# use default non-perturbative parameters 
set MySplitting:Parameters /Herwig/OniumParameters

# set up effectively 0 pT cut (pTmin = 0.001 GeV) 
set MySplitting:Cutoff <PTCutOff>

# set the maximum value of PDF weight (default value is 2)
set MySplitting:PDFmax <PDFMaxWeight>

# set the colour structure. The choices are:
# TripletTripletSinglet or OctetOctetSinglet for colour-singlets
# OctetSinglet for colour-octets
# For diquarks
set MySplitting:ColourStructure <MyColourStructure>

# set the quarkonium structure: ccbar, bbbar, bcbar, cc, bb or bc
set MySplitting:State <MyStructure>

# register the splitting
do SplittingGenerator:AddFinalSplitting q->q,H; MySplitting

# delete the splitting
do SplittingGenerator:DeleteFinalSplitting q->q,H; MySplitting
\end{verbatim}
Non-perturbative inputs, such as colour-singlet wavefunctions and colour-octet LDMEs, are stored in a shared parameter object initialised by reading:
\begin{verbatim}
read Onium.in
\end{verbatim}
These values are bound to each splitting object using:
\begin{verbatim}
set <SplittingObject>:Parameters /Herwig/OniumParameters
\end{verbatim}
These parameters may be overridden at runtime to allow consistent tuning without recompilation:
\begin{verbatim}
cd /Herwig
do OniumParameters:SetWaveFunction <MyStructure> <nL> <value>
do OniumParameters:SetOctetProductionMatrixElement <MyStructure> <nL> <PDGcode> <value>
\end{verbatim}
Here, \texttt{<nL>} refers to the spectroscopic notation of the quarkonium state, where $n$ denotes the principal quantum number and $L$ the orbital angular momentum. For CS configurations, the relevant LDMEs depend only on these quantum numbers. In contrast, the CO MEs are process-dependent and must be specified individually for each physical final state, identified by its PDG code. Additionally, one can enhance the splitting probability of the given splitting object using
\begin{verbatim}
cd /Herwig/Shower
set <SplittingObject>:EnhancementFactor <value>
\end{verbatim}
where \texttt{value} can be a real positive number. Finally, the singlet-triplet mixing of quarkonium states can be configured using the \texttt{SetSingletTripletMixing} command. This allows users to specify the mixing angle (in degrees) for specific orbital states. 
\begin{verbatim}
cd /Herwig
do OniumParameters:SetSingletTripletMixing <nL> <mixing angle>
\end{verbatim}
The default values used in this work are $25.0^\circ$ for the \texttt{1P} and $34.4^\circ$ for the \texttt{1D} mixings.

\bibliography{herwig}
\end{document}